\documentclass[
  journal=large two,
  manuscript=Research-Article,
  year=2026,
  volume=xx,
]{cup-journal}

\usepackage{amsmath}
\usepackage[nopatch]{microtype}
\usepackage{url}
\usepackage{booktabs}
\usepackage{textcomp}
\usepackage[utf8]{inputenc}
\usepackage{newunicodechar}
\usepackage{ragged2e}
\usepackage{gensymb}
\newunicodechar{δ}{\delta}
\newunicodechar{ζ}{\zeta}
\newunicodechar{−}{\textminus}
\newunicodechar{⁰}{$^{0}$}
\newunicodechar{′}{'}
\newunicodechar{≈}{\approx}
\usepackage{multicol}
\usepackage{xcolor}

\title{\textit{\small{(This Work has been accepted to Journal of Atmospheric Sciences. The American Meteorological Society (AMS) does not guarantee that the copy provided here is an accurate copy of the Version of Record (VoR).The version is not proof corrected by the authors and is kept for self-archival purposes.)}\\
Original Article Reference: DOI: https://doi.org/10.1175/JAS-D-24-0093.1} \\
Dynamics of Heatwave Intensification over the Indian Region }

\author{\textcolor{blue}{S. Lekshmi}}
\affiliation{Monsoon Mission, Indian Inst. of Tropical Meteorology, Pune, 411008, MH, India}

\author{\textcolor{blue}{Rajib Chattopadhyay}}
\affiliation{Extended Range Prediction Group, Indian Inst. of Tropical Meteorology, Pune, 411008, MH, India}
\email[Rajib Chattopadhyay]{rajib@tropmet.res.in}

\author{\textcolor{blue}{D.S. Pai}}
\affiliation{India Meteorological Department, New Delhi, India}


\begin{document}

\begin{abstract}
In a warming world, heatwaves over India have become intense, causing severe health impacts. Earlier studies have identified two dominant modes of temperature variability over India. They are related to the development of dry (Mode 1) and moist (Mode 2) heatwaves. The present study has analysed the mechanisms for the regional intensification of the moist heatwave mode (Mode 2). From the observation, it is found that moist modes are associated with midlatitude Rossby waves intruding over the Indian region through a Europe-Middle East-Indian Ocean pathway. Sometimes, moist heatwaves are intense and are found to be associated with a strong warming over the Bay of Bengal (BoB). This BoB intensification mechanism is validated next.
Based on the observed circulation features, a barotropic vorticity model is used. The model has a prescribed mean flow background and is initialized with a large-scale wave pattern. Two types of experiments are conducted. In the control experiment, using an extratropical stationary anticyclonic forcing on the initial and the background forcing, we first simulate the Rossby wave teleconnection pathway that develops an anticyclone over India, leading to heatwaves during summer. In the sensitivity run, an additional BoB cyclonic forcing is used as per observation. Sensitivity run analysis indicates that an intensification of the Indian anticyclone (simulated in the control experiment) occurs as the wave response initiated by the BoB forcing superimposes on this anticyclone.  Further analysis under a climate change scenario shows that a range of initial wave forcing and jet speed can favour such intensifications.
. 
\end{abstract}

\section{Introduction}
Over the Indian region, intense heatwaves and heat stress are significant environmental hazards with severe impacts on different sectors. Increased exposure to excess heat I\& humidity can have adverse effects on health, which cause heat strain and discomfort, leading to heat injury (Sharma et al. 1983; Taylor et al. 2008). The driving mechanism of the intensification of heatwaves is influenced by multi-scale spatiotemporal processes. This comprises large-scale climate variabilities such as ENSO (Kenyon and Hegerl 2008), synoptic-scale features such as Rossby wave activity or blocking highs (Matsueda 2011; Meehl and Tebaldi 2004), and local factors such as land use I\& land cover, and soil moisture conditions (Alexander 2011). Over the Indian region, studies point towards the possible role of mid-latitudinal Rossby waves (Rohini et al. 2016; Ratnam et al. 2016), the impact of the Indian Ocean (Roxy et al. 2015), climate drivers such as El Niño (Murari et al. 2016; Pai and Nair 2022) and depleted soil moisture (Ganeshi et al. 2020, 2023) in driving the heat extremes. The observed trends (Perkins et al. 2012; Perkins-Kirkpatrick and Lewis 2020) as well as projections (IPCC 2013, 2021) show an increase in the intensity, frequency, and duration of heatwaves both globally and regionally.  A similar trend is observed over the Indian region (Pai et al. 2013, 2017; Rohini et al. 2016) which is also evident from the heat-related mortality rate (Guleria and Gupta 2016; Mazdiyasni et al. 2017). 

Previous studies have identified that the extratropical Rossby wave intrusion over the Indian region can cause extremes (Kalshetti et al. 2022; White et al. 2022). Different teleconnection pathways can link the Rossby wave source regions to remote locations (Simmons et al. 1983). The propagation of Rossby waves through the subtropical westerly Jetstream, which acts as a waveguide, is one such pathway (Teng and Branstator 2019). Lekshmi and Chattopadhyay 2022 (hereafter referred to as LC22) have identified two dominant modes of summer temperature variability obtained from the detrended surface temperature anomaly for April-May of 1951-2020. These dominant modes obtained using empirical orthogonal function (EOF) analysis are reproduced in \textcolor{Blue}{\textbf{Fig. 1a and Fig. 1b}}. Based on the regional impact, these large-scale modes were further attributed as the dry (Mode 1) and moist (Mode 2) heat stress modes over the Indian region (Lekshmi et al. 2024).  For Mode 1, the subtropical Rossby wave propagates through the subtropical westerly jet waveguide and reaches over the Indian region. For Mode 2, the extratropical Rossby waves propagate through the ‘Europe-Middle East-Indian Ocean pathway’ (Ambrizzi and Hoskins 1997).  This mode can cause moist heatwave conditions in the northwest Indian region. At the same time, the reverse pattern of this mode can cause moist heatwave conditions in the southeast Indian region. Both the northwest and southeast coastal regions are known for heat stress conditions in India. These modes can sometimes lead to intensified heatwaves. Then, what differentiates an intensified heatwave from a normal one? Or, from a wave propagation perspective, what differentiates the waves from causing ‘normal’ (climatology) to ‘anomalous’ fluctuations (extremes)? One of the primary synoptic-scale features associated with every heatwave is the upper-air anticyclonic circulations, which are often caused by the blocking highs (Matsueda 2011; Perkins 2015) or the stationary/propagating Rossby waves (Kornhuber et al. 2020; Parker et al. 2014). Such systems can cause ‘normal’ fluctuations and sometimes extremes such as heatwaves.  The two possible reasons for such ‘anomalous’ fluctuations are (i) amplification of the circulation or (ii) its persistence over the region for a longer duration. 

Amplification of circulation can occur as a result of many factors, such as (a)High amplitude quasi-stationary Rossby waves, (b) Quasi-resonant amplification (QRA), or (c) Superimposing waves. There has been an increase in the occurrence of concurrent heat extremes in regions in America, Europe, and Asia (Kornhuber et al. 2019, 2020; Coumou et al. 2014) due to amplified Rossby waves 5 and 7. Such ‘phase-locking’ of hemispheric patterns is favored by local topography (Jiménez-Esteve et al. 2022; Jiménez-Esteve and Domeisen 2022). Latitudinal trapping and amplification of quasi-stationary free synoptic-scale Rossby waves can occur through the quasi-resonance of free and forced waves in the midlatitude waveguides (Petoukhov et al. 2013; Kornhuber et al. 2017) which can lead to heatwaves (Rao et al. 2021). It is shown that the response to winter-time El-Niño convective forcing over the tropical Pacific is due to the combined effect of westward propagating and eastward propagating barotropic Rossby waves (Shaman and Tziperman 2016). This leads to a predominantly linear superposition of these oppositely directed waves, which are trapped inside the North African-Asian jet with weaker nonlinear effects. Such a type of linear superposition of waves with the same phase over a region can cause wave amplification, which could also lead to extremes.
Often, regional factors, e.g., the persistence of upper-air anticyclonic circulation along with positive soil moisture-temperature feedback during dry conditions, can intensify heatwaves (Seneviratne et al. 2006). The absence of any atmospheric processes that can weaken this system, such as moist air intrusion, convective activity, or the appearance of another system, can cause such systems to linger around.  The possible reasons for the highs to persist over a region are: (a) formation of stationary/quasi-stationary waves, formation of thermally/ orographically induced waves causing heat extremes (Wolf et al. 2018). Studies show that quasi-stationary waves with wave numbers 5-7 can lead to humid heat extremes in the Northern Hemisphere midlatitudes (Lin and Yuan 2022). (b) slowing down of propagating waves in a weakened jet stream waveguide: This factor is closely related to climate change and the long-term trend in observed zonal mean wind and jet streams. Recent studies have brought out the role of the weakening of the westerly jet due to the weakening of the meridional thermal gradient (Coumou et al. 2014; Francis and Vavrus 2012). This can trigger more waviness in the jet and lead to more mid-latitudinal extreme weather (Moon et al., 2022). Apart from all these factors, the long-term trends in the mean surface temperatures can cause a shift in the mean temperature. Studies have shown that a small positive shift in the average temperature can cause a disproportionate increase in the intensity and frequency of extremes (Mearns et al. 1984; Boer and Lambert 2001). However, the shift in the mean alone cannot fully explain the observed magnitude of recent extremes in the Northern Hemisphere midlatitudes (Petoukhov et al. 2013).

As discussed earlier, over the Indian region, although possible factors are identified, the role of teleconnection pathways and regional factors is not well studied for heatwave intensification. Thus, there is a need to understand the intrinsic dynamical mechanism that can cause the amplification of heatwaves, especially in the context of an increase in summer temperature in a warming scenario. Based on the earlier studies, we hypothesize that the persistence and amplification of anticyclones associated with Rossby wave intrusion over the Indian region can cause intense heatwaves. Once the anticyclone intensifies over the region, sinking air motion, along with more incoming solar radiation due to clear skies, leads to higher temperatures or heatwaves. 
Rossby Wave intrusion and subsequent intensification can be studied based on the modes identified by LC22.  The current analysis will study the relationship of heatwave intensification with teleconnection pathways and regional (local) forcing over the Indian region for Mode 2. Mode 2, which drives the moist heat stress over the Indian region, shows a significant increasing trend (Fig 1c). From Fig S1, the Mode 1 and Mode 2 reconstructed temperature (EOF × PC) anomaly is of a similar range of variability in both modes, and the number of events with PC 1 > 1.0 and PC 2 > 1.0 is almost equal, implying the similar influence of both modes in the extreme heatwave events. Moist heat stress has been reported to have significantly increased in recent years in India (Sojan and Srinivasan 2024) and is a cause of great concern for policymakers (Ghosh 2024). Therefore, the current study will analyze Mode-2, which is related to moist heat stress patterns, and Mode-1 intensification will be reported elsewhere.

\begin{figure*}[hbt!]
\centering
\includegraphics[width=0.8\linewidth]{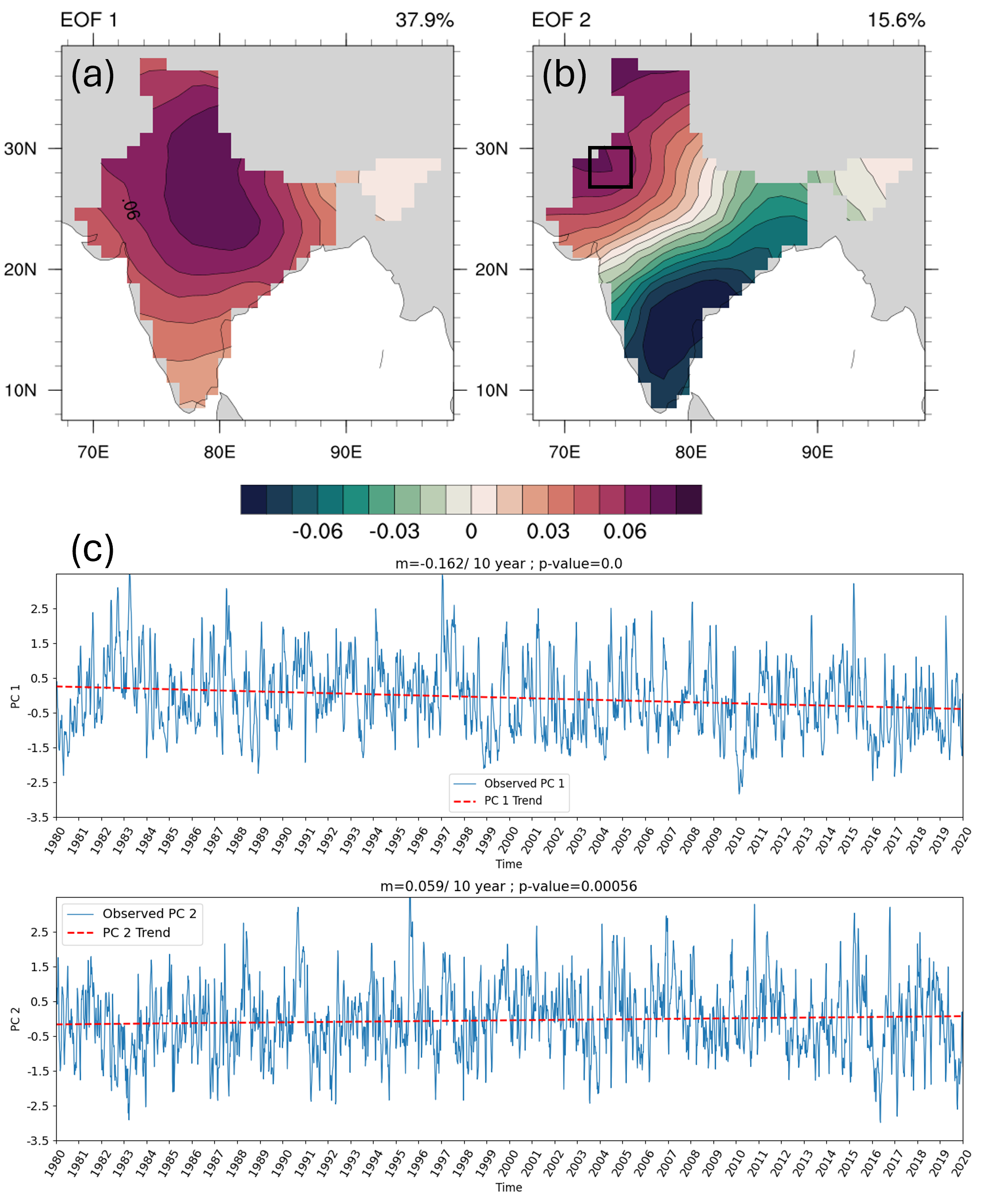}
\caption{Spatial pattern of (a) EOF Mode 1 and (b) EOF Mode 2 of the April-May detrended surface mean temperature obtained for a period from 1951-2020 (c) Time series of the principal components (PCs) obtained by projecting temperature anomaly onto EOF modes from 1980-2020 and their long-term trend. The white box in Fig.1a represents Region-1 defined in Sec. 2.}
\label{fig_wide}
\end{figure*}
In the climate change scenario, a rise in temperature over the Indian region and regional changes in moisture supply from the Bay of Bengal or the Arabian Sea (Roxy et al. 2015) are noted. Especially, the relationship between jets and the amplification of waves has been debated in the warming scenario (White et al. 2022). While some studies say that weaker jets lead to stronger waves (Francis and Vavrus 2012), others propose that weaker jets lead to weaker amplitudes of mid-latitudinal waves (Manola et al. 2013). Such a wide range of scenarios may impact heatwave intensification. In the present scenario, climate models are projecting a rise in the heatwave intensity, duration, and frequency over the Indian region (Dubey and Kumar 2023; Goyal et al. 2023). As mentioned earlier, the increasing trend of mode 2 makes it important to study such types of wave interactions (tropical-extratropical waves as well as wave-waveguide), which help in understanding extreme event dynamics. But what causes heatwave intensification on the subseasonal scale? Factors, such as the Rossby waves, stationary wave pattern, zonal wind, jet speed, wave and waveguides, etc., can impact the subseasonal scale heatwave intensification. 

For understanding the formation, propagation, and intensification of Rossby waves in response to thermal or orographic forcing, studies have relied on the non-divergent equivalent barotropic model for a long time (Hoskins and Karoly 1981; Karoly 1983; Hoskins and Ambrizzi 1993; Grimm and Dias 1995). Especially, downstream amplification of Rossby waves in a barotropic setting with jets as background and stationary as well as dissipative forcing is studied in several literature (Segalini et al. 2024). Based on the ratio of barotropic to baroclinic vorticity defined as | ($\zeta_{850}+\zeta_{200})/(\zeta_{850}-\zeta_{200})$ |, it is found that over most of the Indian region and the midlatitudes, the barotropic mode is dominant compared to the baroclinic mode during the days of active Mode 2 (PC 2 > 1.0) (Fig S2). The vertical vorticity profile over the Northwest Indian region, where heatwave occurs, confirms this observation (\textcolor{Blue}{\textbf{Fig. 4e}} to be discussed later). Hence, in this study, we use the non-divergent barotropic vorticity equation as a toy model to understand the dynamical mechanism associated with modal teleconnection pathways and its intensification due to remote and regional forcing over the Bay of Bengal. From observations and from earlier modelling studies (Ting and Yu 1998), it is known that a local heat source can provide favorable conditions for modal intensification. In this study, the role of such heat sources, which are predominantly barotropic in nature (Fig 2a), in driving the Rossby wave intrusion and heatwave intensification over India will be explored in the context of the Europe-Middle East-Indian Ocean pathway as described earlier.We believe this analysis will throw some light on the possible processes that cause occasional amplification of the modes of sub-seasonal variability, leading to the intensification of heatwaves, and some insights into the same in the context of climate change.

\section{Data}
The observational patterns associated with the mode and its intensification are analyzed using different meteorological parameters. The maximum and minimum temperatures are obtained from the India Meteorological Department gridded data of 1$^{\circ}$ ×1$^{\circ}$ resolution (Srivastava et al. 2009). The daily mean temperature is obtained as the mean of the maximum and minimum temperature data. The maximum temperature and daily mean temperature data are used for analysis. Apart from that, the NOAA Daily Optimum Interpolation Sea Surface Temperature (OI SST) V2 (Huang et al. 2021) of spatial resolution 0.25$^{\circ}$ × 0.25$^{\circ}$ is used for understanding the possible role of regional forcing in the modal amplification over the Indian region. Also, the dynamical features associated with the mode are analyzed based on the upper-level (200hPa) relative vorticity pattern and the vertically integrated (1000-300hPa) moisture flux over the Indian region. The zonal and meridional winds, specific humidity, and mean sea level pressure required for the analysis are obtained from the NCEP DOE Reanalysis 2 dataset, of  2.5$^{\circ}$ × 2.5$^{\circ}$ resolution (Kanamitsu et al. 2002). The heat stress is assessed based on the NOAA National Weather Service (NWS) Heat Index (https://www.weather.gov/safety/heat-index), which is obtained from air temperature and relative humidity (RH). The values are obtained using the inbuilt function in the Python MetPy library: \url{https://unidata.github.io/MetPy/latest/api/generated/metpy.calc.heat\_index.html.} The observed air temperature and RH data for calculating the Heat Index (HI) are obtained from the NCEP DOE Reanalysis 2 dataset. The EOFs are obtained from the detrended temperature data of April-May from 1951-2020 (discussed in detail in LC22). All the other observational analyses have been carried out for the period from 1980 to 2020 for the months of April and May, which are the months when more frequent and intense heatwaves occur over India. The anomalies of the observed variables shown are obtained by removing the daily climatology of the period from 1980-2020. For the model experiments, only the actual values are shown; no climatology was removed. The regions defined as follows are considered for the analysis of intensification occurring over the Indian region: (i) Region 1 showing intensification in observation: 27-30 $^{\circ}$N; 72-75 $^{\circ}$E (refer to the white box in \textcolor{Blue}{\textbf{Fig.1b)}}  (ii) Region 2 showing intensification in model: 15-40 $^{\circ}$N and 70-90 $^{\circ}$E (iii) Vorticity Forcing regions in model Source Region 1 and Source Region 2 (detailed explanation in Section 4).
Before proceeding further, it is important to bear in mind that this study intends to explain only the mode (i.e., moist heatwave mode)-related intensification of circulation (i.e., anti-cyclonic vorticity) and, in turn, the surface temperature and heat stress over India and does not address any other modes of heatwave intensification. Henceforth, ‘moist heatwave mode’ refers to the second dominant mode of summer temperature intraseasonal variability (EOF 2), and the principal component (i.e., PC 2) associated with this mode will be referred to as PC unless specified otherwise.

\begin{figure*}[hbt!]
\centering
\includegraphics[width=0.8\linewidth]{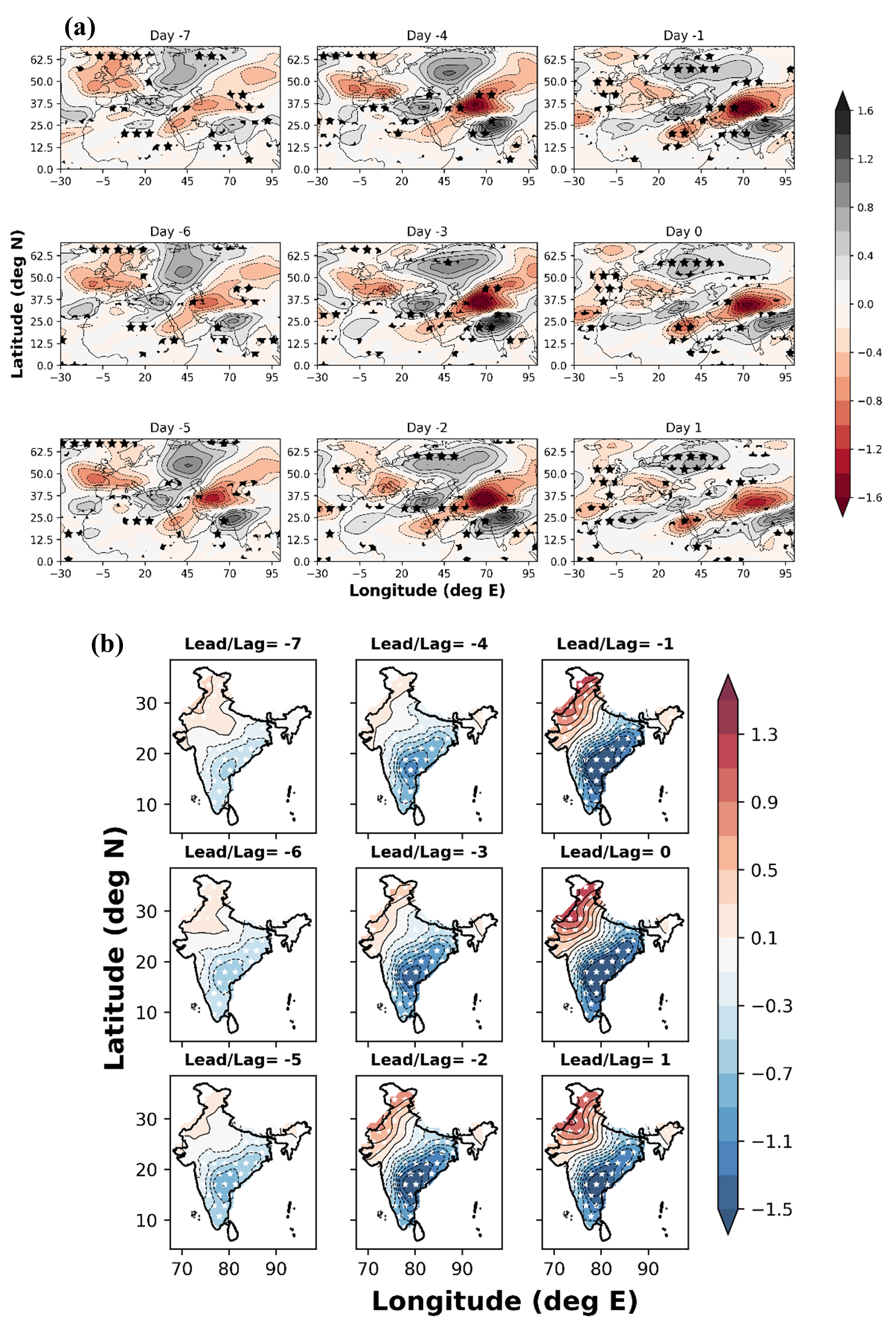}
\caption{(a) Spatial composite of relative vorticity anomaly (s-1) at 200 hPa for active mode days (PC  >1.0) from lag -7 to lead 1 day with respect to active mode days. (b) Spatial composite of surface temperature anomaly (⁰C) for active mode days in the same way as shown in Fig 3a. The star represents the statistically significant regions.}
\label{fig_wide}
\end{figure*}

\begin{figure*}[hbt!]
\centering
\includegraphics[width=0.8\linewidth]{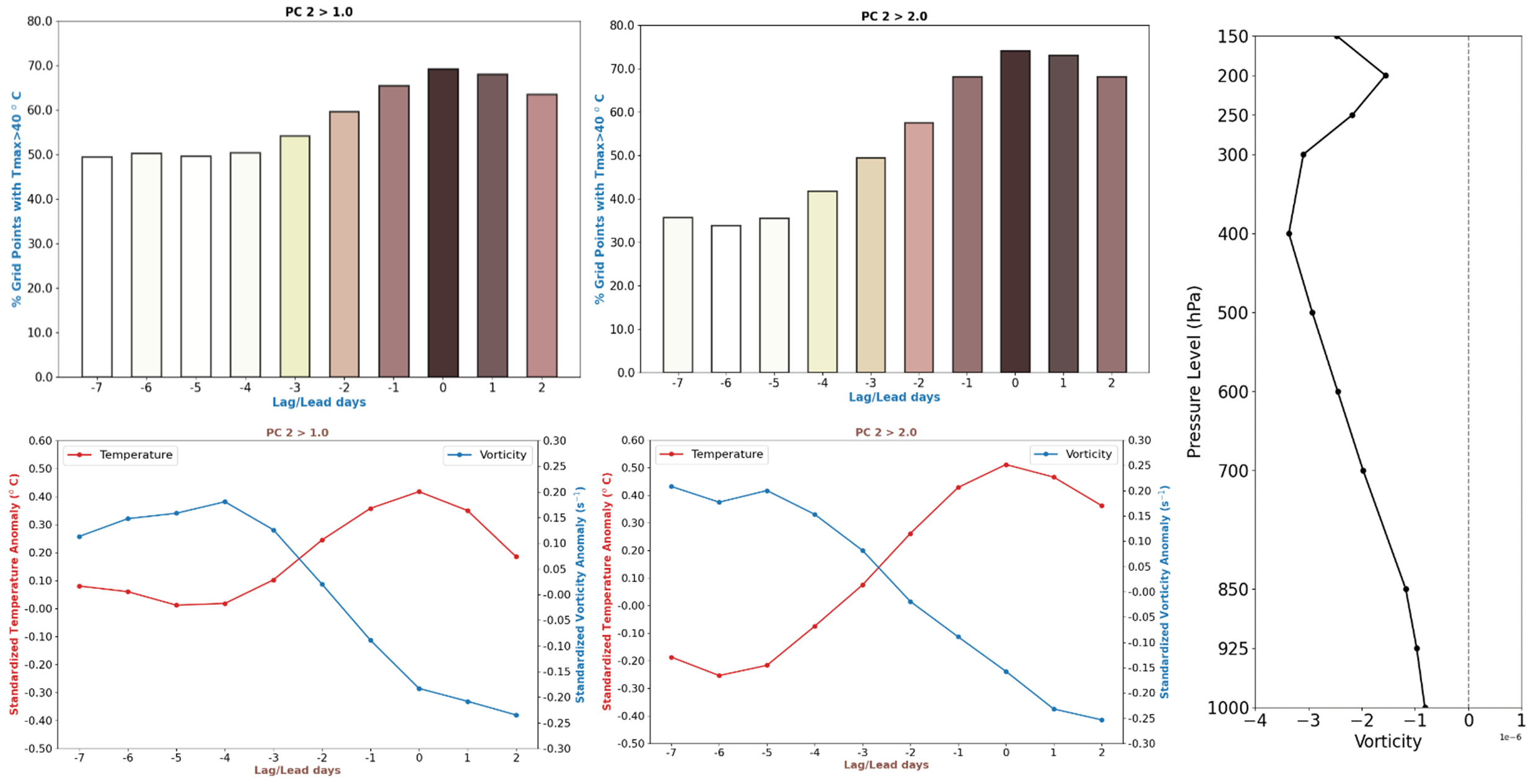}
\caption{Composite of the standardized anomaly of relative vorticity at 200 hPa and maximum temperature for (a) Active mode days (PC  > 1.0) and (b) Extreme mode days (PC  > 2.0) over Region 1 (marked in lag=0 of Fig 3b as a white box) during different lags during April-May from 1980-2020. Percentage number of grid points at different lags with maximum temperature > 40 ⁰C for (c) Active mode and (d) Extreme mode days. (e) Vertical profile of area-averaged observed relative vorticity over Region 1 for active mode days}
\label{fig_wide}
\end{figure*}

\section{Observed Modal Intensification over the Indian Region}
In this section, the modal signatures in the dynamic and thermodynamic parameters, representing the observed intensification over the Indian region, are analyzed. The composite of surface temperature anomaly and relative vorticity anomaly at 200hPa is shown for different lags with respect to PC > 1.0 (active Mode) days. A wave train propagates from midlatitudes to the Indian region (\textcolor{Blue}{\textbf{Fig. 2a}}) and intensifies from lag-5 onwards over the northern parts of the Indian region (Also refer to \textcolor{Blue}{\textbf{Fig. 3a- 3d}}, discussed next). \textcolor{Blue}{\textbf{Fig. 2b}} shows the spatial composite of temperature anomaly, which depicts the rise in temperature over the North-West Indian (NWI) region, especially from lag-4 days onwards, clearly responding to the intensified circulation pattern. Also, the composite of average vorticity over Region 1, as shown in a white square region in \textcolor{Blue}{\textbf{Fig. 1b}} at 200hPa, and standardized maximum temperature anomaly are plotted from lag -7 to lead 2 with respect to the active and extreme mode (PC > 2.0) days (\textcolor{Blue}{\textbf{Fig. 3}}). The selected region is based on the observed EOF pattern (\textcolor{Blue}{\textbf{Fig 1b)}} and the observed composite anomaly patterns, as seen in Fig 2. In both cases, an apparent intensification of upper-level vorticity on the negative side (depicting anticyclonic circulation) and associated rise in standardized maximum temperature anomaly is evident over Region 1 (\textcolor{Blue}{\textbf{Fig. 3a}} and \textcolor{Blue}{\textbf{3b}}) from lag -5 onwards. \textcolor{Blue}{\textbf{Fig 3c}} and \textcolor{Blue}{\textbf{3d}} show the percentage of grid points that showed maximum temperature value greater than 40 ⁰C (Tmax> 40 ⁰C) at different lag/leads for the same region as in \textcolor{Blue}{\textbf{Fig 3a}} and \textcolor{Blue}{\textbf{3b}}.  It shows a gradual rise in the spatial extent of the region with Tmax> 40 ⁰C from lag -7 onwards (\textcolor{Blue}{\textbf{Fig 3c}}), with a steeper rise for extreme days (\textcolor{Blue}{\textbf{Fig 3d}}). Thus, the modal circulation shows a distinct intensification pattern resulting in a rise in temperature over the Indian region. The anticyclonic vorticity structure (\textcolor{Blue}{\textbf{Fig. 2a}}) observed over the Northwest Indian region during active mode days clearly shows barotropic structure as seen in \textcolor{Blue}{\textbf{Fig. 3e}}.  It is also to be noted from \textcolor{Blue}{\textbf{Fig. 2a}} that there is a strong cyclonic circulation associated with the mode along the Eastern Coastal regions in India.
\begin{figure*}[hbt!]
\centering
\includegraphics[width=0.8\linewidth]{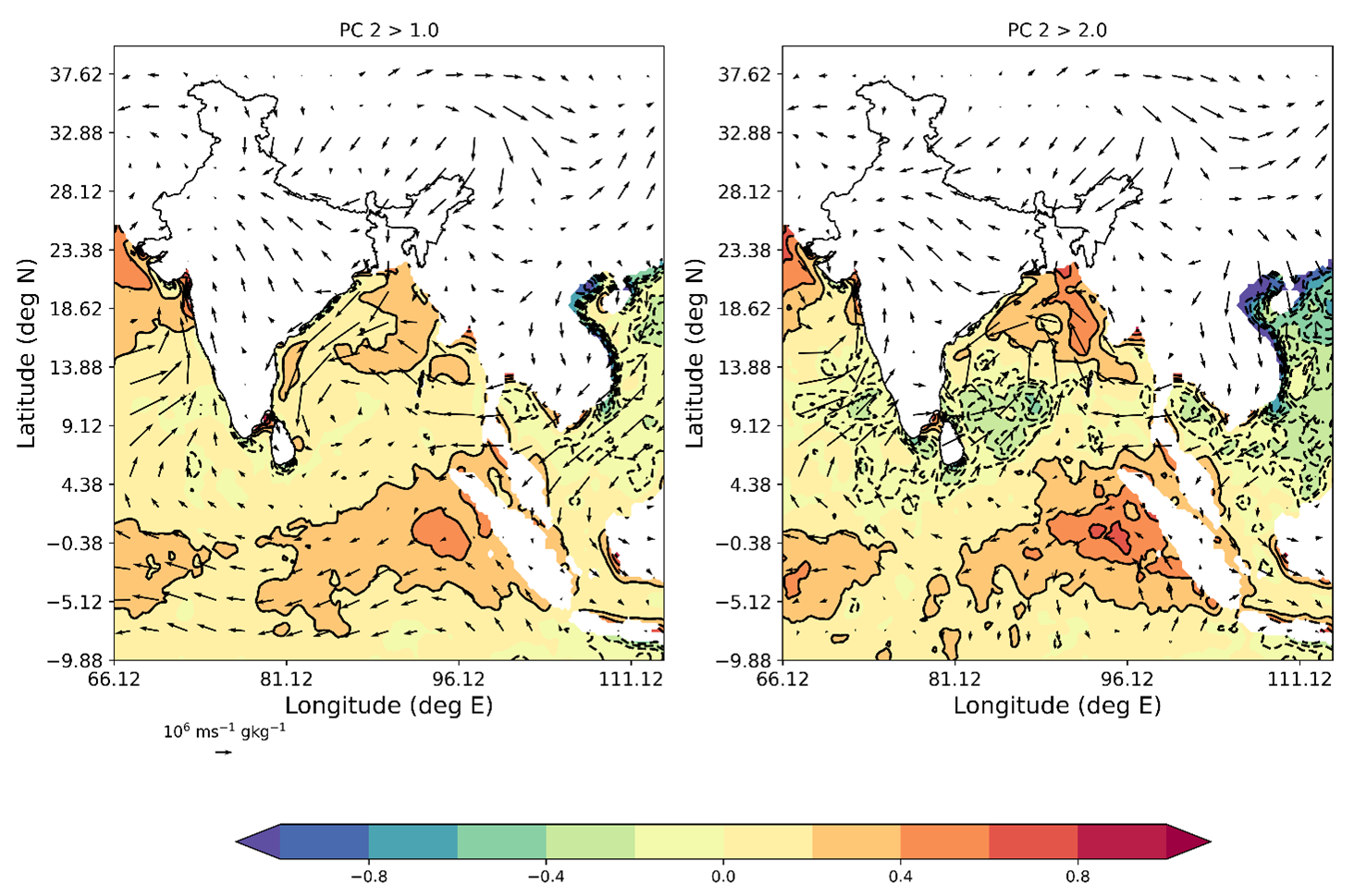}
\caption{Composite of sea surface temperature anomalies (⁰C; shaded contour) superimposed with 1000-300 hPa vertically integrated moisture flux (IMF; vectors) anomalies ($\mathrm{ms}^{-1}\,\mathrm{g}\,\mathrm{kg}^{-1}$) for (a) Active mode days (PC > 1.0) and (b) Extreme mode (PC > 2.0) days during April-May from 1980-2020. The anomalies are calculated with respect to daily climatology from 1980-2020. The statistically significant regions of the SST composite are given in Fig S3.}
\label{fig_wide}
\end{figure*}
The sea-surface temperature (SST) and vertically integrated moisture flux (IMF) anomalies for active and extreme mode days are also given in \textcolor{Blue}{\textbf{Fig. 4}}. The SST shows a distinct anomalous warming over the southernmost BoB, especially close to Indonesia. Apart from that, there is a strong moisture influx towards the eastern coastal and central Indian regions. This hints towards two possible conditions associated with the moist heatwave mode and its intensification: (a) Warm SST (especially in the southernmost BoB) can act as a regional heat source, and (b) strong regional scale circulation pattern (as seen from IMF) can bring moisture supply to Indian landmass which could lead to humid heat stress (given the high-temperature conditions). To understand the modal teleconnection pathway and its intensification over the Indian region, the role of vorticity forcing in midlatitudes (F1) and the one in the BoB (F2 ), as described in detail in Section 4, which is consistent with the warm SST pattern, will also be considered. We describe this scenario (i.e., intensification related to the teleconnection pathway along with a regionally forced circulation) based on circulation obtained using a non-divergent barotropic model aided by a regression-based model to explore moisture and temperature, as discussed in detail in the following section. 

\begin{figure*}[hbt!]
\centering
\includegraphics[width=0.8\linewidth]{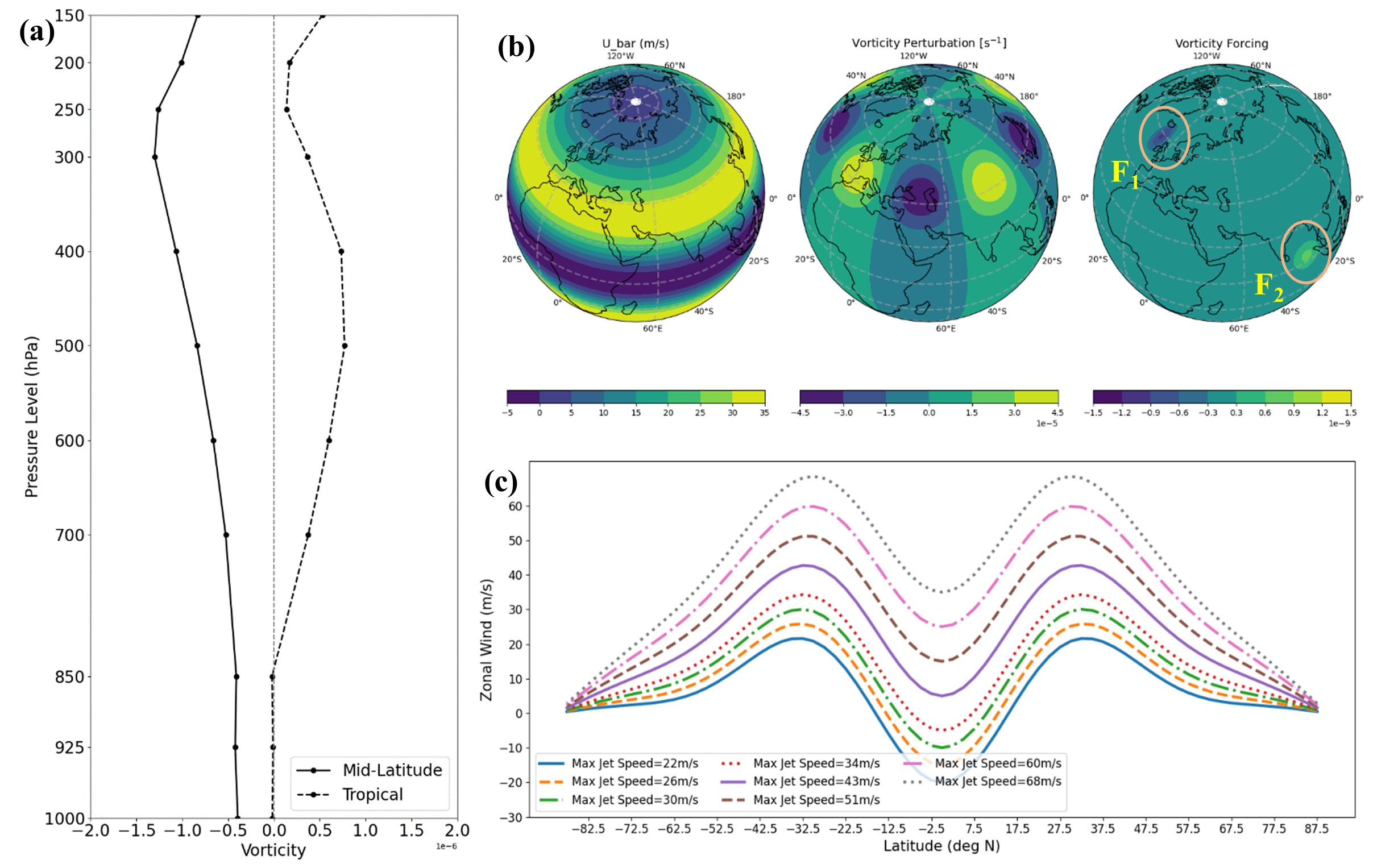}
\caption{(a) Vertical profile of the relative vorticity anomaly (s$^{-1}$) composite for those days when PC 2 $>$ 1.0 for two regions, Source Region 1 (25--5$^{\circ}$W; 50--70$^{\circ}$N) and Source Region 2 (0-20$^{\circ}$N; 80--95$^{\circ}$E). (b) Set of background initial conditions: zonal mean wind ($\bar{u}$; m/s), vorticity perturbation ($\zeta^{\prime}$) and the locations of the vorticity (s$^{-1}$) forcing F1 (Source Region 1) and F2 (Source Region 2) applied in the equivalent barotropic model for the experiments. (c) Mean zonal wind configurations $\bar{u}(j_{\max})$ used for sensitivity experiments TS (z, $j_{\max}$) and IS (z, $j_{\max}$).}
\label{fig_wide}
\end{figure*}

\section{Model}
\subsection{Model Description}
As mentioned earlier, the dynamical mechanism by which the Rossby wave associated with the mode is driven and propagates towards the Indian region is investigated using a simple equivalent barotropic model framework and an empirical model (based on linear regression). The whole model framework has (i) A dynamical model in which vorticity forcing over different regions and subsequent response of the forcing(s) over the Indian region, especially in the intensification of anticyclonic vorticity, is considered using the equivalent barotropic vorticity equation with prescribed initial conditions. (ii) A simple Multilinear Regression model is used next, through which the equivalent barotropic model response is evaluated for the temperature, RH, and HI. The whole model is set up in such a way as to give a step-by-step strategy in a simplified framework, beginning from understanding the propagation of the mode-related anticyclonic circulation and its persistence over the Indian region to the intensification of circulation and, finally, its impact on the regional heat stress. Such a hybrid dynamical-statistical modelling framework makes evaluating the temperature field and further heat stress possible without explicitly considering a baroclinic dynamical model.

\begin{figure*}[hbt!]
\centering
\includegraphics[width=0.8\linewidth]{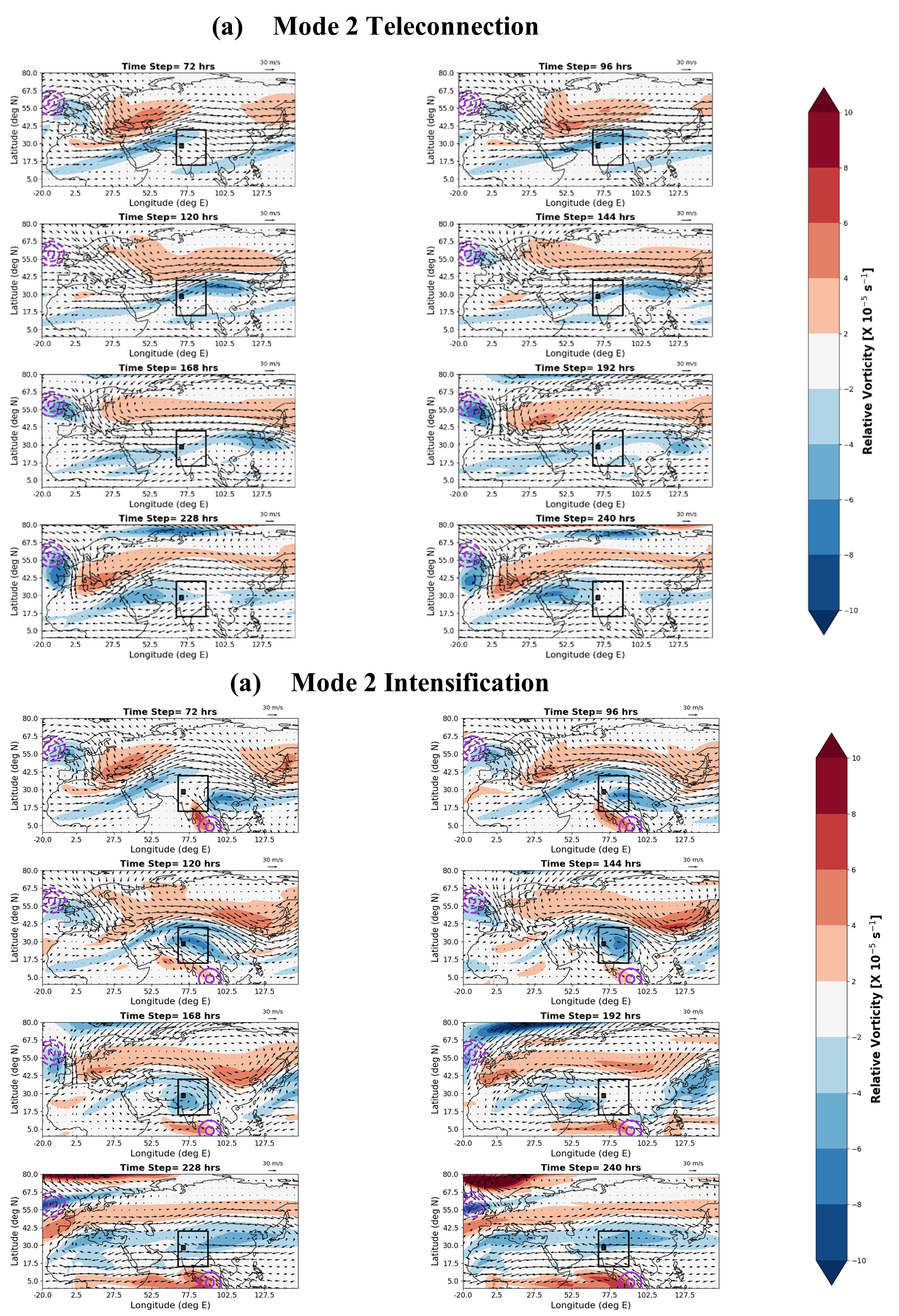}
\caption{Spatial pattern of equivalent barotropic model predicted relative vorticity and wind vectors at different time steps from t=72 hrs to t=192 hrs for the control experiments (a) T(4,34) Teleconnection and (b) I(4,34) Intensification experiment. The dotted circle shows the anticyclonic forcing F1, and the closed circle represents cyclonic forcing F2. The shaded part represents only the negative relative vorticity (anti-cyclonic vorticity). The black outlined box represent the Region 2 and the inner small black filled box in northwest India inside the large box represent the area averaged in Fig 3.}
\label{fig_wide}
\end{figure*}
	\section*{(i) Dynamical Model}

The dynamical model here takes the zonal and meridional wind as input and calculates its non-divergent component. The atmospheric flow evolves based on the barotropic vorticity equation from the input vorticity.  The barotropic vorticity equation, given in Eq. 1, is integrated using a prescribed set of initial conditions (ICs) such as the zonally mean zonal ($\overline{u}$) and meridional ($\overline{v}$) wind (first panel in Fig 5b), initial vorticity perturbations ($\zeta'$) from which perturbation winds \textit{u}′ and \textit{v}′ are obtained with a preferred zonal wave number pattern (\textit{z}) and any additional stationary vorticity forcing (Fig 5b). The zonal wind shows a jet at midlatitudes, and the wave perturbations are imposed in these mean patterns (see below). For all the experiments $\overline{v}=0$ (which is not shown in Fig 5b). Here, the overbar denotes the zonal mean.

\begin{equation*}
\frac{\partial \zeta}{\partial t}=-J(\psi, f+\zeta)=-\beta v-J(\psi, \zeta)=-\frac{2 \Omega}{a^{2}} \frac{\partial \psi}{\partial \lambda}-J(\psi, \zeta) \tag{1}
\end{equation*}

where $f=2 \Omega \sin \theta, \beta=\frac{2 \Omega \cos \theta}{a}$ is the meridional gradient of $f$ and Jacobian, $J$ is defined as

$$
J(A, B)=\frac{1}{a^{2} \cos \theta}\left(\frac{\partial A}{\partial \lambda} \frac{\partial B}{\partial \theta}-\frac{\partial B}{\partial \lambda} \frac{\partial A}{\partial \theta}\right)
$$

In the above equations, $a, \psi, \zeta, \Omega, \theta$ and $\lambda$ are the radius of Earth, stream function, relative vorticity, angular velocity, latitude, and longitude, respectively. A biharmonic diffusion term is added to the vorticity tendency, and eq (1) becomes,

\begin{equation*}
\frac{\partial \zeta}{\partial t}=-\frac{2 \Omega}{a^{2}} \frac{\partial \psi}{\partial \lambda}-J(\psi, \zeta)-v(-1)^{n} \nabla^{2 n} \zeta \tag{2}
\end{equation*}

where $\mathrm{n}=2$ and $v=1.338 \times 10^{16} \mathrm{~m}^{4} \mathrm{~s}^{-1}$ is the coefficient.\\
A zonally symmetric mean flow \textcolor{Blue}(Fig. 5c) of the form of Eq. (3) is imposed in the model.

\begin{equation*}
\bar{u}(\theta)=A \cos \theta-B \cos ^{3} \theta+C \sin ^{2} \theta \cos ^{6} \theta \tag{3}
\end{equation*}

where $\mathrm{A}=25$, $\mathrm{B}=30$ and $\mathrm{C}=300 \mathrm{~m} \mathrm{~s}^{-1}$ as used by (Held 1985). Along with that, a vorticity perturbation $\zeta^{\prime}(z)$ of the form given below (Held 1985) is provided as an initial condition (middle panel of \textcolor{Blue}{Fig. 5b}).

\begin{equation*}
\zeta^{\prime}(z, t=0)=D * \cos (\theta) e^{-\left[\left(\theta-\theta_{0}\right) / \theta_{w}\right]^{2}} e^{i z \lambda} \mathrm{~s}^{-1} \tag{4}
\end{equation*}

where $\theta_{0}=45^{0}, \theta_{w}=15^{0}$ and $z=4$ for the control experiment. This zonal wave number 4 eddy can evolve freely on a stable mid-latitudinal zonal jet. $\boldsymbol{D}$ is a constant and is fixed for all the experiments and is taken to be $6 \times 10^{-5}$, so that $\mathrm{D}^{*} \cos \theta$ at $\theta=45^{\circ}=\theta_{0}$ is maximum up to $\sim 4 \times 10^{-5}$. The value is chosen based on the maximum amplitude of the anticyclonic relative vorticity pattern between $30^{\circ}-45^{\circ} \mathrm{N}$ during strong heatwave days over India, as shown in \textcolor{Blue}{Fig. 8 c} of Lekshmi and Chattopadhyay (2022). Also, we selected z=4 based on \textcolor{Blue}{Fig. 3} of the same study and some other studies which have shown the role of stationary wave 4 as a precursor to heatwaves, e.g.,(Petoukhov et al. 2013, 2018; Jiménez-Esteve et al. 2022; Yang et al. 2024). The aim of this study is to understand the intensification of the circulation that can lead to intensified heatwaves, and thus, the zonal wavenumber pattern 4 is suitable to be used in the control experiment. Other experiments with $z$ varying from 1 to 8 are also performed.

Apart from the initial vorticity perturbation $\left(\zeta^{\prime}\right)$, a stationary circular Rossby wave forcing, $\boldsymbol{F}_{1}$, in the midlatitudes (Source Region 1) and similar vorticity forcing in the tropics over the Bay of Bengal (BoB), $\boldsymbol{F}_{2}$ (Source Region 2; also mentioned as 'regional forcing' as it is close to the Indian domain under consideration) are used for different experiments (refer to the last panel of Fig 5b). Both the forcings are circular vorticity forcing, where $F_{1}$ is anticyclonic and $F_{2}$, which acts as a local source for the Indian region, is cyclonic in nature. Considering these sources, the vorticity tendency is obtained as,

\begin{equation*}
\frac{\partial \zeta}{\partial t}=-\frac{2 \Omega}{a^{2}} \frac{\partial \psi}{\partial \lambda}-J(\psi, \zeta)-v(-1)^{n} \nabla^{2 n} \zeta+\sum_{i=1,2} F_{i} \tag{5}
\end{equation*}

The stationary circular forcing $F_{1}$ placed at the midlatitude, centered at $58.75{ }^{\circ} \mathrm{N}, 13.75{ }^{\circ} \mathrm{W}$, has its amplitude defined using Eq. (6) with $\mu$ as the center of the forcing location, $r$ as the radial distance in degrees from the center with $0^{\circ}<\mathrm{r} \leq 22.5^{\circ}$ ( 10 grid points) and $\sigma=0.5^{\circ}$ for all the experiments. Similarly, the stationary circular forcing $F_{2}$, which is in the tropical region over the southernmost BoB , is centered at $3.75^{\circ} \mathrm{N}, 91.25^{\circ} \mathrm{E}$, and the amplitude is defined using Eq. (6).

\begin{figure*}[hbt!]
\centering
\includegraphics[width=0.8\linewidth]{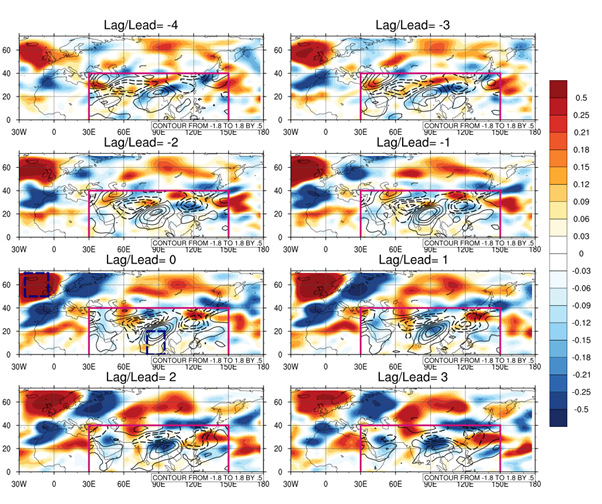}
\caption{Spatial pattern of equivalent barotropic model predicted relative vorticity and wind vectors at different time steps from t=72 hrs to t=192 hrs for the control experiments (a) T(4,34) Teleconnection and (b) I(4,34) Intensification experiment. The dotted circle shows the anticyclonic forcing F1, and the closed circle represents cyclonic forcing F2. The shaded part represents only the negative relative vorticity (anti-cyclonic vorticity). The black outlined box represent the Region 2 and the inner small black filled box in northwest India inside the large box represent the area averaged in Fig 3}
\label{fig_wide}
\end{figure*}

\begin{equation*}
F_{i}=P \times e^{-\frac{(r-\mu)^{2}}{2 \sigma^{2}}} \mathrm{~s}^{-2} \tag{6}
\end{equation*}

where $\mathrm{P}=-1.0 \times 10^{-9}$ for $F_{1}$ and $\mathrm{P}=0.7 \times 10^{-9}$ for $F_{2}$. It is to be noted that $F_{1}, F_{2}$, and $\zeta^{\prime}$ are prescribed directly to the model (i.e., not derived from winds), and the zonally mean winds ( $\bar{u}$ and $\bar{v}$ ) are prescribed from which the model derives mean vorticity conditions. For the integration, apart from the first-time step, the model uses the leapfrog scheme followed by the Robert-Asselin time filter to control time splitting, as in eq (7), where $\mathrm{r}=0.2$ is the coefficient for the Robert-Asselin filter. The model has a grid resolution of $2.5^{0} \times 2.5^{0}$.

\section*{(ii) Linear Regression Model}

A multilinear regression model is used to estimate the maximum surface temperature (T) and surface relative humidity (RH) from the relative vorticity patterns. In the regression, along with the relative vorticity at a particular grid point, the spatial gradients of the relative vorticity of that grid from its nearest four neighbors are also used as predictors. The predictands are the T and RH at that grid point. Mathematically, the T and RH at any grid point (x, y) at a time step, t is obtained as,

$$
\begin{aligned}
& T(x, y)=\beta_{1}^{T}[\zeta(x, y)-\zeta(x-1, y)]+\\
&\beta_{2}^{T}[\zeta(x, y)-\zeta(x+1, y)]+
\beta_{3}^{T}[\zeta(x, y)-\zeta(x, y) \\
& \quad+1)]+\beta_{4}^{T}[\zeta(x, y)-\zeta(x, y-1)]+
\beta_{5}^{T} \zeta(x, y)+C_{T}(x, y) \\
& R H(x, y)=\beta_{1}^{R H}[\zeta(x, y)-\zeta(x-1, y)]+\\
&\beta_{2}^{R H}[\zeta(x, y)-\zeta(x+1, y)]+\beta_{3}^{R H}[\zeta(x, y) \\
& \quad-\zeta(x, y+1)]+\beta_{4}^{R H}[\zeta(x, y)-\zeta(x, y-1)]\\
&+\beta_{5}^{R H} \zeta(x, y)+C_{R H}(x, y)
\end{aligned}
$$

where $\beta_{1,2,3,4,5}^{T}$ and $C_{T}$ are the regression coefficients and constant for maximum temperature $\beta_{1,2,3,4,5}^{R H}$ and $C_{R H}$ are the same for RH .

\begin{figure*}[hbt!]
\centering
\includegraphics[width=0.8\linewidth]{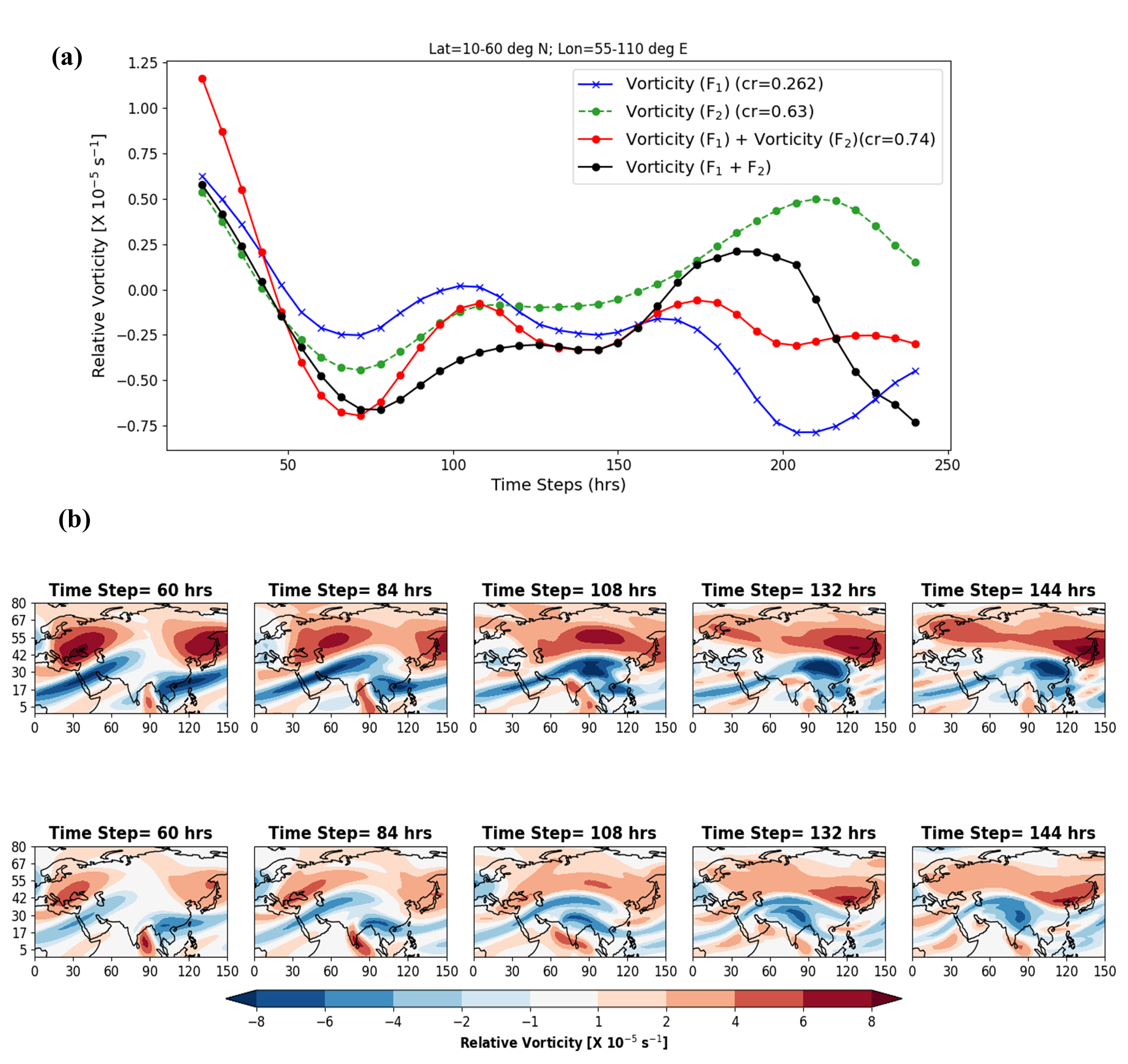}
\caption{(a) Area-averaged (5-60 ⁰N; 55-110 ⁰E) relative vorticity (× 10-5 s-1) predicted by model experiment I(4,34), i.e., Vorticity (F1+F2) and compare with Vorticity (superposition), i.e., Vorticity (F1) + Vorticity (F2), Vorticity (F1) and Vorticity (F2) separately at different model time steps. The correlation of these experiments with Vorticity (F1+F2) is mentioned in the plot.  (b) The spatial pattern of the same as in Fig 8a at different model time steps. The upper panel shows Vorticity (superposition), and the lower panel shows Vorticity(I(4,34)).}

\label{fig_wide}
\end{figure*}

The regression coefficients and constants are obtained from the observed relative vorticity at the upper level ( 200 hPa ), the surface maximum temperature, and RH for a period from 1980 to 2020, from those days with PC $2>1.0$. The F-test was used to measure the accuracy and significance of the regression method. The null hypothesis is that the predictand 'temperature' / 'RH' cannot be estimated from the predictors, i.e., the $\zeta$ and gradient of $\zeta$. The ratio of the variance explained by the regression to the unexplained variance was found based on the overall F-test. The average F-ratio value over India is $\sim 8$ (for T ) and $\sim 4$ (for RH ) and was found to be significant with $90 \%$ confidence (plots not shown). Thus, the predictand (T \& RH) can be estimated using $\zeta$ and spatial gradients of $\zeta$. Using these coefficients and constants, the model predicted maximum temperature and RH are estimated from the barotropic modelpredicted vorticity pattern. Once the maximum temperature and RH are estimated, the NOAA NWS Heat Index is then calculated.

\subsection{Model Experiment Design}
To understand the intensification of Indian heatwaves associated with the midlatitude teleconnection pathway and local factors such as a stationary forcing over the Bay of Bengal, four sets of experiments are designed. The basic purpose of these four experiments is to show how the “Europe-Middle East-Indian Ocean teleconnection pathway” (Ambrizzi and Hoskins 1997) is amplified and is sensitive to large-scale and imposed stationary forcing. They are described below: 

\textcolor{blue}{(a) Midlatitude Teleconnection Experiment} T(4,34), where the mode-related Rossby wave is generated by placing the mid-latitudinal (Source Region 1) vorticity forcing (F1) above the model base state ($\bar{u}$) with j$_{max}$ (maximum jet speed)=34m/s, which corresponds to A=25 in Eq (7), and vorticity perturbation ($\zeta^{\prime}$) with z (zonal wave number)=4. This identifies the extratropical to tropical teleconnections (mentioned as the ``Europe-Middle East-Indian Ocean teleconnection pathway'' by Ambrizzi and Hoskins 1997). 

\textcolor{blue}{(b) Teleconnection Sensitivity Experiments} TS(z, j$_{max}$), in which a set of experiments is designed that study the role of different jmax (i.e., $\bar{u}$) and z (i.e., $\zeta^{\prime}$) in driving the Rossby wave associated with the mode and its persistence over the Indian region is done. 

\textcolor{blue}{(c) Teleconnection mediated local intensification Experiment} I(4,34), where a regional forcing in the southernmost BoB (F2) is added to T(4,34) experiment.

\textcolor{blue}{(d) Intensification Sensitivity Experiments} IS(z, j$_{max}$), where the role of different $\bar{u}$ and z in causing the intensification of the modal circulation over the Indian region is done. Different $\bar{u}$ configurations are achieved by varying the constant A in eq (7), which is shown in Fig 4c.

\textcolor{blue}{(e) Regional Forcing Experiment:} A regional forcing experiment R(4,34) in which only the cyclonic vorticity forcing over the southernmost BoB is given. All the model parameters are similar to I(4,34) except that no mid-latitudinal forcing F1 exists, but only BoB forcing F2 is given. 

Experiments (a) and (c) are control experiments, while sets (b) and (d) are the respective sensitivity experiments. In the control experiment (all combinations of T(z,jmax)) the previously identified teleconnection pathway is simulated and the sensitivity experiments (all combinations of I(z,jmax)) study the intensification of this pathway in presence of a regional forcing in BoB Experiment (e) will be considered for the explanation of the intensification process over the Indian region. The summary of the set of experiments is provided in Table 1. The model is integrated with a time step of 900 s, and the output is saved every 6 hrs. The maximum jet speed for control T(4,34) is assumed based on the observed climatological maximum jet speed over the subtropical region during the April-May period and the choice of z=4 as a starting point is due to the presence of z=4 like wave number in the mean zonal wind pattern when the wind at 200hPa is regressed with temperature anomaly over India, shown in Fig.3 of Lekshmi and Chattopadhyay (2022). 

\begin{table*}[hbt!]
\begin{center}
\captionsetup{labelformat=empty}
\caption{Table 1: Set of experiments conducted using the equivalent barotropic vorticity equation model framework.}
\begin{tabular}{|l|l|l|l|l|}
\hline
Experiment Name & \begin{tabular}{l}Stationary Vorticity \\ Forcing (F)\end{tabular} & Vorticity Perturbation $\zeta^{\prime}(z, t=0)$ & \begin{tabular}{l}Maximum Jet Speed ($\boldsymbol{j}_{\text{max}}$; \\Corresponding to \\ zonal mean wind)\end{tabular} & No. of Experiments \\
\hline
\begin{tabular}{l}
(a) Midlatitude \\
Teleconnection \\
Experiment $\boldsymbol{T} \boldsymbol{(} \mathbf{4 , 3 4} \boldsymbol{)}$ \\
\end{tabular} & \begin{tabular}{l}
$F_{1}$ at Source \\
Region 1 \\
(Midlatitude) \\
\end{tabular} & $z=4$ & \begin{tabular}{l}
$34 \mathrm{~m} / \mathrm{s}$ \\
( $\mathrm{A}=25$ ) \\
\end{tabular} & 1 \\
\hline
\begin{tabular}{l}
(b) Teleconnection \\
Sensitivity Experiment \\
$\boldsymbol{T S}\left(\boldsymbol{z}, \boldsymbol{j}_{\text {max }}\right)^{*}$ \\
\end{tabular} & \begin{tabular}{l}
$F_{1}$ at Source \\
Region 1 \\
(Midlatitude) \\
\end{tabular} & \begin{tabular}{l}
$z$ \\
=1,2,3,4,5,6,7, \\
8 \\
\end{tabular} & \begin{tabular}{l}
22,26,30,34,42,50,60,70 \\
$\mathrm{m} / \mathrm{s}$; \\
(A= \\
10,15,20,25,35,45,55,65 \\
) \\
\end{tabular} & \begin{tabular}{l}
64 \\
(including \\
$\boldsymbol{T} \boldsymbol{(} \mathbf{4 , 3 4 )} \boldsymbol{)}$ \\
\end{tabular} \\
\hline
\begin{tabular}{l}
(c) Teleconnection mediated \\ local intensification \\
Experiment $\boldsymbol{I} \mathbf{( 4 , 3 4 )}$ \\
\end{tabular} & \begin{tabular}{l}
$F_{1}$ at Source \\
Region $1+F_{2}$ at Source \\
Region 2 \\
(Midlatitude and BoB) \\
\end{tabular} & $z=4$ & \begin{tabular}{l}
$34 \mathrm{~m} / \mathrm{s}$ \\
( $\mathrm{A}=25$ ) \\
\end{tabular} & 1 \\
\hline
\begin{tabular}{l}
(d) Intensification \\
Sensitivity \\
Experiments \\
$\boldsymbol{I S} \boldsymbol{(} \mathbf{z} \boldsymbol{,} \boldsymbol{j}_{\text {max }} \boldsymbol{)} \boldsymbol{*}^{\boldsymbol{*}}$ \\
\end{tabular} & \begin{tabular}{l}
$F_{1}$ at Source \\
Region $1+F_{2}$ \\
at Source \\
Region 2 \\
(Midlatitude and BoB) \\
\end{tabular} & \begin{tabular}{l}
$z$ \\
=1,2,3,4,5,6,7, \\
8 \\
\end{tabular} & \begin{tabular}{l}
22,26,30,34,42,50,60,70 \\
m/s \\
(A= \\
10,15,20,25,35,45,55,65 \\
) \\
\end{tabular} & \begin{tabular}{l}
64 \\
(including \\
I(4,34)) \\
\end{tabular} \\
\hline
\begin{tabular}{l}
(e) Regional Forcing \\
Experiment \\
$\boldsymbol{R}(4,34)^{* *}$ \\
\end{tabular} & \begin{tabular}{l}
$F_{2}$ at Source \\
Region 2 \\
(BoB) \\
\end{tabular} & $z=4$ & \begin{tabular}{l}
$34 \mathrm{~m} / \mathrm{s}$ \\
( $\mathrm{A}=25$ ) \\
\end{tabular} & 1 \\
\hline
\end{tabular}
\end{center}

*For each stationary wave number pattern ( $z=1,2,3,4,5,6,7$ and 8 ), the model is prescribed maximum jet speeds $j_{\text {max }}=22,26,30,34,42,50,60$ and $70 \mathrm{~m} / \mathrm{s}$. Together with $(e)$, total number of experiments $=129$.
\end{table*}

\subsection{Methods for evaluation of model experiments}

This study measures the intensification of anti-cyclonic vorticity as a function of amplification and persistence in a prescribed background condition and initial conditions. This can be looked at as follows. The mode can amplify sharply due to Rossby wave-related anticyclone amplification, which can cause large departures of temperatures above normal for a short period. Similarly, the mode may not be amplified much in other cases but can persist over a region longer. This can also cause temperature departures above normal for many days. Or it may be a mixture of both above scenarios. Considering the possibility of both these factors, intensification in this study is considered a function of both amplification and persistence. For persistence, the percentage of time steps in each run where the model predicted anticyclonic vorticity with magnitude below a threshold value ( -2.0 × 10-5 s-1) is measured. This value is ~ (Mean – 1 standard deviation) of the observed climatological relative vorticity distribution over the Indian region during April-May. The ratio of vorticity in intensification experiments with respect to the corresponding teleconnection experiments is calculated to define the amplification process. 

\section{Results}
\subsection{Modal Teleconnection Pathway and Its Intensification in the Equivalent Barotropic Model}
\subsubsection{Mid-latitudinal Forcing Driven Modal Teleconnection over the Indian Region}
Based on the observational evidence, the modal propagation and the persistence of anticyclones over the Indian region are studied using the equivalent barotropic model. In the teleconnection T(4,34) experiment (Refer to Section 4), an anticyclonic stirring in the form of a stationary circular anticyclonic vorticity forcing below Iceland, close to the west coast of Europe (F1) is provided (\textcolor{Blue}{Fig. 5b}). This region is a known Rossby wave source, especially due to the land-sea thermal contrast. Studies have shown that this is a dominant Rossby wave source during summer ( Refer to Fig 4 in Shimizu and de Albuquerque Cavalcanti 2011). The model response from t=72 hrs. up to t=192 hrs. is shown in \textcolor{Blue}{Fig. 6a}. The propagating anticyclone-cyclone-anticyclone tilted pattern from observation at lag 0 (\textcolor{Blue}{Fig. 2a}) qualitatively matches the simulated vorticity tilted pattern at t=72 hrs. It is seen that a wave train is propagating from the Source Region 1 (marked by a dotted circle), which is further elongating and propagating over the northern Indian region, peaking during t=72 to 120 hrs. and later weakening but persisting over North India. Thus, the anti-cyclonic vorticity forcing (F1) in Source Region 1, consistent with the observed pattern (\textcolor{Blue}{Fig. 2a}), can drive the modal teleconnection pathway.
\subsubsection{Modal Intensification: Response to Imposed Forcing over the Bay of Bengal (BoB)}

The occasional intensification of the moist heatwave mode can lead to heatwaves and act as a significant driver for local temperature extremes, especially the moist heat stress. As discussed earlier, a local heat source can provide favorable conditions for modal intensification. 
From observation, the cyclonic Source Region 2 (F2) in the Bay of Bengal (BoB), which is barotropic in nature (Fig 4a), has already been identified. We would like to see the impact of this F2, along with F1 in Source Region 1, on the anticyclone over the Indian region. The model vorticity response for this experiment, i.e., I(4,34), is shown in Fig 6b. The time steps t=120, 144, and 168 hrs. show signatures of the superimposing of the anticyclonic phase of the eastward propagating mid-latitudinal Rossby wave and the north-westward propagating pattern over the Indian region. The intensification occurs around the same time steps as seen from the box-averaged plot in Fig S3c. Earlier studies have identified similar mechanisms as a major cause of Rossby wave amplification (Shaman and Tziperman (2016). While Shaman and Tziperman (2016) considered the waves originating from the same source, here, the waves propagate from two different sources and superimpose with each other over India.

\begin{figure*}[hbt!]
\centering
\includegraphics[width=0.8\linewidth]{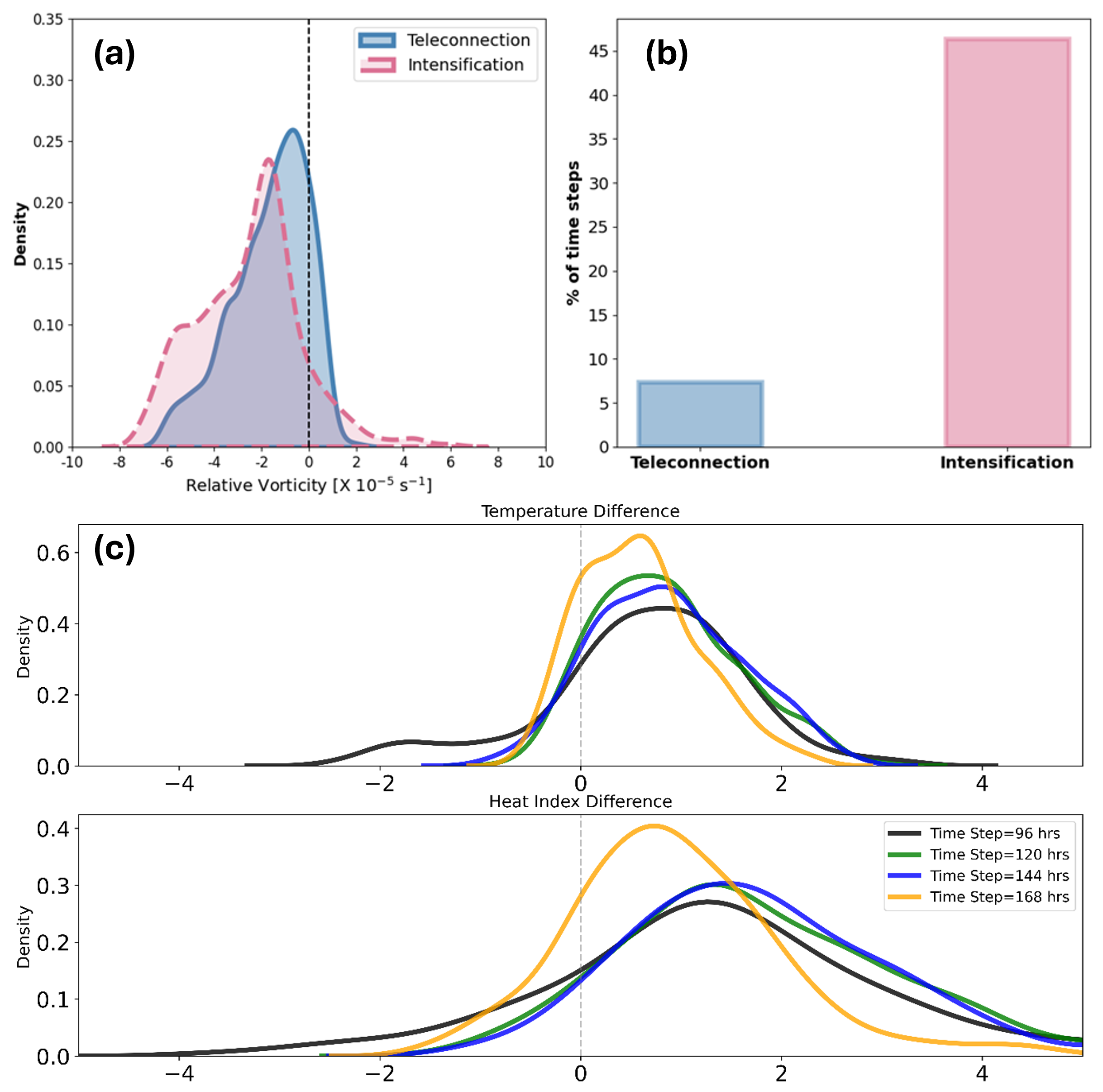}
\caption{(a) Density distribution of relative vorticity (\(\times 10^{-5}\,\mathrm{s}^{-1}\)) over the Indian region (Region 2) for model time steps from \(t=126\) to \(t=174\) hrs for Teleconnection experiment T(4,34) and Intensification Experiment I(4,34). (b) Percentage of time steps when the average of vorticity over the Indian region falls below \(-2.0 \times 10^{-5}\,\mathrm{s}^{-1}\) for T(4,34) and I(4,34) experiments. (c) Density distribution of the difference of maximum temperature (upper panel) and HI (lower panel) estimated from model predicted vorticity between I(4,34) and T(4,34) experiments shown for different time steps.}

\label{fig_wide}
\end{figure*}

\subsubsection{	Mechanism of Modal Intensification: Large-Scale Wave Superposition}
A more quantitative approach for understanding the superposition of waves is presented for both observation and the model. \textcolor{Blue}{Fig. 7} shows the relative vorticity anomaly pattern at different lags with respect to PC > 1.0 days regressed with (1) area-averaged vorticity anomaly in the Source Region 1 and (2) area-averaged vorticity anomaly over the warm Source Region 2 during PC > 1.0. The first is displayed using shaded contours for a larger region, and the latter, in contour lines, is overlaid in it for a smaller region, denoted by the pink box. Here in the Source Region 1, an anticyclonic relative vorticity forcing is provided. So, in the regressed pattern, negative values (blue shade) represent cyclonic and positive values (red shade) represent anticyclonic circulation. At the same time, the forcing provided in Source Region 2 is cyclonic vorticity, and so in the regressed pattern, positive values represent cyclonic and negative values represent anticyclonic vorticity. It is seen that the anticyclonic pattern of both the contours (red shade and dotted contour lines) superimposes over the northwest Indian region.

The wave superposition in the model is analyzed by comparing Vorticity (superposition) and Vorticity (F1+ F2), which is I(4,34). The Vorticity (superposition) is obtained as the algebraic sum of Vorticity (F1 only) and Vorticity (F2 only). That is, over the model’s basic state and the initial vorticity perturbation, the model is run separately by placing F1 only (i.e., T(4,34)) and F2 only (i.e., R(4,34)). The sum of the wave responses from these two is compared with I(4,34) and is shown in \textcolor{Blue}{Fig. 8}. In \textcolor{Blue}{Fig. 8a}, the area-averaged relative vorticity over a large box including the Indian region (5-60 ⁰N; 55-110 ⁰E) of T(4,34), R(4,34), T(4,34)+R(4,34) and I(4,34) is provided. A large spatial region, which is of typical Rossby wave scale (Lu and Boyd 2008), which includes the Indian region, is considered the appropriate scale of averaging here. It is seen that vorticity obtained from both superposition and I(4,34) follow the same pattern with slight changes in magnitude, especially until t=150 hrs., and later show large deviations from each other. The spatial pattern of the same is provided in \textcolor{Blue}{Fig. 8b}, which also shows the similarity between the superposed waves and the I(4,34) experiment. The evolution of the waves and intensification of the anticyclone over the Indian region show a remarkable similarity between the I(4,34) and the linearly superposed wave response, especially up to t=108 hrs., when the waves start to interact closely. After t=108 hrs., a shift in the phase location of the merged waves by ~10⁰ can be seen, suggesting that the non-linear interaction between waves can also occur in I(4,34), which cannot be entirely neglected. 
Now, does superimposing of the heat sources F1 and F2 lead to the intensification of the moist heatwave mode?  In \textcolor{Blue}{Fig. 9a}, the density distribution of vorticity over Region 2 (see Section 6) during the model time steps from t=72 to t=216 hrs is provided for T(4,34) and I(4,34). The distribution of I(4,34) shows a shift in the vorticity towards more negative values than the same for T(4,34). This shows that amplification of model anticyclonic vorticity over the central-northwest Indian region occurs by placing a cyclonic forcing over the BoB. Also, it is seen that amplified anticyclonic vorticity (below a reference threshold of $-2.0 \times 10^{-5}\ \mathrm{s}^{-1}$, which is approximately the mean-1 standard deviation of the observed climatological relative vorticity for April-May months over the Indian region) is observed for a higher percentage of model time steps in I(4,34) than in T(4,34), indicating its persistence over the northwest Indian region (\textcolor{Blue}{Fig. 9b}). Thus, one of the dynamic mechanisms by which the moist heatwave mode-related Rossby wave amplifies over the northwest Indian region is the superposition of anticyclonic phases of the mid-latitudinal Rossby wave and north-westward propagating wave pattern from the tropics. So, the regional forcing over BoB, i.e., F2, plays a dominant role in the intensification and persistence of anticyclonic patterns over the northwest Indian region associated with the modal teleconnection pathway. The westward propagation of the pattern suggests possible moisture influx to the east coast, as observed for the moist heatwave mode (\textcolor{Blue}{Fig. 4}). That is, this dynamical mechanism of the modal Rossby wave intensification also explains the behaviour of this mode as a cause of humid heat stress over the Indian region.
\subsubsection{The Intensified Moist Heatwave Mode and the Resultant Heat Stress over the Indian Region}

Using the multilinear regression model (Refer to Section 4.1), the temperature at different lead times is estimated from the spatial pattern of the model-predicted relative vorticity. A similar estimation is done for the RH, then using temperature and RH, HI is calculated. The maximum temperature, RH, and HI are estimated for the teleconnection (T(4,34)) as well as intensification (I(4,34)) experiments. The density distribution of the difference between them (I(4,34)-T(4,34)) for maximum temperature and heat stress (HI) over Region 2 is provided in \textcolor{Blue}{Fig. 9c}. It can be found that both variables show a shift in the positive direction, depicting the intensification of the heat stress over the northwest Indian region associated with the intensified anticyclone. 
\subsubsection{The Role of Background Condition}
The background conditions (the mean wind and the initial vorticity perturbation at t=0) prescribed in the experiments were selected based on observational considerations and recent studies on the role of stationary waves on heatwaves, as reviewed in the introduction. It is found that the large-scale stationary wave-like forcing ($\zeta^{\prime}$) at t=0, is necessary for intensification to occur over the Indian region in this model on the prescribed mean flow. We did several experiments to understand the role of the initial stationary wave. These results are summarized in the Supplementary material (supplementary material SW) as a separate section. It is found that the prescribed background condition provides a realistic spatial pattern of cyclone and anticyclone during intensification. The pattern correlation between the intensified cyclone-anticyclone pattern in the model with the composite based on the observation is 0.47(plot not shown). Without this background mean flow or initial vorticity perturbation, the vorticity forcings $F_1$ and $F_2$ could produce only a very localized system with minimum spatial propagation. Both the background mean flow and initial vorticity perturbation aid in the zonal intensification that takes place in the presence of $F_2$. The intensification in amplitude is quite generally seen for different phase shifts and wavenumber of the initial wave, and also different jet speed configurations (Supplementary text SW, \textcolor{Blue}{Fig.SW-3, Fig.SW-8; Fig. SW-14, SW-15, and SW-16}). This indicates the validity of the chosen configuration and meridional propagation of the system generated by $F_1$ towards the Indian region, and the background condition in the current analysis.

\begin{figure*}[hbt!]
\centering
\includegraphics[width=0.8\linewidth]{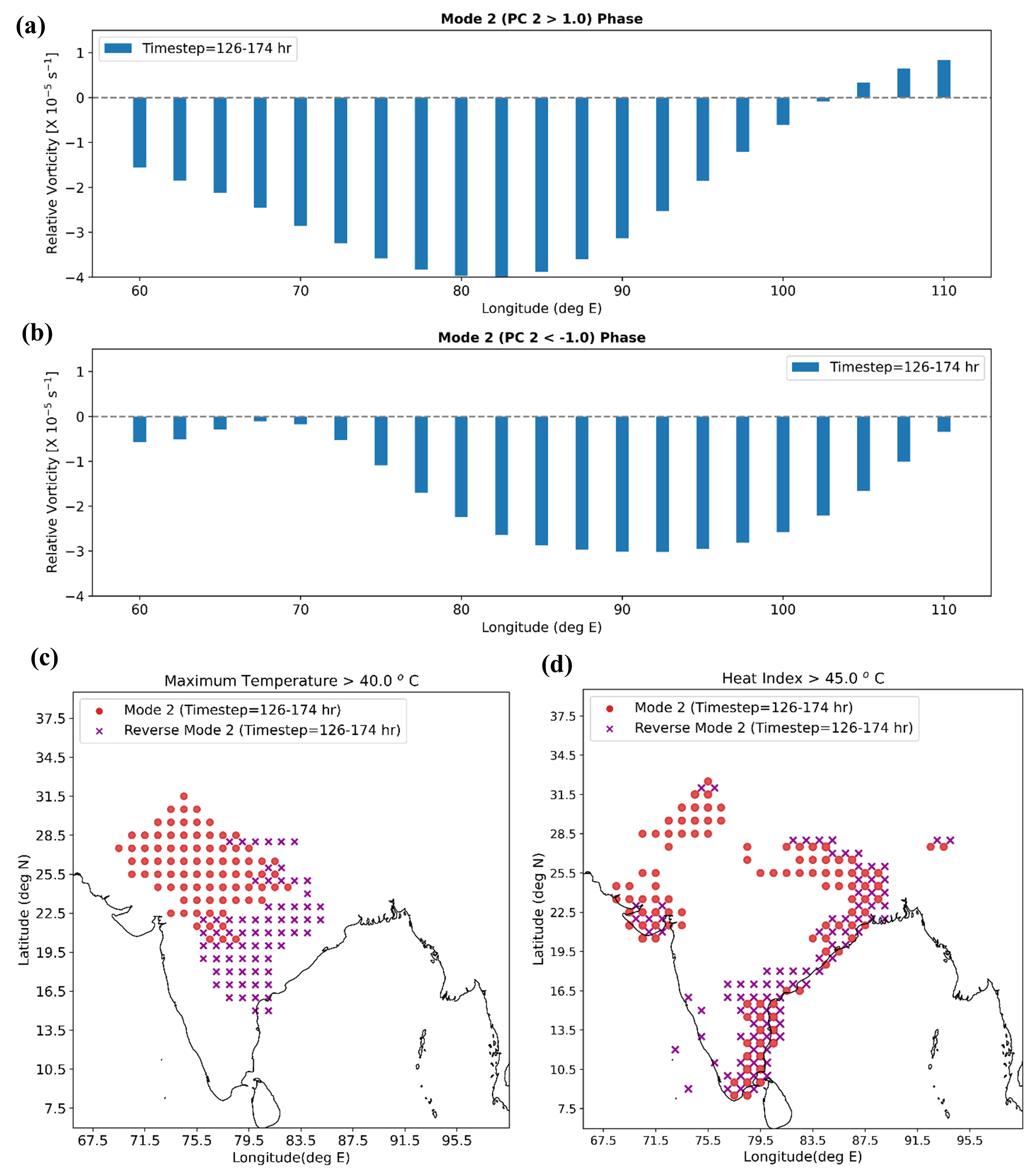}
\caption{(a) Zonal variation of the model predicted relative vorticity averaged over 15-40 ⁰N and for t=126-174 hrs from the intensification experiment of the mode (I(4,34)) (b) Same as in Fig 10a averaged for time steps t=126-174 hrs for intensification of R-Mode Experiment with z=4 and jmax=34 m/s (c) Spatial pattern of average maximum temperature > 40 ⁰C for time steps t=126-174 hrs for the mode \& R-Mode intensification experiments over the Indian region. (d) Same as Fig 10 c for HI > 45 ⁰C.}

\label{fig_wide}
\end{figure*}

\begin{figure*}[hbt!]
\centering
\includegraphics[width=0.8\linewidth]{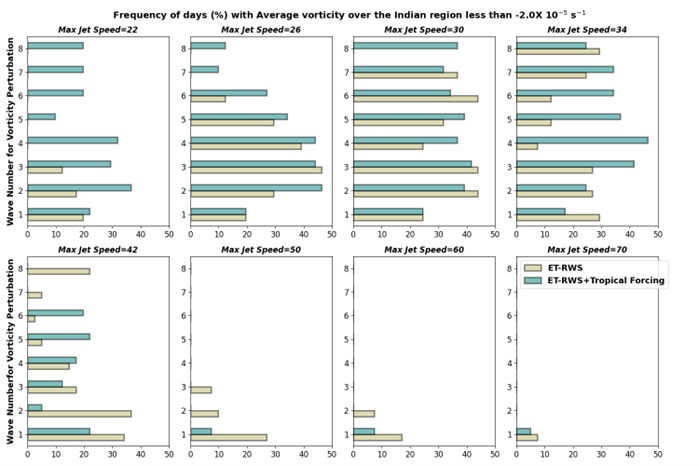}
\caption{Percentage frequency of model time steps when the average relative vorticity (s-1) over the Indian region (Region 2) falls below \(-2.0 \times 10^{-5}\,\mathrm{s}^{-1}\) for each TS (ET-RWS) and IS (ET-RWS + Tropical Forcing) experiments. Each subplot represents a different zonal wind (m/s) configuration (jmax), and the y-axis represents different wave numbers for vorticity perturbations.}

\label{fig_wide}
\end{figure*}

\subsection{Reverse Pattern of the Modal Teleconnection and Intensification}

From the spatial pattern of the moist heatwave mode (i.e., Mode 2) EOF (\textcolor{Blue}{Fig. 1b}), it is apparent that the spatially reverse pattern of the dipole structure can also exist. That is the pattern corresponding to the anticyclonic circulation anomaly over the east coast, and the cyclonic circulation anomaly over northwest India also exists. Such a spatial pattern of this mode, when intensified, can lead to positive temperature anomalies over central India to the east coast, where the anticyclone lies. The observed vorticity and the temperature anomaly pattern associated with this reverse pattern of the moist heatwave mode (hereafter mentioned as R-moist heatwave mode) are obtained by taking the composite for those days when PC <-1.0 and are given in \textcolor{Blue}{Fig. S4}. The SST pattern associated with this mode is also shown in \textcolor{Blue}{Fig. S5}, which shows a similar warming pattern over the BoB as in the moist heatwave mode. From the composite of the vorticity anomaly (\textcolor{Blue}{Fig. S4a}), it is seen that over the mid-latitudinal Rossby wave source region, there exists a cyclonic circulation anomaly in the upper level, which is barotropic in nature (figure not shown), rather than an anticyclone as seen for the moist heatwave mode. The set of experiments for understanding R-moist heatwave mode teleconnection and its intensification was carried out using the same model with the control experiment configuration z=4 and jmax=34m/s. The model setup and initial conditions are identical to those for the moist heatwave mode, apart from the cyclonic F1.

From \textcolor{Blue}{Fig. S6}, it can be seen that the anticyclonic pattern evolved in the same manner as that in the moist heatwave mode, but is shifted in its location towards southeast India. Together with the impact from F2 (\textcolor{Blue}{Fig. S6b}), the superposition of the anticyclonic pattern peaks over east-southeast India. The intensification of the R-moist heatwave mode occurs as a result of amplified (\textcolor{Blue}{Fig. S7a}) and persistent (\textcolor{Blue}{Fig. S7b}) anticyclonic circulation over India. Using a similar (Section 4.1) multilinear regression model (obtained from observation based on days with PC < -1.0), the maximum temperature and heat stress (HI) are obtained. The density distribution of the difference between the intensification and teleconnection experiment for maximum temperature and heat stress (HI) over the entire Indian region shows intensified heat stress over India at different time steps (\textcolor{Blue}{Fig. S7c}). This indicates that an intensification of the circulation anomalies and a corresponding intensification of the temperature and heat stress also resulted from the R-moist heatwave mode.

Even though the spatial pattern of the relative vorticity pattern (\textcolor{Blue}{Fig. S5}) indicates the intensification as taking place over the east-southeast Indian region, a more quantitative analysis of the distinction of R-moist heatwave mode from the moist heatwave mode is presented in \textcolor{Blue}{Fig. 10}. The zonal pattern of the model predicted relative vorticity averaged over 15-40 ⁰N for the moist heatwave mode and R-moist heatwave mode (t=126 to 174 hrs) are given in \textcolor{Blue}{Fig. 10}. There is a gradual transition from an anticyclone to a cyclonic pattern for the moist heatwave mode and a very weak or no anticyclone to intense anticyclonic circulation in the R-moist heatwave mode, indicating a shift in the spatial location of intensification from the west (in the moist heatwave mode) to the east (in R-moist heatwave mode). The impact caused by the shift in the circulation pattern on the maximum temperature and heat stress is also compared for both in \textcolor{Blue}{Fig. 10c-d}. The locations of average maximum temperature > 40 ⁰C and the HI > 45 ⁰C during t=126-174 hrs. of the moist heatwave mode and R-moist heatwave mode are provided. In temperature, the extreme values show a significant difference in the spatial distribution, with the moist heatwave mode concentrating more in the northwest and R-moist heatwave mode in southeast India, with both overlapping over a few grids in the central Indian region. In the HI, both modes have extreme values in the east-southeast region. As HI is impacted by both RH and temperature, and due to the relatively high RH observed over the east coast, the regression relation obtained from observation results in high HI on the east coast for both the modes (\textcolor{Blue}{Fig. 10d}). But over the northwest, where the RH is comparatively lower, the HI is high only for the moist heatwave mode. Thus, a symmetric response of the mid-latitudinal Rossby wave forcing associated with the moist heatwave mode is observed over the Indian region. The anticyclonic and cyclonic forcing in the midlatitude shifts the spatial location of intensification from northwest to southeast, respectively, with the dynamics of intensification being the same. 
The long-term trend of the moist heatwave mode-reconstructed temperature (EOF × PC) for both the period from 1980 to 2020 and from 2000 to 2020 is shown in \textcolor{Blue}{Fig. S7}. The trend pattern has been reversed in recent years, with an increasing trend along the east coast. Since this mode is identified as the causal factor for humid heat stress (S et al. 2024) it is consistent with the fact that recent years show an increase in moist heatwave mode  (i.e., humid heat stress), as found in many other studies (Sojan and Srinivasan 2024) especially over the East Coast.

\subsection{Sensitivity of the Modal Teleconnection and its Intensification to the Initial Zonal Wavenumber Perturbation and Mean Zonal Wind State}

Now we investigate the model sensitivity to different initial conditions and mean state in driving the modal intensification over the Indian region. This is important because climate change can impact heat extremes by manifesting large-scale circulation features, such as stationary wave patterns and jetstreams, apart from its direct impact on temperature. Studies have shown that the summertime stationary waves in the Northern Hemisphere tropics are projected to weaken (Wills et al. 2019). However, in the subtropics, the stationary circulations are projected to strengthen for a high-emission scenario, especially over northwestern North America (Zhang et al. 2023). Thus, it is necessary to understand the impact that the background conditions will have on intensification. Considering the recent literature, two aspects are studied here based on the current modelling framework. One is the impact caused by the different zonal wave numbers via the initial vorticity perturbation, and the second is the weakening/ strengthening of the zonal wind speed.

The probability distributions of the difference between the intensification and teleconnection experiments of the respective (zonal wave number z, maximum jet speed jmax) configurations are provided in \textcolor{Blue}{Fig. S9}. Also, to look at the persistence of the circulation over the Indian region, the frequency of model timesteps (\%) when the average vorticity over Region 2 was below $-2.0 \times 10^{-5} \text{s}^{-1}$ is plotted (\textcolor{Blue}{Fig. 11}). The frequency is calculated for all the TS and IS experiments. It is clear from \textcolor{Blue}{Fig. 11} that the anticyclonic vorticity associated with the moist heatwave mode persisted over the central-northern Indian region for the experiments with jmax=26,30 and 34 m/s for almost all z patterns (TS Experiments). This is also true for the IS experiments, with the persistence of anticyclonic vorticity extending to more than 30-40\% of model time steps. This decreases as jmax increases for both TS and IS experiments and becomes nil when $j_{max}$ rises to 70 m/s. For $j_{max}$=26,30,34 m/s configurations, the persistence shows wide variations among different values of z. At the same time, the same for the IS experiments are maximum when $j_{max}$=22,26,30 and 34 m/s with z in the range of 2-6. That is, the formation and intensification of anticyclonic vorticity in Region 2 is robust for different sets of wave numbers and jmax configurations, especially when $j_{max}$ < 50 m/s.
A summary of the favorable conditions that can lead to intensified anticyclonic vorticity over the Indian region caused by superposition is provided in \textcolor{Blue}{Fig. 12}. The runs are generated with the same amplitude of imposed forcing (F1, F2, and background), in all cases. The figure shows the ratio of the average vorticity over the Indian region between the IS and the TS experiments. Here, while calculating the ratio between average relative vorticity over Region 2 (Intensification/Teleconnection), the values are plotted only if the average relative vorticity obtained from the Intensification experiment is negative (anticyclone). Others are kept blank (white boxes). That is, the ratio measures how intensified the anticyclone is in the experiment (ignoring the cyclone). The magnitude shows how often the anticyclonic vorticity intensified over Region 2 in the presence of the regional forcing in the BoB. The combinations of z and $\overline{u}$  (or $j_{max}$) favorable for such a superimposition and intensification of Rossby waves associated with the moist heatwave mode are evident from \textcolor{Blue}{Fig. 12}, which plots the amplitude ratio of intensification to the superposition experiment. A ratio value greater than one suggests intensification.

Thus, by considering the amplification of anticyclonic circulation associated with the moist heatwave mode (\textcolor{Blue}{Fig. 12}) and also its persistence over the Indian region (\textcolor{Blue}{Fig. 11}), we can summarize as follows. Generally, the intensification is favored more for the configurations where jmax= 26-34 m/s and z=2,3,4,5 and 6. Even though the maximum jet speed in some experiments is above realistic limits, the experiments are done to understand the dynamical evolution of a range of jets from small to large speeds, which helps us understand the intensification process in climate change scenarios. Interestingly, it is found that the modal intensification, even in the simplistic model version used here, occurs within the realistic jet speed limits. Above and below these limits, the Rossby wave intensification associated with the moist heatwave mode is not favored in the current modelling framework. In the context of the superposition, the results suggest that the amplitude ratio is often very large, and sometimes amplifications do not occur (white boxes). Thus, when the superposition is very large, it is not always linear over the Indian region. The background conditions may play an important role in the amplification in such cases, as shown in Supplementary Material SW, in terms of shift in phase and amplitude. However, it is not discussed in detail in the context of the current study.

\begin{figure*}[hbt!]
\centering
\includegraphics[width=0.8\linewidth]{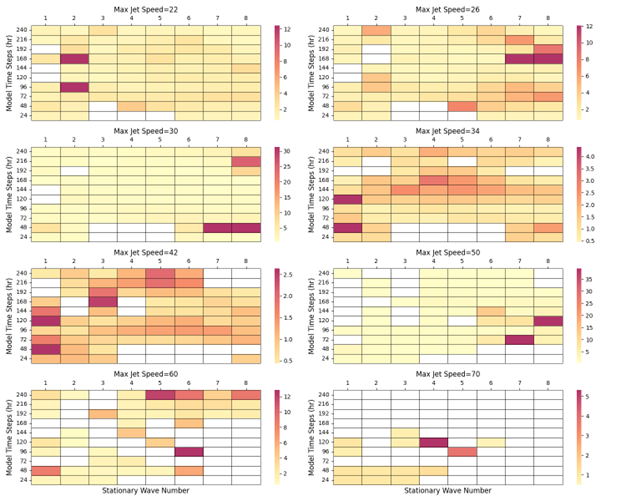}
\caption{Heatmap of the ratio of the average anticyclonic vorticity over the Indian region (Region 2) between the IS(z, jmax) and the TS(z, jmax)  experiments at different time steps (y-axis) and different zonal wavenumbers (x-axis) for vorticity perturbations. White boxes represent the non-existence of anticyclonic vorticity in the respective IS(z, jmax) experiment at different time steps. Each panel represents a different zonal wind configuration (m/s) with the jmax mentioned at the top.}

\label{fig_wide}
\end{figure*}

\section{Summary and Discussions}
Temperature variability over India has two dominant modes during summer, driven by mid-latitudinal Rossby waves (LC22). While the first mode drives the dry heat stress, Mode 2, along with a strong regional response, can drive moist heat stress in India, which can be fatal (Lekshmi et al. 2024). The current study analyses the intensification of moist heat stress associated with Mode 2 patterns. An intensification of these Rossby wave modes occurs over the Indian region, driving a significant number of extreme temperature events. However, it was unclear if the intensified heatwave modes arise due to teleconnection modes alone or if some regional forcing contributes to the intensification. Also, the moist heatwave (second) mode shows a significantly increasing trend over the Indian region, thus making the interpretation imperative. The evidence from observation leads to the hypothesis that local forcing may contribute to the intensification of heatwaves over India through locally mediated amplification of the extratropical Rossby wave mode intruding over the Indian region under certain background conditions. With the use of the non-divergent equivalent barotropic model, initialized with a stationary large-scale wave like flow on a mean background and forced by a BoB and extratropical forcing, this work focuses on the dynamical process behind the amplification of the second mode (hereafter mentioned as ‘moist heatwave mode’) and associated circulation, which can lead to the intensification of the heatwaves over the Indian region.

Two sets of control experiments for understanding (a) the propagation of Rossby waves associated with the moist heatwave mode and the persistence of anticyclones over the Indian region, and (b) the intensification of the anticyclones over India, are conducted using the model. For that, over the model basic state ($\overline{\zeta}$), and the initial vorticity perturbation ($\zeta^{\prime}(t=0)$), a mid-latitudinal Rossby wave source (F1) and a tropical source (F2) were provided, respectively. In the first control experiment (a), only F1 is prescribed.  F1 drives the moist heatwave mode-related Rossby waves and their propagation towards the Indian region. In the second control experiment (b), in addition to F1, another forcing over BoB (F2) is added. It was found that F2 causes the intensification of the anti-cyclonic vorticity over the north-west Indian region by the superposition of the westward spatial phase extension of the anticyclonic pattern from the BoB and the eastward propagating mid-latitudinal Rossby wave (\textcolor{Blue}{Figs. 6-9}). The intensified anticyclonic circulation can cause subsidence motion of air and further formation of clear skies, leading to adiabatic warming of air. 

With the aid of a multilinear regression model based on observation, the maximum temperature and the RH are estimated from the equivalent barotropic model predicted vorticity. The distribution of HI and maximum temperature shows intensified heat stress over India (\textcolor{Blue}{Fig. 9}). This is the novelty of the study in which the dynamical mechanism of the intensification of the circulation features (which is predominantly barotropic in nature) is simplified by using the equivalent barotropic model, but is still able to evaluate the temperature patterns and regional heat stress without actually involving the complexities of a baroclinic model. An anticyclonic forcing over the midlatitudes, along with tropical cyclonic forcing, can cause the intensification of circulation over the northwest Indian region. At the same time, a cyclonic forcing over the midlatitudes, along with a cyclonic tropical forcing, can shift the location of intensification towards the southeast Indian region. This pattern is consistent with the reverse pattern of the observed EOF 2 mode (\textcolor{Blue}{Fig. 1b}), which shows an increasing trend in recent years (\textcolor{Blue}{Fig. S7}). Thus, a symmetric response of the mid-latitudinal forcing is observed over the Indian region, with the dynamics of the process being the same.

In addition to the F1 and F2, the role of the initial forcing ($\zeta^{\prime}(t=0)$) is worth mentioning for the moist heatwave mode intensification. The spatial pattern of the initial condition drives the intensification and is responsible for creating the spatial pattern. Without the mean state and initial wave perturbation, the intensification does not occur over the Indian region. Rather, a very localized forced waves appear in the region where F1 and F2 are applied, depending on the intensity of the stationary forcing. Further sensitivity analysis is carried out to validate the role of the initial wave ($\zeta^{\prime}(t=0)$), range of wavenumber and phases, and is reported in a separate supplementary material (Supplementary SW). The detailed analysis suggests that there exists a range of wavenumbers and jet speeds that could lead to moist heatwave intensification, and such intensification occurs due to superpositions of local forcings for a wide range of choices of initial wavenumber, its phase or jet speed.  Hence, our analysis with wavenumber and jet speed combination (4,34) is general in nature, reflecting a wide variety of possible configurations for intensification. In general, for the zonal and meridional propagation of the mid-latitudinal Rossby wave towards the Indian region, and its superposition with the anticyclonic phase from the BoB, both the zonal mean state and the initial wave number vorticity perturbation play a prominent role.

A similar set of analyses for the reverse pattern of moist heatwave mode (R-moist heatwave mode) is also done (Section 6.2). All the control and sensitivity experiments were conducted for the reverse mode with the same set of zonal mean state and initial wavenumber perturbation. The only difference was that the mid-latitude Rossby wave forcing F1 was replaced with a cyclonic vorticity forcing, rather than an anticyclone used in the moist heatwave mode experiment. A spatial shift in the region of heat stress is noted in the reverse mode with maximum over the south-east coastal region (\textcolor{Blue}{Fig. 10}). This spatial shift is consistent with the observed pattern of moist heatwaves over India. That is, by keeping an anticyclonic or cyclonic forcing in the mid-latitudinal Rossby wave forcing region (Source region 1), a symmetric response in heatwave intensification is noted over the Indian region with respect to the moist heatwave mode EOF pattern (EOF 2). 

Climate change plays both a direct and indirect role in driving the temperature extremes over any region. While the direct role can be attributed to the rise in mean temperature, the indirect role is through the changes in the background stationary wave patterns, mean zonal wind pattern, location and speed of jetstream, etc. The interaction of these features and the resultant extreme event is highly non-linear in nature and requires further understanding (Palmer 2013; White et al. 2022). In the aspect of wave amplification, this understanding usually provides us with a set of ‘weather regimes’ (defined here as patterns associated with eddies and zonal mean wind combinations) where the Rossby wave amplification is more favored. Here, we found that the superimposing waves can lead to more intensified Rossby waves, especially when z=3,4,5 and 6 and $j_{max}$=26,30 and 34 m/s (\textcolor{Blue}{Fig S8, 11 and 12}). Such preferential selections are likely to occur because of highly non-linear processes. A summary of the weather regimes that prove to be favorable for the intensification of the modal Rossby waves is shown (\textcolor{Blue}{Fig. 12}). However, care must be taken while linking the intensified (amplified \& persisting) Rossby waves with extreme events. Because intensified waves can cause anomalously warm conditions in particular locations, and at the same time, they can cause anomalous cooling at other locations as well. That is, global warming can favor local cooling  (Palmer 2013). From \textcolor{Blue}{Fig. 12}, along with the regimes where the anticyclonic circulation intensifies (in the presence of tropical forcing), a glimpse of other regimes where the cyclonic circulation could intensify is also seen (which is given as the white boxes). Such type of conditions can lead to intense cyclonic circulation and the formation of local-scale thunderstorms because of the moisture availability as part of the moist heatwave mode over the eastern coastal states. These types of scenarios are not uncommon during the pre-monsoon season in the coastal states of India, but are out of the scope for the present study. The current study, though based on a very simple Rossby wave superposition paradigm, provides a simple dynamical mechanism in which the Rossby wave associated with modal teleconnection intensifies, intensifying heatwaves in the Indian region. Although the analysis is based on a simplified dynamics and ignores the role of other local factors, such as soil moisture, the model is able to simulate a few basic features of the observed moist heatwave mode such as: (a) The midlatitude teleconnection pathway, (b) the role of the jet and the waveguide, (c) persistence conditions, (d) shifting of anticyclone in the modal patterns in the direct and the reverse forcing, (e) initial large-scale conditions. Though a more realistic basic state, including multiple local factors and multi-level baroclinic response, would give more realistic results, the basic structural dynamics are well evident in the simplified solutions.

\paragraph{Acknowledgments}
The authors thank the anonymous Reviewers for their constructive comments and suggestions. The core equivalent barotropic model used in the study is based on the Held-Suarez Barotropic model adapted from \url{https://github.com/lmadaus/Barotropic-Python} and further developed for the current study. We thank the Python code developers for the barotropic model. LS acknowledges the Ministry of Earth Sciences (MoES), Govt. of India, for the research fellowship support from the MRFP Project. Research support from the India Meteorological Department (IMD) and the Indian Institute of Tropical Meteorology (IITM), an autonomous institute under MoES, is also acknowledged. 

\paragraph{Data Availability} 
All the input datasets used for the analysis are freely available, as mentioned in Section 2. Any computed data/ codes used to generate the plots will be made available on request.

\paragraph{conflict of interest}

The authors declare no competing interests.

\paragraph{\textbf{References\\ \\}}

Alexander, L., 2011: Extreme heat rooted in dry soils. Nat. Geosci., 4, 12–13, \url{https://doi.org/10.1038/ngeo1045}.\\

Ambrizzi, T., and B. J. Hoskins, 1997: Stationary rossby-wave propagation in a baroclinic atmosphere. Q. J. R. Meteorol. Soc., 123, 919–928, \url{https://doi.org/https://doi.org/10.1002/qj.49712354007}.\\

Boer, G. J., and S. J. Lambert, 2001: Second-order space-time climate difference statistics. Clim. Dyn., 17, 213–218, https://doi.org/10.1007/PL00013735.\\

Coumou, D., V. Petoukhov, S. Rahmstorf, S. Petri, and H. J. Schellnhuber, 2014: Quasi-resonant circulation regimes and hemispheric synchronization of extreme weather in boreal summer. Proc. Natl. Acad. Sci., 111, 12331–12336, \url{https://doi.org/10.1073/pnas.1412797111}.\\

Dubey, A. K., and P. Kumar, 2023: Future projections of heatwave characteristics and dynamics over India using a high-resolution regional earth system model. Clim. Dyn., 60, 127–145, https://doi.org/10.1007/s00382-022-06309-x.\\

Francis, J. A., and S. J. Vavrus, 2012: Evidence linking Arctic amplification to extreme weather in mid-latitudes. Geophys. Res. Lett., 39, \url{https://doi.org/https://doi.org/10.1029/2012GL051000}.\\

Ganeshi, N. G., M. Mujumdar, R. Krishnan, and M. Goswami, 2020: Understanding the linkage between soil moisture variability and temperature extremes over the Indian region. J. Hydrol., 589, 125183, https://doi.org/10.1016/j.jhydrol.2020.125183.\\

Ganeshi, N. G., M. Mujumdar, Y. Takaya, M. M. Goswami, B. B. Singh, R. Krishnan, and T. Terao, 2023: Soil moisture revamps the temperature extremes in a warming climate over India. npj Clim. Atmos. Sci., 6, https://doi.org/10.1038/s41612-023-00334-1.\\

Ghosh, S., 2024: India reels under a third straight year of severe heatwaves. nature india, June.\\

Goyal, M. K., S. Singh, and V. Jain, 2023: Heat waves characteristics intensification across Indian smart cities. Sci. Rep., 13, 14786, https://doi.org/10.1038/s41598-023-41968-8.\\

Grimm, A. M., and P. L. S. Dias, 1995: Use of Barotropic Models in the Study of the Extratropical Response to Tropical Heat Sources. J. Meteorol. Soc. Japan. Ser. II, 73, 765--780, \url{https://doi.org/10.2151/jmsj1965.73.4_765}.\\

Guleria, S., and A. Gupta, 2016: Heat Wave in India Documentation of State of Telangana and Odisha (2016). 86 pp.\\

Held, I. M., 1985: Pseudomomentum and the Orthogonality of Modes in Shear Flows. J. Atmos. Sci., 42, 2280–2288, \url{https://doi.org/10.1175/1520-0469(1985)042<2280:PATOOM>2.0.CO;2}.\\

Hoskins, B. J., and D. J. Karoly, 1981: The Steady Linear Response of a Spherical Atmosphere to Thermal and Orographic Forcing. J. Atmos. Sci., 38, 1179–1196, \url{https://doi.org/10.1175/1520-0469(1981)038<1179:TSLROA>2.0.CO;2}.\\

——, and T. Ambrizzi, 1993: Rossby Wave Propagation on a Realistic Longitudinally Varying Flow. J. Atmos. Sci., 50, 1661–1671, \url{https://doi.org/10.1175/1520-0469(1993)050<1661:RWPOAR>2.0.CO;2}.\\

Huang, B., C. Liu, V. Banzon, E. Freeman, G. Graham, B. Hankins, T. Smith, and H.-M. Zhang, 2021: Improvements of the Daily Optimum Interpolation Sea Surface Temperature (DOISST) Version 2.1. J. Clim., 34, 2923–2939, https://doi.org/10.1175/JCLI-D-20-0166.1.\\

IPCC, 2013: IPCC, 2013: Climate Change 2013: The Physical Science Basis. Contribution of Working Group I to the Fifth Assessment Report of the Intergovernmental Panel on Climate Change. T.F. Stocker et al., Eds. Cambridge University Press,.\\

——, 2021: Summary for Policy Makers. Climate Change 2021: The Physical Science Basis. Contribution of Working Group I to the Sixth Assessment Report of the Intergovernmental Panel on Climate Change, B. Masson-Delmotte, V. and Zhai, P. and Pirani, A. and Connors, S.L. and Péan, C. and Berger, S. and Caud, N. and Chen, Y. and Goldfarb, L. and Gomis, M.I. and Huang, M. and Leitzell, K. and Lonnoy, E. and Matthews, J.B.R. and Maycock, T.K. and Waterfield, T, Ed., Cambridge University Press, p. 3−32.\\

Jiménez-Esteve, B., and D. I. V. Domeisen, 2022: The role of atmospheric dynamics and large-scale topography in driving heatwaves. Q. J. R. Meteorol. Soc., 148, 2344–2367, https://doi.org/10.1002/qj.4306.\\

Jiménez-Esteve, B., K. Kornhuber, and D. I. V Domeisen, 2022: Heat Extremes Driven by Amplification of Phase-Locked Circumglobal Waves Forced by Topography in an Idealized Atmospheric Model. Geophys. Res. Lett., 49, e2021GL096337, https://doi.org/https://doi.org/10.1029/2021GL096337.\\

Kalshetti, M., R. Chattopadhyay, K. Hunt, M. K. Phani, S. Joseph, D. Pattanaik, and A. Sahai, 2022: Eddy transport, Wave‐mean flow interaction, and Eddy forcing during the 2013 Uttarakhand Extreme Event in the Reanalysis and S2S Retrospective Forecast Data. Int. J. Climatol., \url{https://doi.org/10.1002/joc.7706}.\\

Kanamitsu, M., W. Ebisuzaki, J. Woollen, S.-K. Yang, J. J. Hnilo, M. Fiorino, and G. L. Potter, 2002: NCEP–DOE AMIP-II Reanalysis (R-2). Bull. Am. Meteorol. Soc., 83, 1631–1644, https://doi.org/10.1175/BAMS-83-11-1631.\\

Karoly, D. J., 1983: Rossby wave propagation in a barotropic atmosphere. Dyn. Atmos. Ocean., 7, 111–125, \url{https://doi.org/https://doi.org/10.1016/0377-0265(83)90013-1}.\\

Kenyon, J., and G. C. Hegerl, 2008: Influence of Modes of Climate Variability on Global Temperature Extremes. J. Clim., 21, 3872–3889, \url{https://doi.org/https://doi.org/10.1175/2008JCLI2125.1}.\\

Kornhuber, K., V. Petoukhov, S. Petri, S. Rahmstorf, and D. Coumou, 2017: Evidence for wave resonance as a key mechanism for generating high-amplitude quasi-stationary waves in boreal summer. Clim. Dyn., 49, 1961–1979, \url{https://doi.org/10.1007/s00382-016-3399-6}.\\

Kornhuber, K., S. Osprey, D. Coumou, S. Petri, V. Petoukhov, S. Rahmstorf, and L. Gray, 2019: Extreme weather events in early summer 2018 connected by a recurrent hemispheric wave-7 pattern. Environ. Res. Lett., 14, 54002, \url{https://doi.org/10.1088/1748-9326/ab13bf}.\\

——, D. Coumou, E. Vogel, C. Lesk, J. F. Donges, J. Lehmann, and R. M. Horton, 2020: Amplified Rossby waves enhance risk of concurrent heatwaves in major breadbasket regions. Nat. Clim. Chang., 10, 48–53, \url{https://doi.org/10.1038/s41558-019-0637-z}.\\

Lekshmi, S., and R. Chattopadhyay, 2022: Modes of summer temperature intraseasonal oscillations and heatwaves over the Indian region. Environ. Res. Clim., 1, 025009, \url{https://doi.org/10.1088/2752-5295/ac9fe7}.\\

 Lekshmi S., R. Chattopadhyay, and D. S. Pai, 2024: Attribution of subseasonal temperature modes as the drivers of dry and moist heat discomfort for the heat hazard monitoring over the Indian region. Clim. Dyn., 
 \url{https://doi.org/10.1007/s00382-024-07236-9.}\\
 
Lin, Q., and J. Yuan, 2022: Linkages between Amplified Quasi-Stationary Waves and Humid Heat Extremes in Northern Hemisphere Midlatitudes. J. Clim., 35, 8245–8258, https://doi.org/10.1175/JCLI-D-21-0952.1.\\

Lu, C., and J. P. Boyd, 2008: Rossby Wave Ray Tracing in a Barotropic Divergent Atmosphere. J. Atmos. Sci., 65, 1679–1691, https://doi.org/10.1175/2007JAS2537.1.\\

Manola, I., F. Selten, H. de Vries, and W. Hazeleger, 2013: “Waveguidability” of idealized jets. J. Geophys. Res. Atmos., 118, 10,410-432,440, \url{https://doi.org/https://doi.org/10.1002/jgrd.50758.}\\

Matsueda, M., 2011: Predictability of Euro-Russian blocking in summer of 2010. Geophys. Res. Lett., 38, \url{https://doi.org/https://doi.org/10.1029/2010GL046557}.\\

Mazdiyasni, O., and Coauthors, 2017: Increasing probability of mortality during Indian heat waves. Sci. Adv., 3, e1700066, https://doi.org/10.1126/sciadv.1700066.\\

Mearns, L. O., R. W. Katz, and S. H. Schneider, 1984: Extreme High-Temperature Events: Changes in their probabilities with Changes in Mean Temperature. J. Appl. Meteorol. Climatol., 23, 1601–1613, https://doi.org/10.1175/1520-0450(1984)023<1601:EHTECI>2.0.CO;2.\\

Meehl, G. A., and C. Tebaldi, 2004: More Intense, More Frequent, and Longer Lasting Heat Waves in the 21st Century. Science (80-. )., 305, 994–997, \url{https://doi.org/10.1126/science.1098704}.\\

Moon, W., B.-M. Kim, G.-H. Yang, and J. S. Wettlaufer, 2022: Wavier jet streams driven by zonally asymmetric surface thermal forcing. Proc. Natl. Acad. Sci., 119, e2200890119, https://doi.org/10.1073/pnas.2200890119.\\

Murari, K. K., A. S. Sahana, E. Daly, and S. Ghosh, 2016: The influence of the El Niño Southern Oscillation on heat waves in India. 713, 705–713, https://doi.org/10.1002/met.1594.\\

Pai, D. S., N. Smitha Anil, and A. N. RAMANATHAN, 2013: Long term climatology and trends of heat waves over India during the recent 50 years (1961-2010). MAUSAM, 64, 585–604, \url{https://doi.org/10.54302/mausam.v64i4.742}.\\

——, A. K. Srivastava, and S. A. Nair, 2017: Heat and Cold Waves Over India. Observed Climate Variability and Change over the Indian Region, M.N. Rajeevan and S. Nayak, Eds., Springer Singapore, 51–71.\\

Pai, D. S., and S. Nair, 2022: Impact of El-Niño-Southern Oscillation (ENSO) on extreme temperature events over India. MAUSAM, 73, 597–606, \url{https://doi.org/10.54302/mausam.v73i3.5932.}\\

Palmer, T. N., 2013: Climate extremes and the role of dynamics. Proc. Natl. Acad. Sci., 110, 5281–5282, \url{https://doi.org/10.1073/pnas.1303295110}.\\

Parker, T. J., G. J. Berry, and M. J. Reeder, 2014: The Structure and Evolution of Heat Waves in Southeastern Australia. J. Clim., 27, 5768–5785, https://doi.org/10.1175/JCLI-D-13-00740.1.\\

Perkins-Kirkpatrick, S. E., and S. C. Lewis, 2020: Increasing trends in regional heatwaves. Nat. Commun., 11, 3357, https://doi.org/10.1038/s41467-020-16970-7.\\

Perkins, S. E., 2015: A review on the scientific understanding of heatwaves-Their measurement, driving mechanisms, and changes at the global scale. Atmos. Res., 164–165, 242–267, \url{https://doi.org/10.1016/j.atmosres.2015.05.014}.\\

Perkins, S. E., L. V Alexander, and J. R. Nairn, 2012: Increasing frequency, intensity and duration of observed global heatwaves and warm spells. Geophys. Res. Lett., 39, \url{https://doi.org/https://doi.org/10.1029/2012GL053361}.\\

Petoukhov, V., S. Rahmstorf, S. Petri, and H. J. Schellnhuber, 2013: Quasiresonant amplification of planetary waves and recent Northern Hemisphere weather extremes. Proc. Natl. Acad. Sci., 110, 5336–5341, \url{https://doi.org/10.1073/pnas.1222000110}.\\

——, S. Petri, K. Kornhuber, K. Thonicke, D. Coumou, and H. J. Schellnhuber, 2018: Alberta wildfire 2016: Apt contribution from anomalous planetary wave dynamics. Sci. Rep., 8, 12375, https://doi.org/10.1038/s41598-018-30812-z.\\

Rao, V. B., K. K. Rao, B. Mahendranath, T. V Lakshmi Kumar, and D. Govardhan, 2021: Large-scale connection to deadly Indian heatwaves. Q. J. R. Meteorol. Soc., 147, 1419–1430, https://doi.org/https://doi.org/10.1002/qj.3985.\\

Ratnam, J. V, S. K. Behera, S. B. Ratna, M. Rajeevan, and T. Yamagata, 2016: Anatomy of Indian heatwaves. Sci. Rep., 6, 24395, https://doi.org/10.1038/srep24395.\\

Rohini, P., M. Rajeevan, and A. K. Srivastava, 2016: On the Variability and Increasing Trends of Heat Waves over India. Sci. Rep., 6, 26153, https://doi.org/10.1038/srep26153.\\

Roxy, M. K., K. Ritika, P. Terray, R. Murtugudde, K. Ashok, and B. N. Goswami, 2015: Drying of Indian subcontinent by rapid Indian Ocean warming and a weakening land-sea thermal gradient. Nat. Commun., 6, 7423, https://doi.org/10.1038/ncomms8423.\\

Segalini, A., J. Riboldi, V. Wirth, and G. Messori, 2024: A linear assessment of barotropic Rossby wave propagation in different background flow configurations. Weather Clim. Dyn., 5, 997–1012, https://doi.org/10.5194/wcd-5-997-2024.\\

Seneviratne, S. I., D. Lüthi, M. Litschi, and C. Schär, 2006: Land–atmosphere coupling and climate change in Europe. Nature, 443, 205–209, https://doi.org/10.1038/nature05095.
Shaman, J., and E. Tziperman, 2016: The Superposition of Eastward and Westward Rossby Waves in Response to Localized Forcing. J. Clim., 29, 7547–7557, https://doi.org/10.1175/JCLI-D-16-0119.1.\\

Sharma, V. M., G. Pichan, and M. R. Panwar, 1983: Differential effects of hot-humid and hot-dry environments on mental functions. Int. Arch. Occup. Environ. Health, 52, 315–327, https://doi.org/10.1007/BF02226897.\\

Shimizu, M. H., and I. F. de Albuquerque Cavalcanti, 2011: Variability patterns of Rossby wave source. Clim. Dyn., 37, 441–454, https://doi.org/10.1007/s00382-010-0841-z.\\

Simmons, A. J., J. M. Wallace, and G. W. Branstator, 1983: Barotropic Wave Propagation and Instability, and Atmospheric Teleconnection Patterns. J. Atmos. Sci., 40, 1363–1392, \url{https://doi.org/https://doi.org/10.1175/1520-0469(1983)040<1363:BWPAIA>2.0.CO;2.}\\

Sojan, J. M., and J. Srinivasan, 2024: A comparative analysis of accelerating humid and dry heat stress in India. Environ. Res. Commun., 6, 21002, https://doi.org/10.1088/2515-7620/ad2490.\\

Srivastava, A. K., M. Rajeevan, and S. R. Kshirsagar, 2009: Development of a high resolution daily gridded temperature data set (1969–2005) for the Indian region. Atmos. Sci. Lett., 10, 249–254, https://doi.org/https://doi.org/10.1002/asl.232.\\

Taylor, N. A. S., N. Kondo, and W. L. Kenney, 2008: The physiology of acute heat exposure, with implications for human performance in the heat. Physiological Bases of Human Performance During Work and Exercise, Elsevier.\\

Teng, H., and G. Branstator, 2019: Amplification of Waveguide Teleconnections in the Boreal Summer. 2010, 421–432.\\

Ting, M., and L. Yu, 1998: Steady Response to Tropical Heating in Wavy Linear and Nonlinear Baroclinic Models. J. Atmos. Sci., 55, 3565–3582, https://doi.org/10.1175/1520-0469(1998)055<3565:SRTTHI>2.0.CO;2.\\

White, R. H., K. Kornhuber, O. Martius, and V. Wirth, 2022: From Atmospheric Waves to Heatwaves: A Waveguide Perspective for Understanding and Predicting Concurrent, Persistent, and Extreme Extratropical Weather. Bull. Am. Meteorol. Soc., 103, E923–E935, \url{https://doi.org/https://doi.org/10.1175/BAMS-D-21-0170.1}.\\

Wills, R. C. J., R. H. White, and X. J. Levine, 2019: Northern Hemisphere Stationary Waves in a Changing Climate. Curr. Clim. Chang. Reports, 5, 372–389, \url{https://doi.org/10.1007/s40641-019-00147-6}.\\

Wolf, G., D. J. Brayshaw, N. P. Klingaman, and A. Czaja, 2018: Quasi-stationary waves and their impact on European weather and extreme events. Q. J. R. Meteorol. Soc., 144, 2431–2448, https://doi.org/https://doi.org/10.1002/qj.3310.\\

Yang, X., G. Zeng, S. Zhang, V. Iyakaremye, C. Shen, W.-C. Wang, and D. Chen, 2024: Phase-Locked Rossby Wave-4 Pattern Dominates the 2022-Like Concurrent Heat Extremes Across the Northern Hemisphere. Geophys. Res. Lett., 51, e2023GL107106, \url{https://doi.org/https://doi.org/10.1029/2023GL107106}.\\

Zhang, X., and Coauthors, 2023: Increased impact of heat domes on 2021-like heat extremes in North America under global warming. Nat. Commun., 14, 1690, \url{https://doi.org/10.1038/s41467-023-37309-y}.

\newpage

\printendnotes
\defbibnote{preamble}{By default, this template uses \texttt{biblatex} and adopts the Chicago referencing style. However, the journal you’re submitting to may require a different reference style; specify the journal you're using with the class' \texttt{journal} option --- see lines 1--8 of \emph{sample.tex} for a list of options and instructions for selecting the journal.}

\printbibliography[prenote={preamble}]

\begin{appendix}
\onecolumn
\section\centering{Supplementary Figures}
\centering{The supplementary figures are shown below.}\\

\setcounter{figure}{0} 
\renewcommand{\thefigure}{S\arabic{figure}} 
\begin{figure*}[hbt!]
\centering
\includegraphics[width=0.8\linewidth]{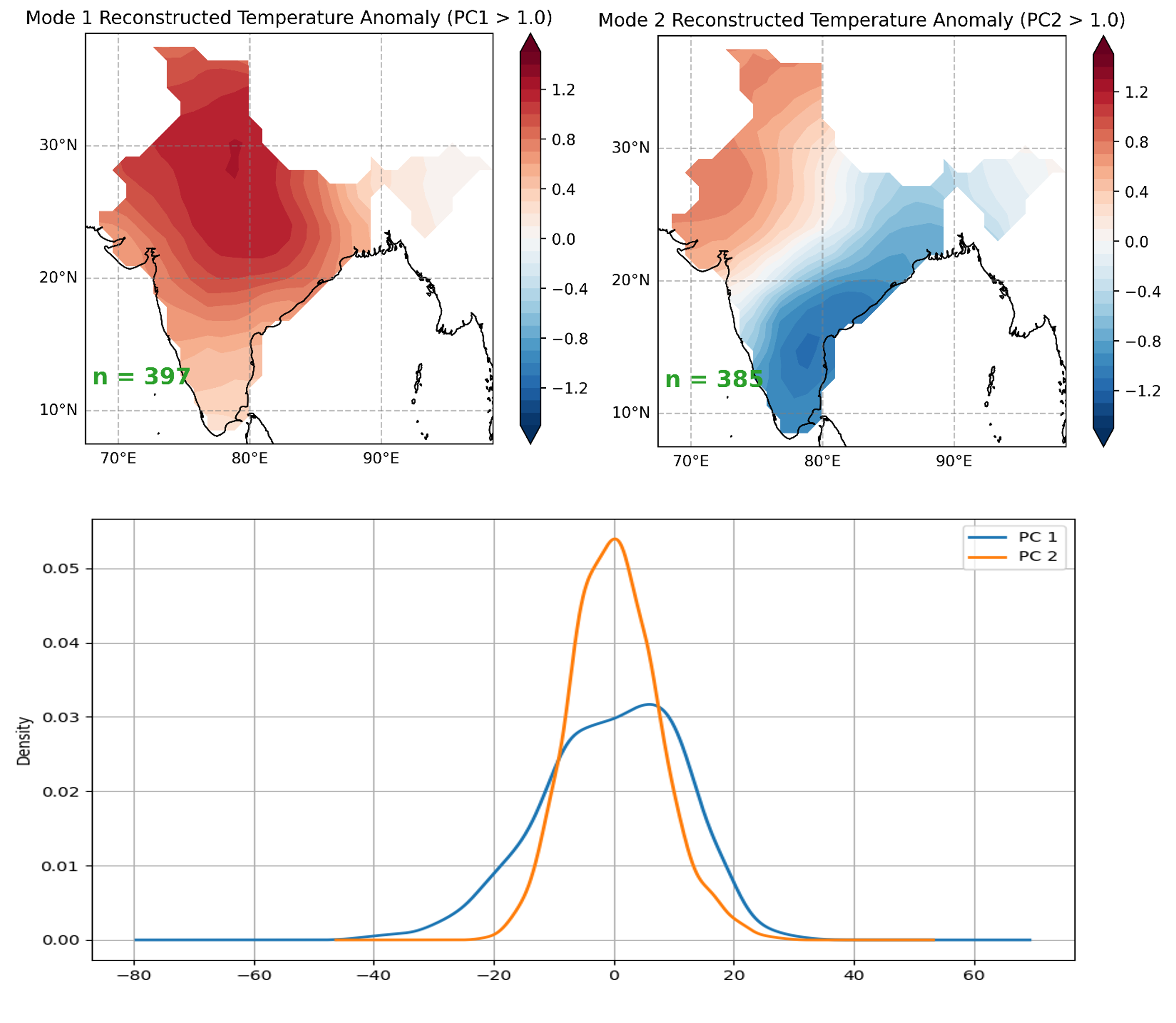}
\caption{Composite of Reconstructed Observed Temperature Anomaly for Mode 1 (EOF 1 x PC 1) and for Mode 2 (EOF 2 x PC 2) for those days with (a) PC 1 > 1.0 and (b) PC 2 > 1.0. The number of days satisfying the criteria is provided as n. (c) Density distribution of the principal components corresponding to EOF 1 and 2, i.e., PC 1 and PC 2.}

\label{fig_wide}
\end{figure*}

\begin{figure*}[hbt!]
\centering
\includegraphics[width=0.8\linewidth]{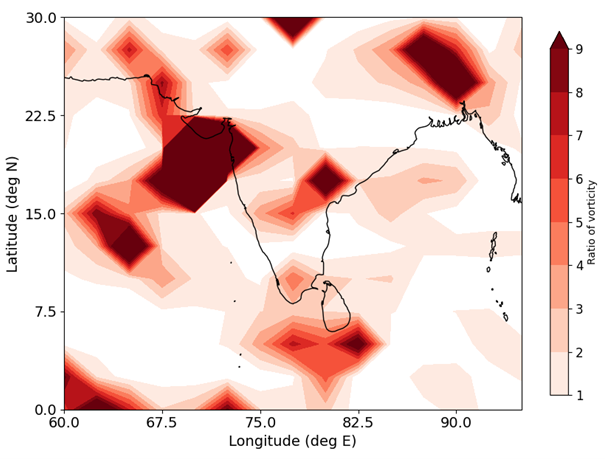}
\caption{Ratio of observed barotropic to baroclinic vorticity ($| (\zeta_{850}+\zeta_{200})/(\zeta_{850}-\zeta_{200}) |$) for those days when PC 2 $>$ 1.0 over the Indian region.}

\label{fig_wide}
\end{figure*}

\begin{figure*}[hbt!]
\centering
\includegraphics[width=0.8\linewidth]{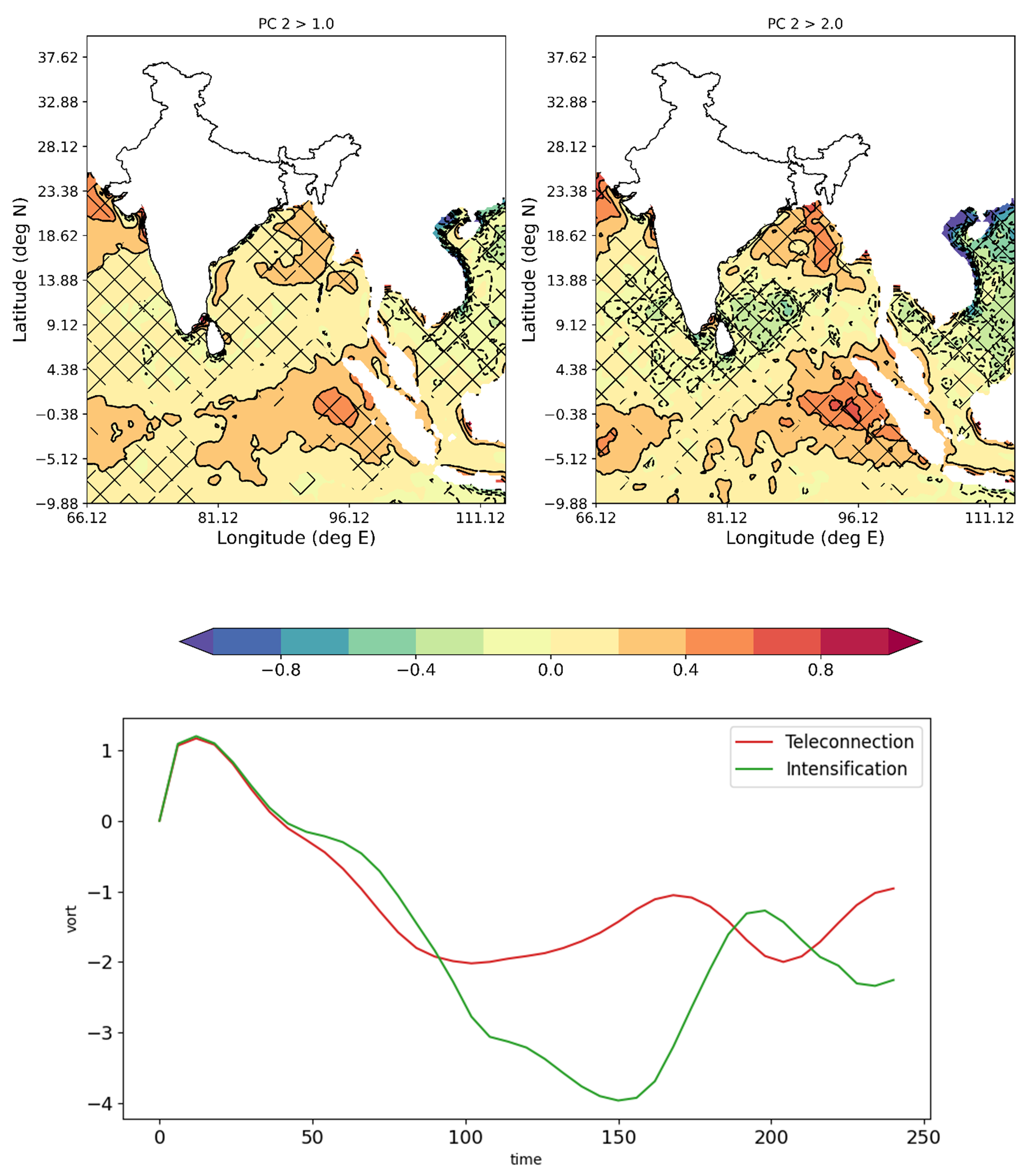}
\caption{ Composite of sea surface temperature anomalies (⁰C; shaded contour) for (a) Active mode days (PC 2 > 1.0) and (b) Extreme mode (PC 2 > 2.0) days during April-May from 1980-2020. The anomalies are calculated with respect to daily climatology from 1980-2020. The hatched regions represent the statistically significant regions (c) Time Series of area-averaged relative vorticity (x 10-5 s-1) over Region 2 in both T(4,34) and I(4,34) experiments.}

\label{fig_wide}
\end{figure*}

\begin{figure*}[hbt!]
\centering
\includegraphics[width=0.8\linewidth]{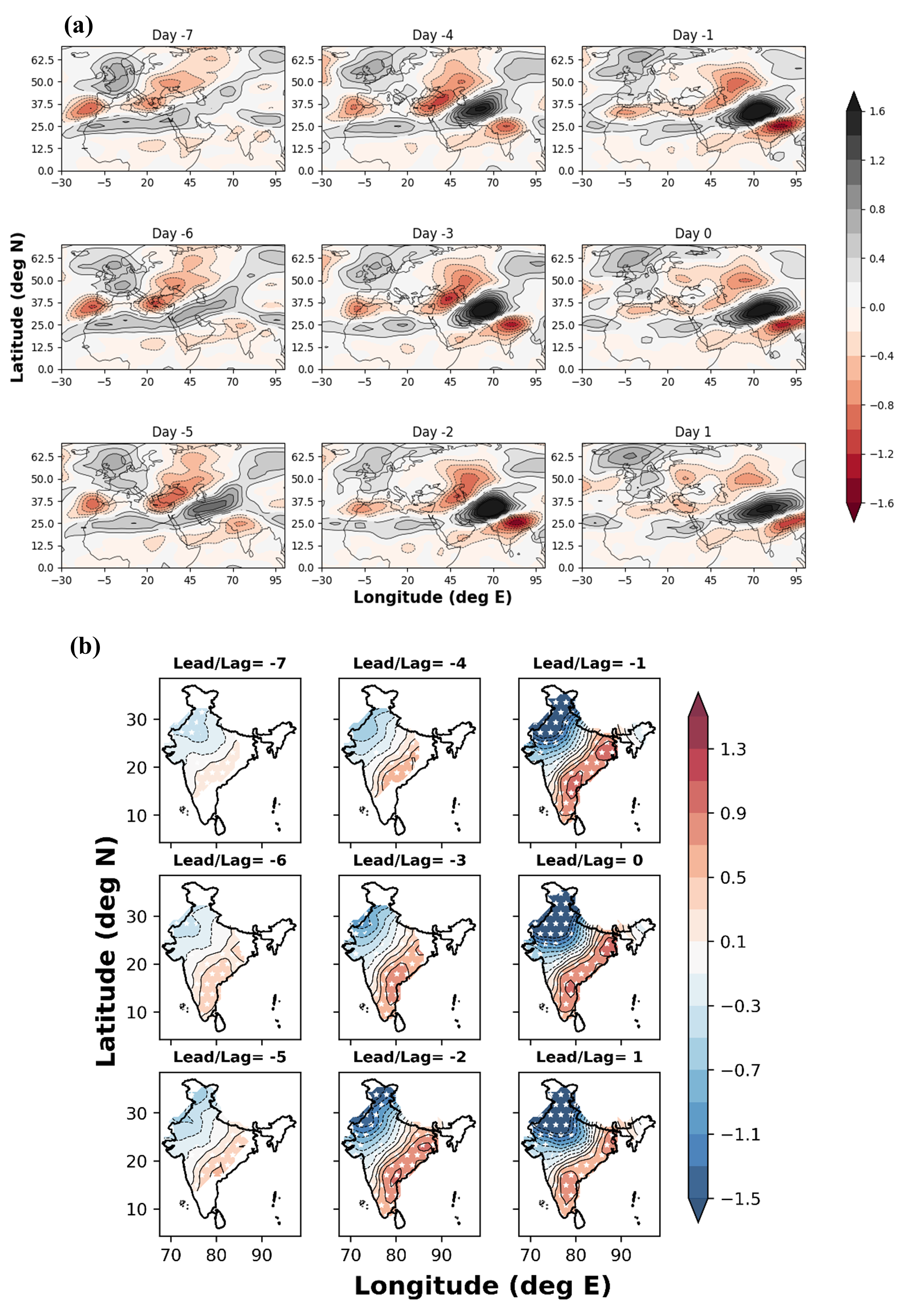}
\caption{(a) Spatial composite of relative vorticity anomaly at 200 hPa for days in R-mode (PC 2 <-1.0) from lag -7 to lead 1 day (b) Spatial composite of surface temperature anomaly for the same as in Fig S3a.}

\label{fig_wide}
\end{figure*}

\begin{figure*}[hbt!]
\centering
\includegraphics[width=0.8\linewidth]{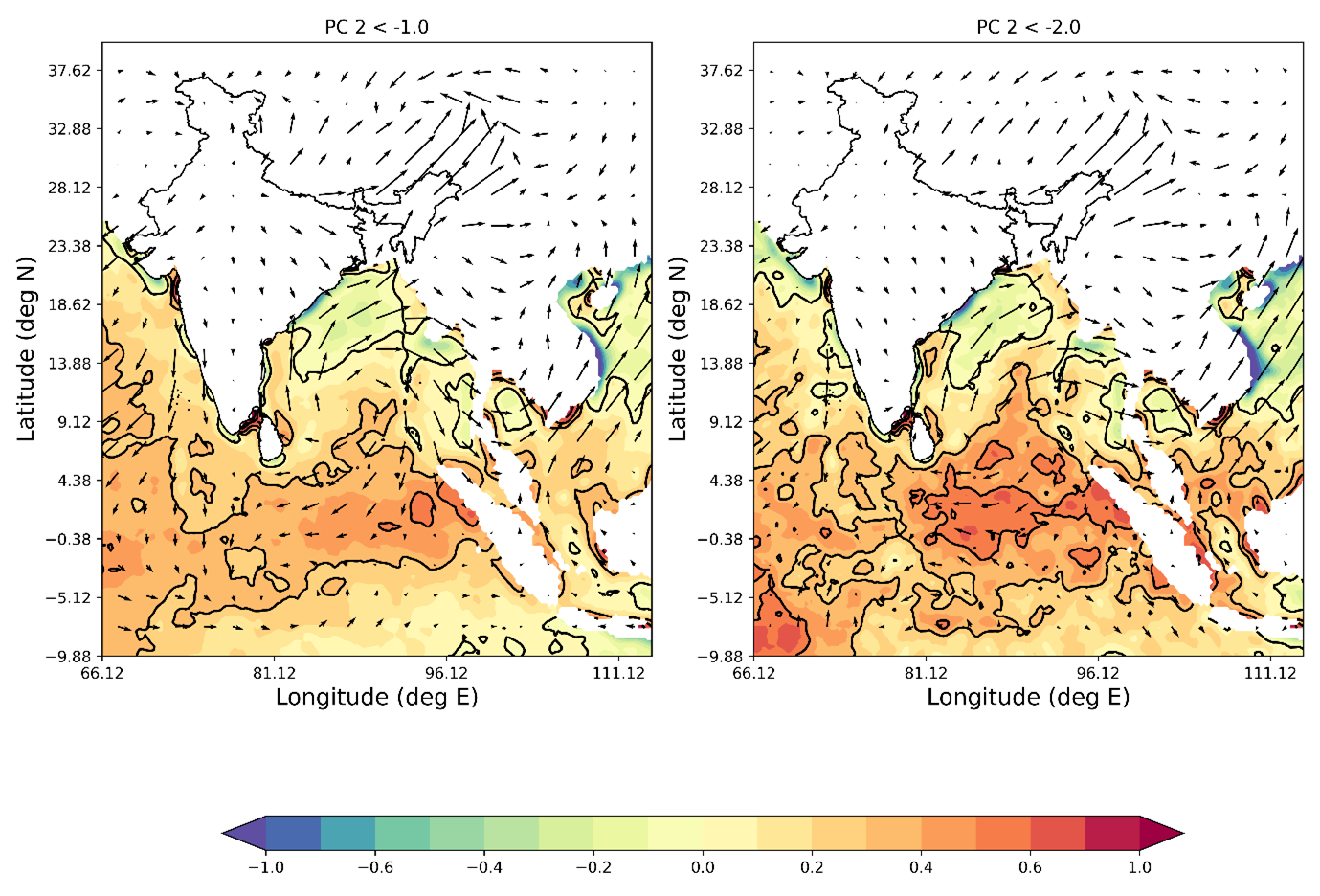}
\caption{Composite of sea surface temperature anomalies (⁰C; shaded contour) superimposed with 1000-300 hPa vertically integrated moisture flux (IMF; vectors) anomalies ($\mathrm{ms}^{-1}\,\mathrm{g}\,\mathrm{kg}^{-1}$) for (a) Active R-Mode days (PC 2 < - 1.0) and (b) Extreme R-Mode (PC 2 < -2.0) days during April-May from 1980-2020. The anomalies are calculated with respect to daily climatology from 1980-2020. }

\label{fig_wide}
\end{figure*}

\begin{figure*}[hbt!]
\centering
\includegraphics[width=0.8\linewidth]{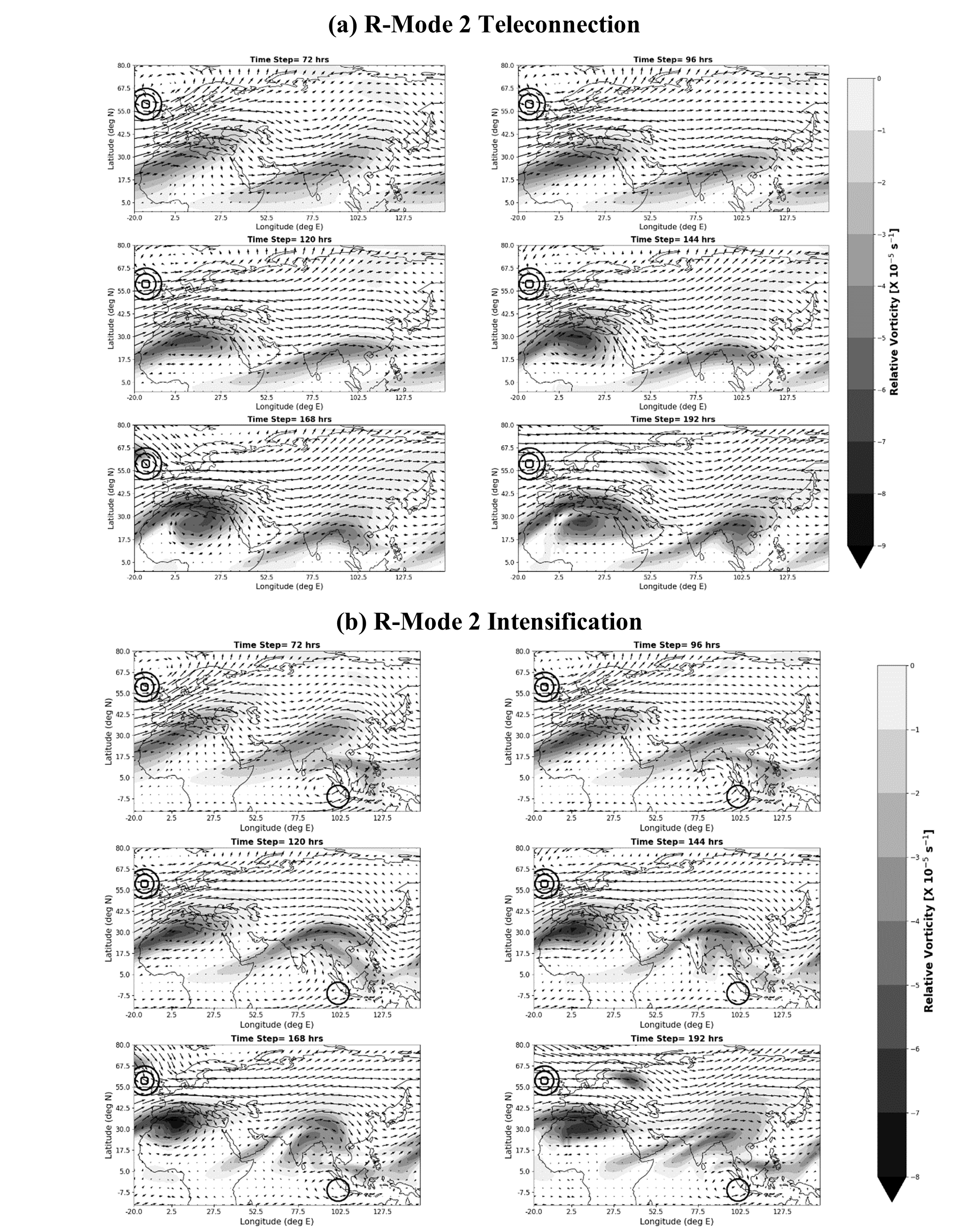}
\caption{Spatial pattern of equivalent barotropic model predicted relative vorticity and wind vectors at different time steps from t=72 hrs to t=192 hrs for the R-Mode control experiments (a) T(4,34) Teleconnection and (b) I(4,34) Intensification experiment. The closed circles show the mid-latitudinal cyclonic forcing F1 in Source Region 1 and tropical cyclonic forcing F2 in Source Region 2.}

\label{fig_wide}
\end{figure*}

\begin{figure*}[hbt!]
\centering
\includegraphics[width=0.8\linewidth]{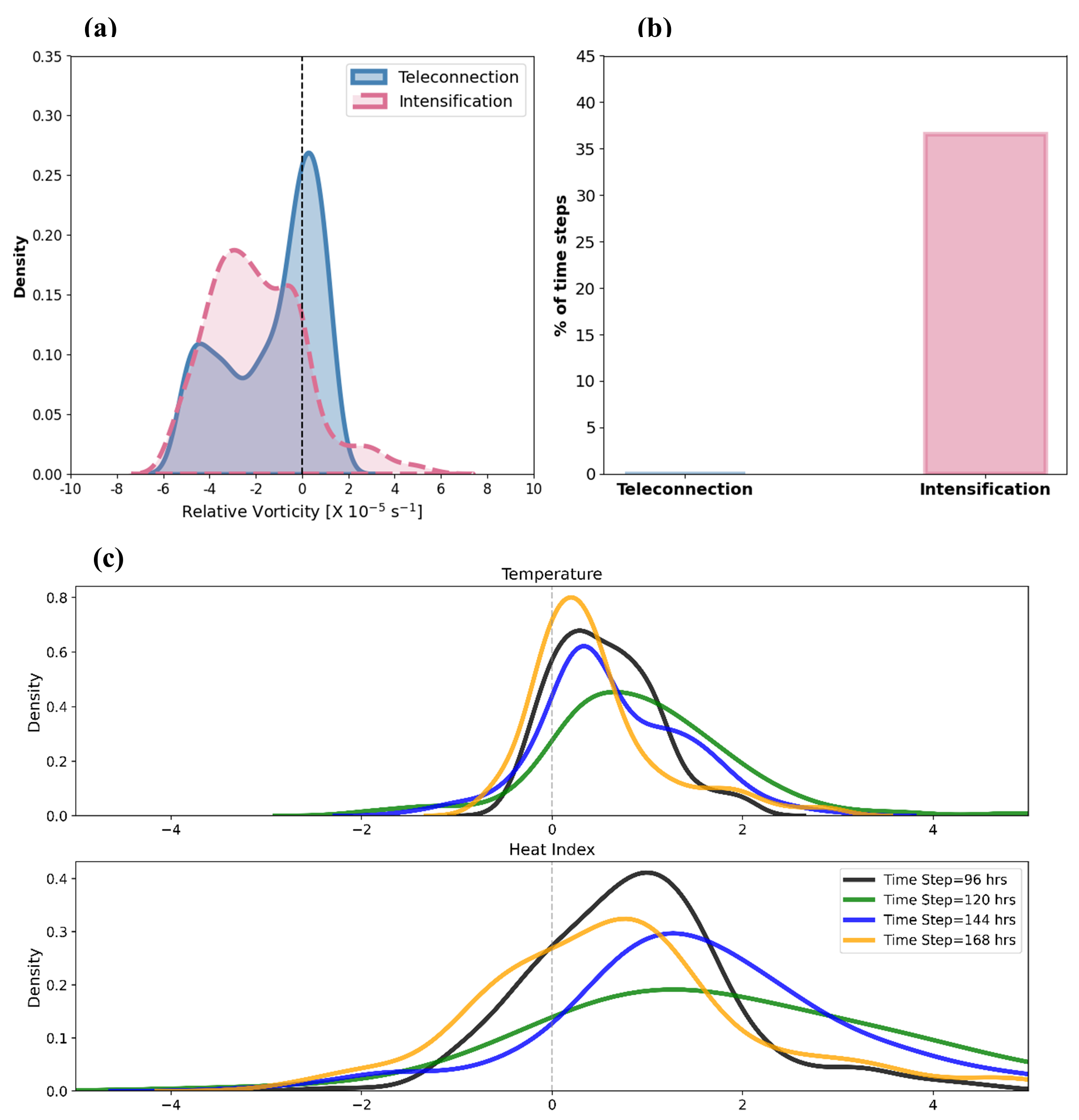}
\caption{(a) Density distribution of relative vorticity (\( \times 10^{-5}\,\mathrm{s}^{-1}\))  over the Indian region (Region 2) for model time steps from t=126 to t=174 hrs for R-Mode Teleconnection experiment and R-Mode Intensification Experiment. (b) Percentage of time steps when the average of vorticity over the Indian region falls below (\(2 \times 10^{-5}\,\mathrm{s}^{-1}\)) for the same set of R-Mode experiments. (c) The density distribution of the maximum temperature difference (upper panel) and HI (lower panel) estimated from model predicted vorticity between Intensification and Teleconnection experiments of R-Mode is shown for different time steps.}

\label{fig_wide}
\end{figure*}

\begin{figure*}[hbt!]
\centering
\includegraphics[width=0.8\linewidth]{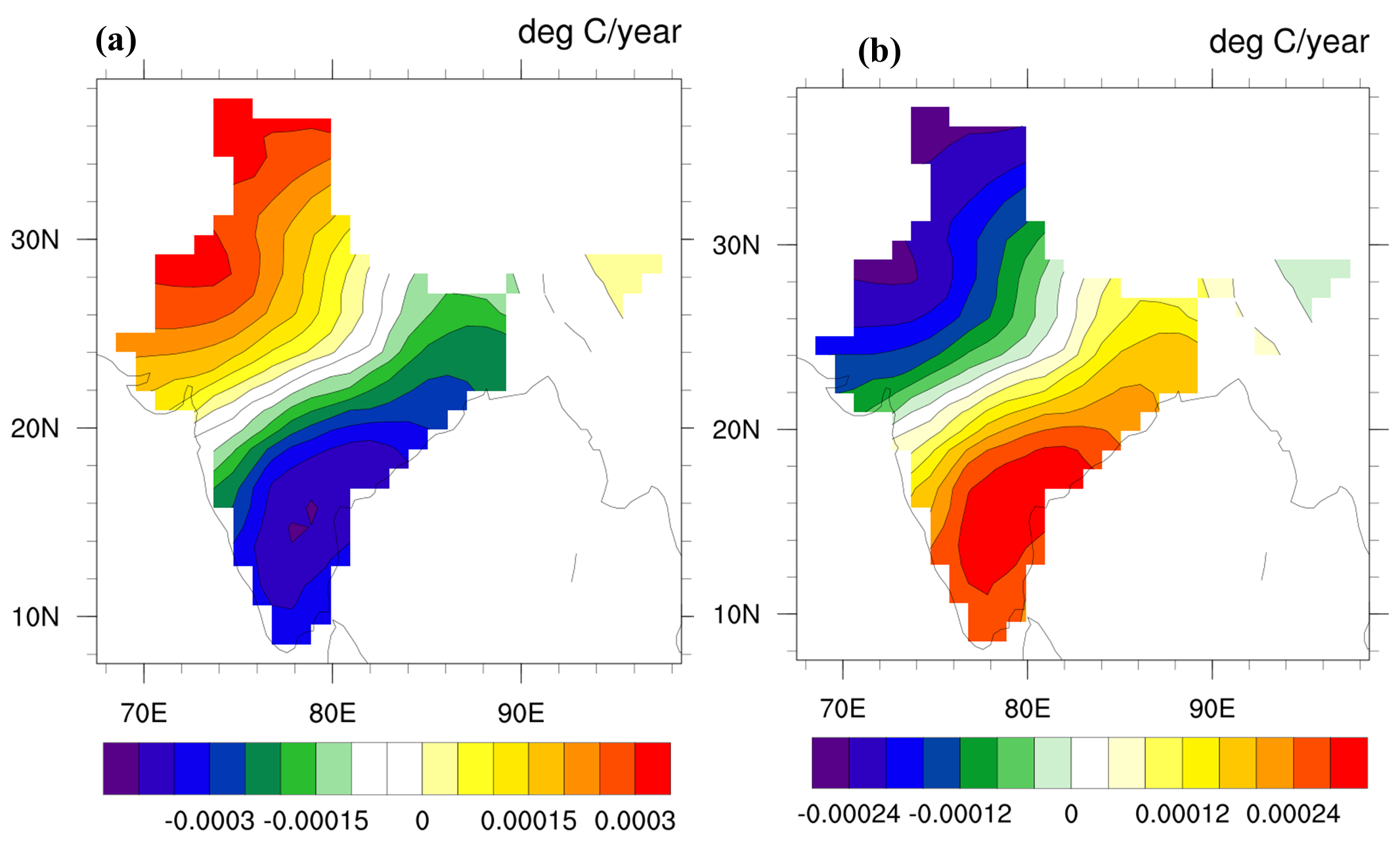}
\caption{Long-term trend in the reconstructed mode 2 (EOF 2 × PC 2) temperature pattern over the Indian region during (a) 1980-2020 and (b) 2000-2020. }

\label{fig_wide}
\end{figure*}

\begin{figure*}[hbt!]
\centering
\includegraphics[width=0.8\linewidth]{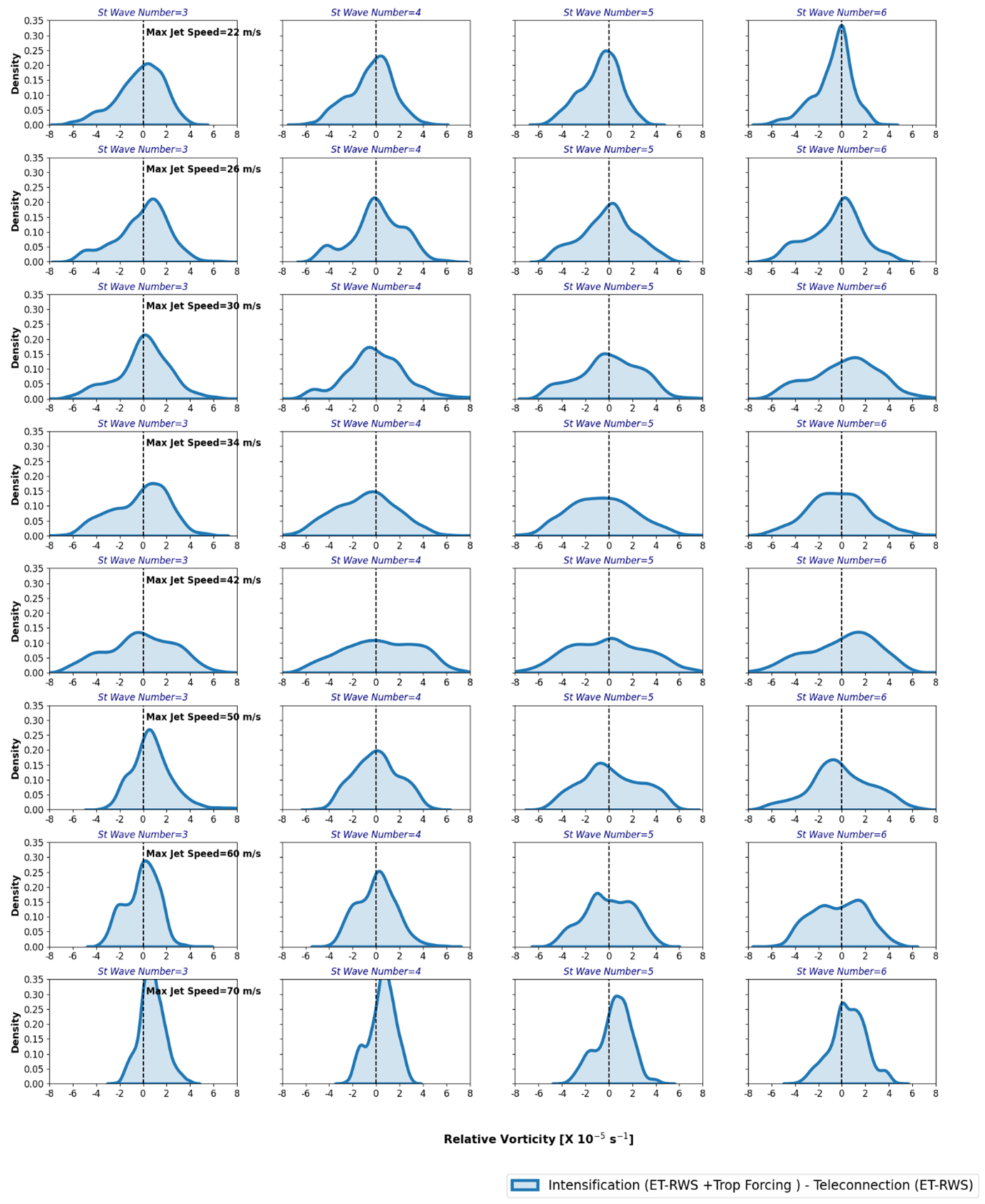}
\caption{Density distribution of the difference in relative vorticity between different TS (ET-RWS) and IS (ET-RWS + Tropical Heat Source) experiments averaged over a box in the Indian region (Region 2) for model time steps from t=72 to t=216 hrs. Each row represents a set of zonal wind configurations, with each column representing different wave numbers of vorticity perturbation configuration.}

\label{fig_wide}
\end{figure*}


\twocolumn
\justify
\section*{ 2. Supplementary Material SW: Supplementary Materials on Stationary Wave (SW) experiments}
We have seen the process of intensification of circulation related to Mode 2 teleconnection over the Indian region in the presence of the initial wave perturbation, mid-latitudinal Rossby wave source, and a regional vorticity forcing over the Bay of Bengal. Each of these forcing plays a significant role. To understand the role of the phase (location of cyclonic and anticyclonic circulation of the wave) of the initial wave perturbation (of different wavenumbers, additional analysis is performed and is reported in this supplementary material.

In this section, we provide plots related to the additional experiments related to the Initial largescale waves (representing the stationary wave) to highlight the importance of large-scale background, superpositions, and intensification. The structure of this large-scale forcing is as follows:

\begin{equation*}
\zeta^{\prime}(z, t=0)=\cos (\theta) e^{-\left[\left(\theta-\theta_{0}\right) / \theta_{w}\right]^{2}} e^{i z \lambda} \tag{4}
\end{equation*}

In the supplementary text below, we will discuss the following things:

\begin{enumerate}
  \item Impact of a spatial phase shift of the initial wave $\zeta^{\prime}(z, t=0)$ pattern on the amplification. We will add spatial phase shifts $\delta=0.25,0.5,0.75$, and 1 (radian) from the initial given locations $\delta=0$ as shown in Fig.SW-1 and show the amplification ratio and spatial pattern. For $\delta=0$, the anticyclone peaks are located at $\sim 50^{\circ}, 140^{\circ}, 230^{\circ}$, and $320^{\circ}$ longitude points. Thus, the phase shifts are given accordingly.
  \item Comparison of the spatial pattern of intensification for the superposition runs, i.e., TR $(4,34)=[\mathrm{T}(4,34)+\mathrm{R}(4,34)]$ and the sum of all forcings $\zeta^{\prime}, F 1$ and $F 2$ i.e. $I(4,34)$, for different phase shifts of $\zeta^{\prime}(z, t=0),[\delta=0,0.25 .5 .0 .75]$. Refer to Table 1 in the main text for details on $\boldsymbol{I}(4,34), \boldsymbol{T}(4,34)$, and $\mathrm{R}(4,34)$.
  \item Comparison of the amplitude of intensification for the temporal evolution of the superposed runs TR $(4,34)$ and $\mathrm{I}(4,34)$, and other forcing combinations are possible. Taking a case of $(\mathrm{z}=4$, $\bar{u}=34$ ), the evolution is shown. We are interested in looking at how close the temporal evolution of TR and I runs go together. This is measured by RMSE for the sum of amplitudes and correlation for phase.
  \item A discussion on the amplitude of vorticity over India due to initial forcing for different jet speeds and different wavenumbers, and comparison with the amplitude of different forcings.
  \item A discussion on the linearity of superposition, taking the case of ( $\mathrm{z}=4, \bar{u}=34$ ).
  \item Experiments showing the heatmap of intensification for different magnitudes of the initial amplitude of forcing.
\end{enumerate}

\section*{2.1 Phase shifts of Stationary Wave and amplification: }
Consider an initial structure as given in Fig.5b, and is reproduced below as Figure SW1:\\
\setcounter{figure}{0} 
\renewcommand{\thefigure}{SW\arabic{figure}} 
\begin{figure}[hbt!]
\vspace{-1.5cm}
\includegraphics[width=0.8\linewidth]{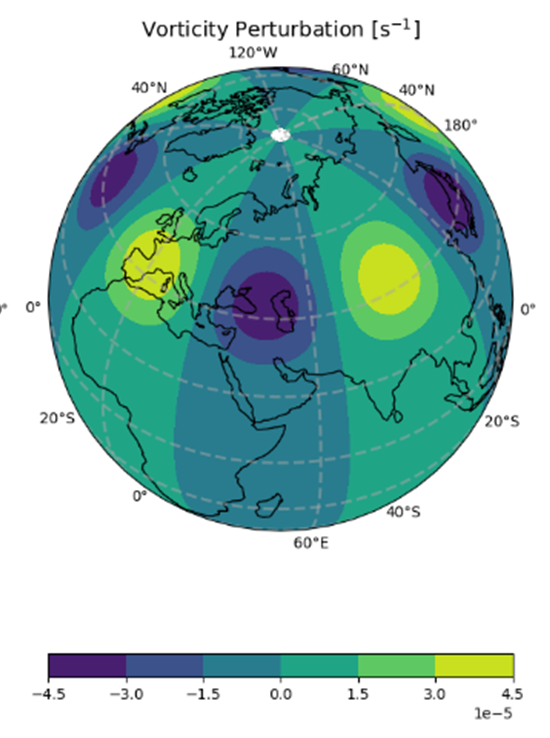}
\caption{The initial structure of wave amplitude used in the paper.}

\label{fig_wide}
\end{figure}

Let’s say that the phase shift of this wave is $\delta$=0. Then we shift the phase of this wave as $\delta$=0.25, 0.5, 0.75 and 1 radian. The figure below shows the zoomed-in version of the original phase $\delta$=0 (shaded) and the shifted patterns:\\

\begin{figure}[hbt!]
\vspace{-0.5cm}
\includegraphics[width=1.15\linewidth,height=0.27\textheight]{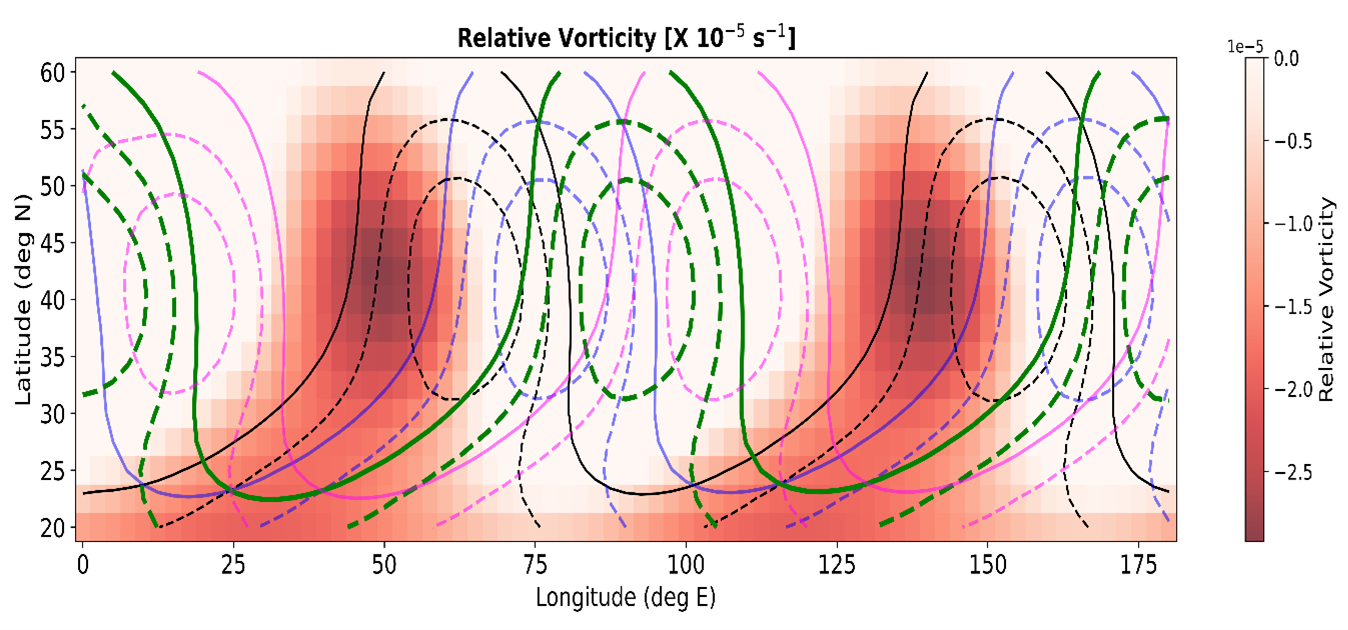}
\caption{Zoomed in version of phase locations $\delta$=0 (shaded), 0.25(black contours),0.5(blue),0.75(green) and 1(pink). Each roughly reflects a phase shift of 25\degree in longitude.}

\label{fig_wide}
\end{figure}
\onecolumn
How does the intensification occur in the presence of these phase-shifted waves at t=0? To understand this, we compared the amplification ratio of T(4,34) and I(4,34) run over the Indian region (like Fig.9ab in the text).
\end{appendix}  

\setcounter{figure}{2} 
\renewcommand{\thefigure}{SW\arabic{figure}} 
\begin{figure*}[hbt!]
\vspace{-1.5cm}
\includegraphics[width=0.95\linewidth,height=0.7\textheight]{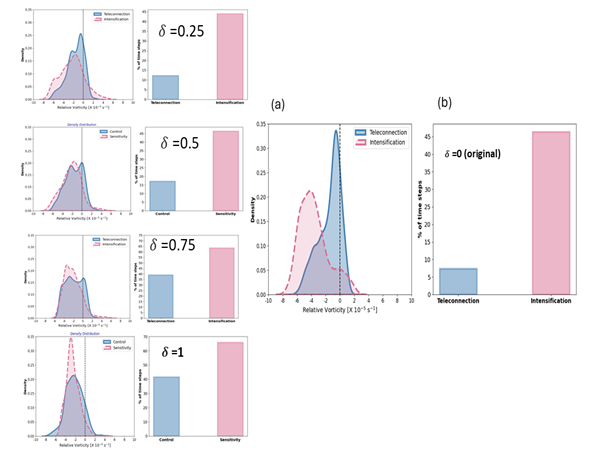}
\caption{The probability distribution function of the amplitude of relative vorticity for different phase-shifted initial waves and their amplification. This is drawn in the same way as Fig.9a and 9b in the text. }

\label{fig_wide}
\end{figure*}

The figure indicates that the amplification occurs for a wide range of phase shifts of the initial wave over the Indian region. Since the averaging area is kept fixed, and if there is a slight displacement in the superposition, the location of maximum superposition and amplification would change. Hence, to have a clear idea, the next few figures will show the spatial map. From the map plot, we also see that superposition occurs and hence amplification. 

\begin{figure*}[hbt!]
\vspace{-0.5cm}
\includegraphics[width=0.95\linewidth,]{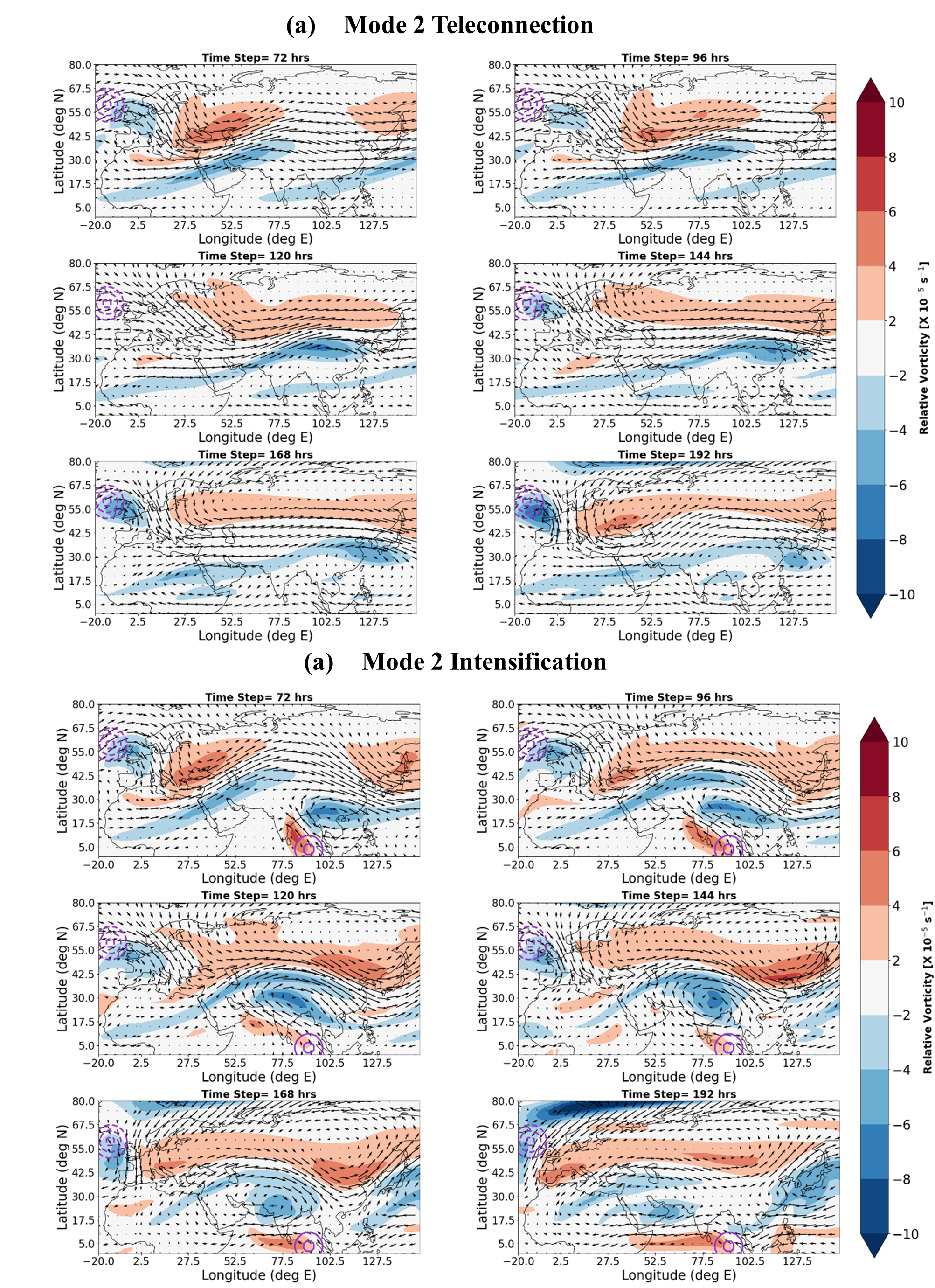}
\caption{Spatial pattern of Mode-2 (moist mode) teleconnection (T(T(4,34) and Intensification I(4,34) experiments show the evolution of relative vorticity and wind for $\delta$=0. The above figure is the colour version of Fig.6 in main text. }

\label{fig_wide}
\end{figure*}

\begin{figure*}[hbt!]
\vspace{-0.5cm}
\includegraphics[width=0.95\linewidth,]{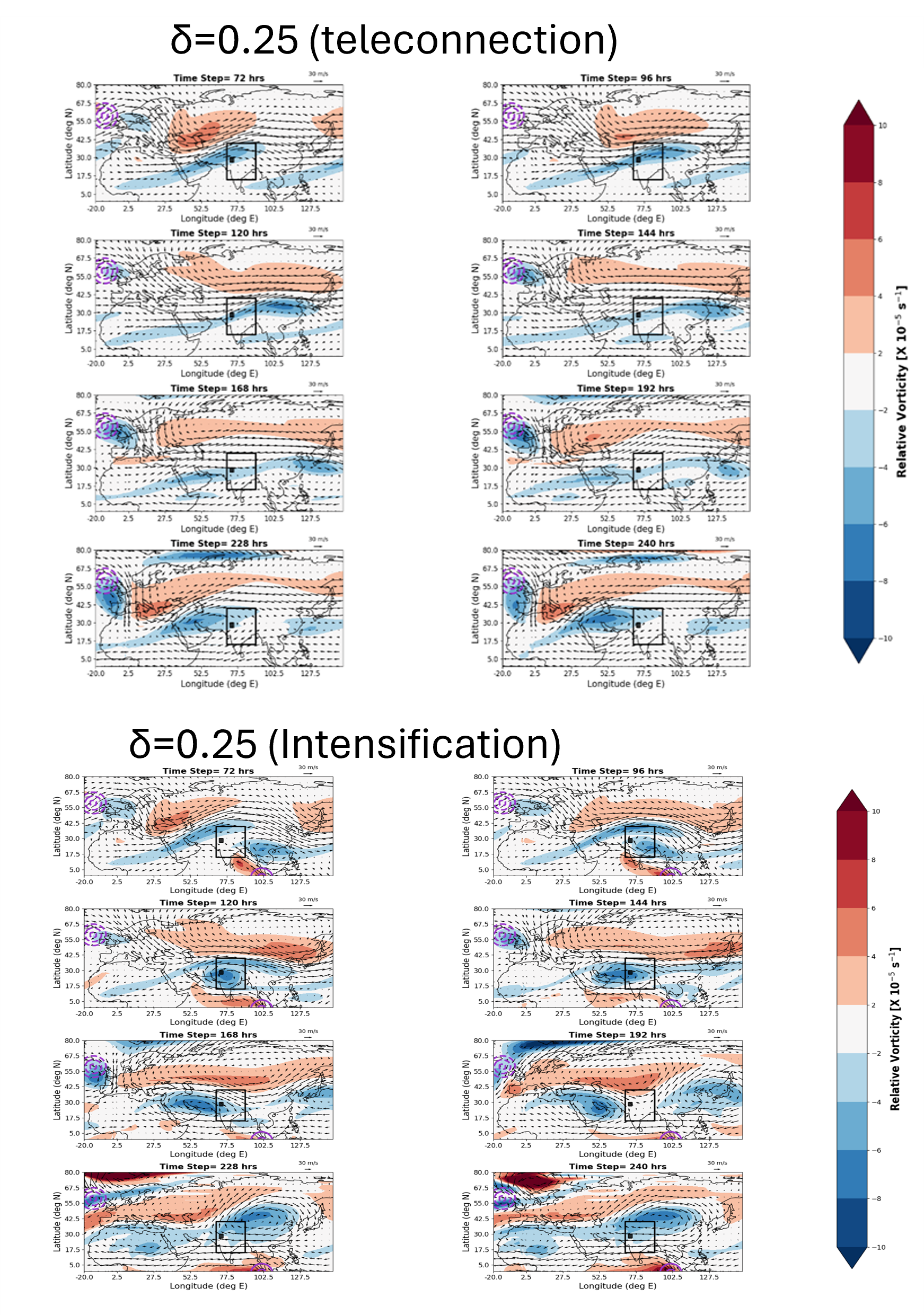}
\caption{Same as Fig. SW-4 but for $\delta$=0.25. }

\label{fig_wide}
\end{figure*}

\begin{figure*}[hbt!]
\vspace{-0.5cm}
\includegraphics[width=0.95\linewidth,]{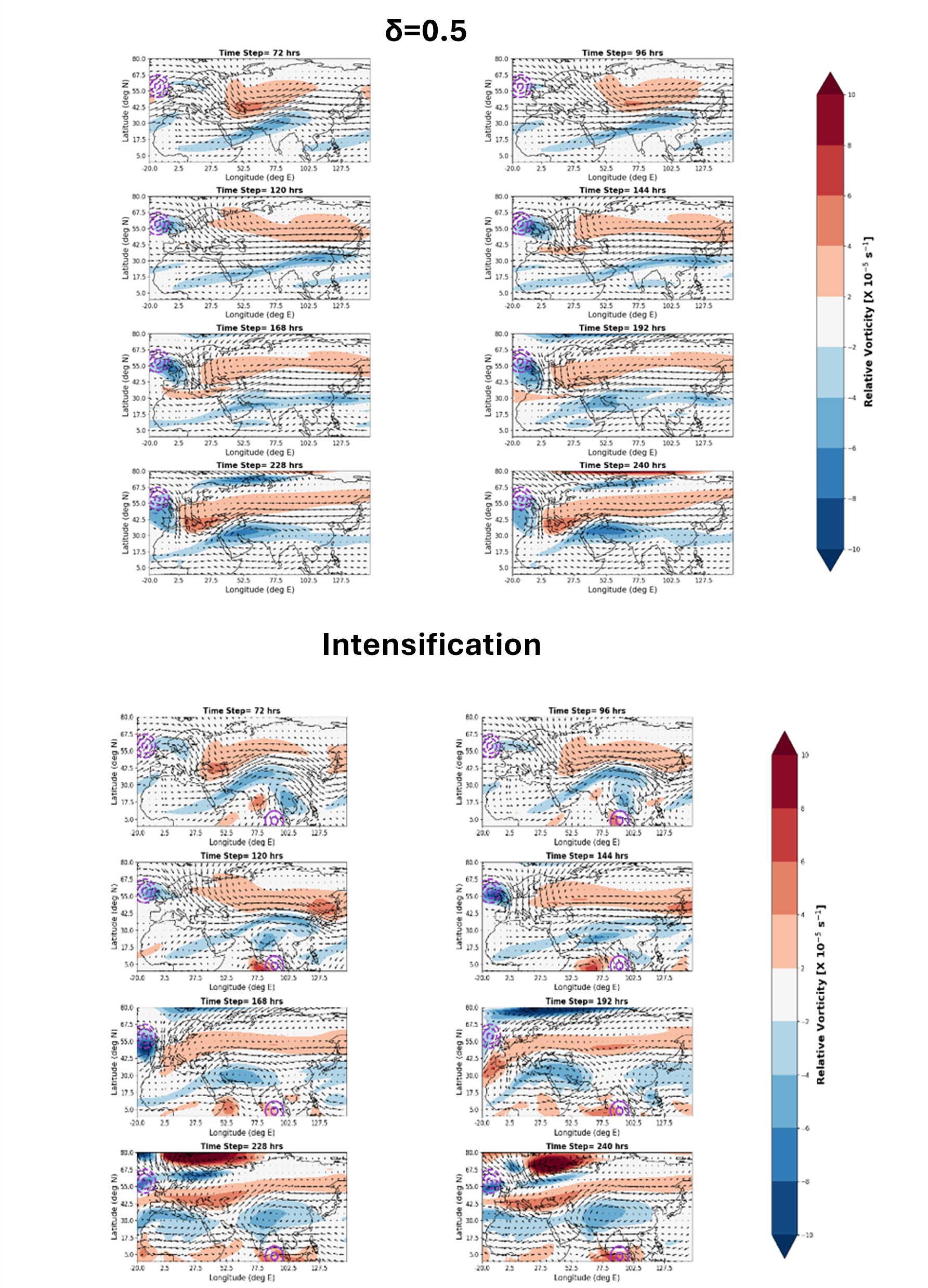}
\caption{Same as Fig. SW-4  $\delta$=0.5. }

\label{fig_wide}
\end{figure*}

\begin{figure*}[hbt!]
\vspace{-0.5cm}
\includegraphics[width=0.95\linewidth,height=0.9\textheight]{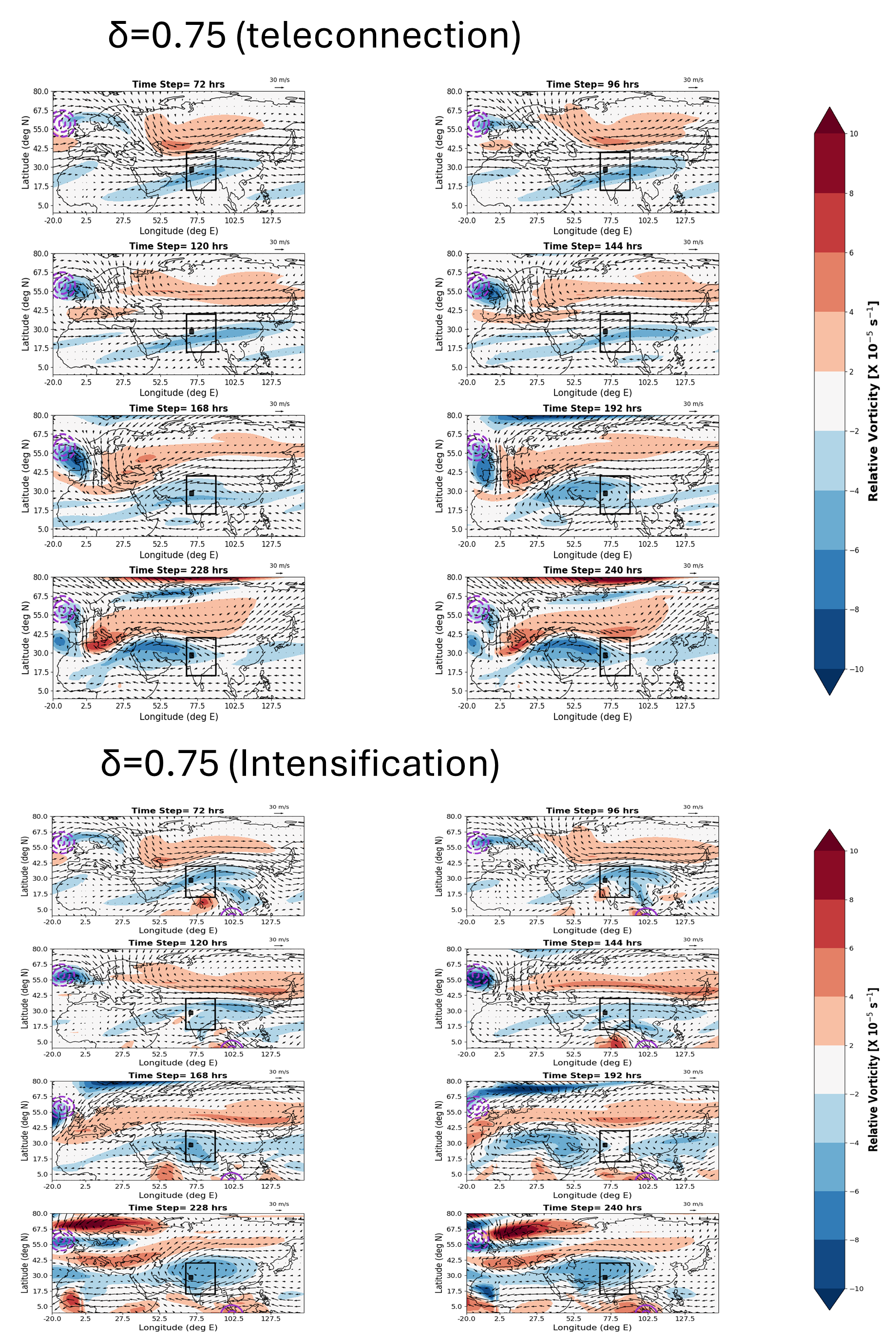}
\caption{Same as Fig. SW-4  $\delta$=0.75. }

\label{fig_wide}
\end{figure*}

\begin{figure*}[hbt!]
\vspace{-0.5cm}
\includegraphics[width=0.95\linewidth,height=0.9\textheight]{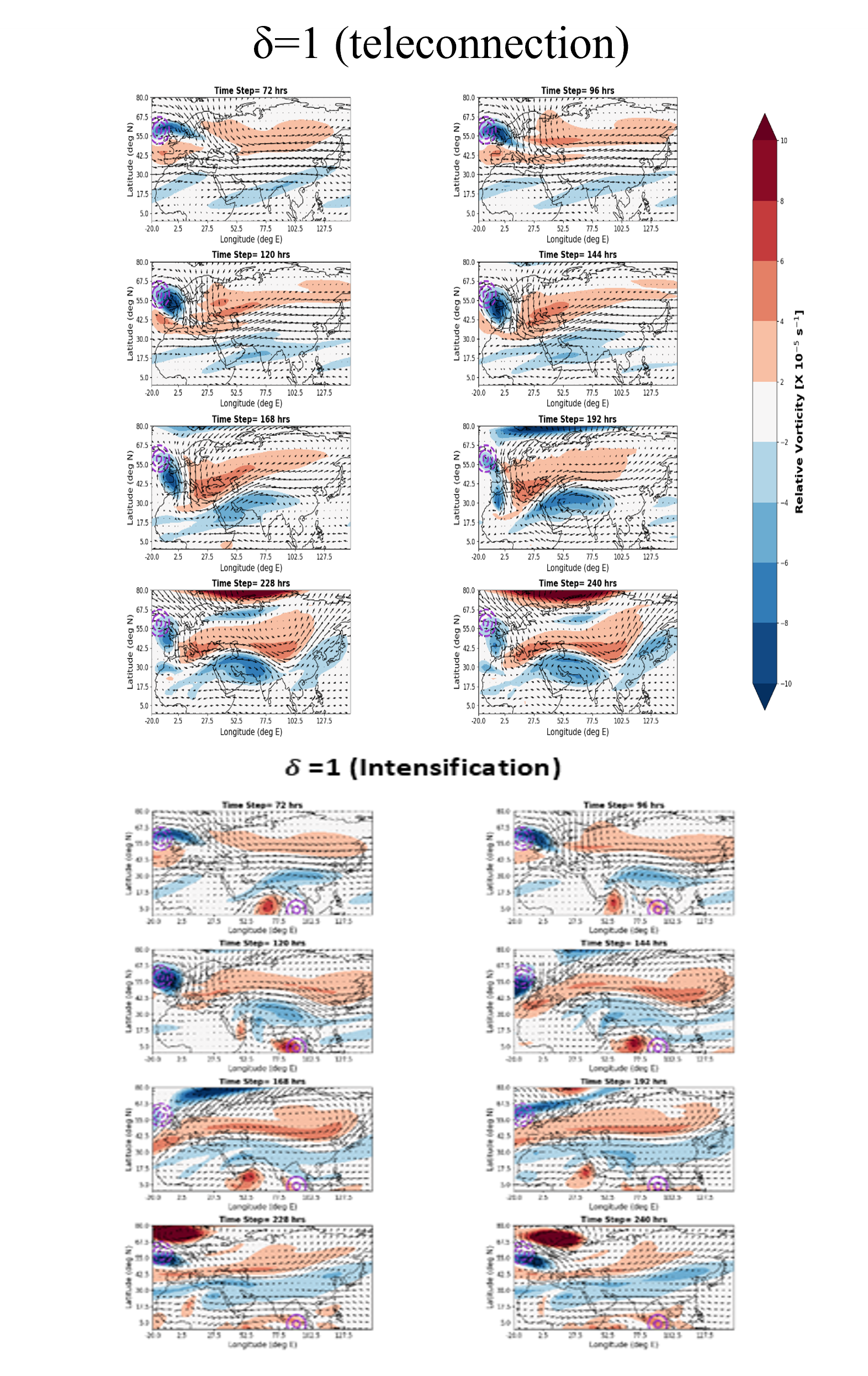}
\caption{Same as Fig. SW-4  $\delta$=1. }

\label{fig_wide}
\end{figure*}
\clearpage
\section*{2.2 Phase shifts of Stationary Wave and amplification: The superposition plots for different phase shifts (top) and comparison with Intensification (I(4,34) run:}

\begin{figure*}[hbt!]
\vspace{-0.5cm}
\includegraphics[width=0.95\linewidth,height=0.4\textheight]{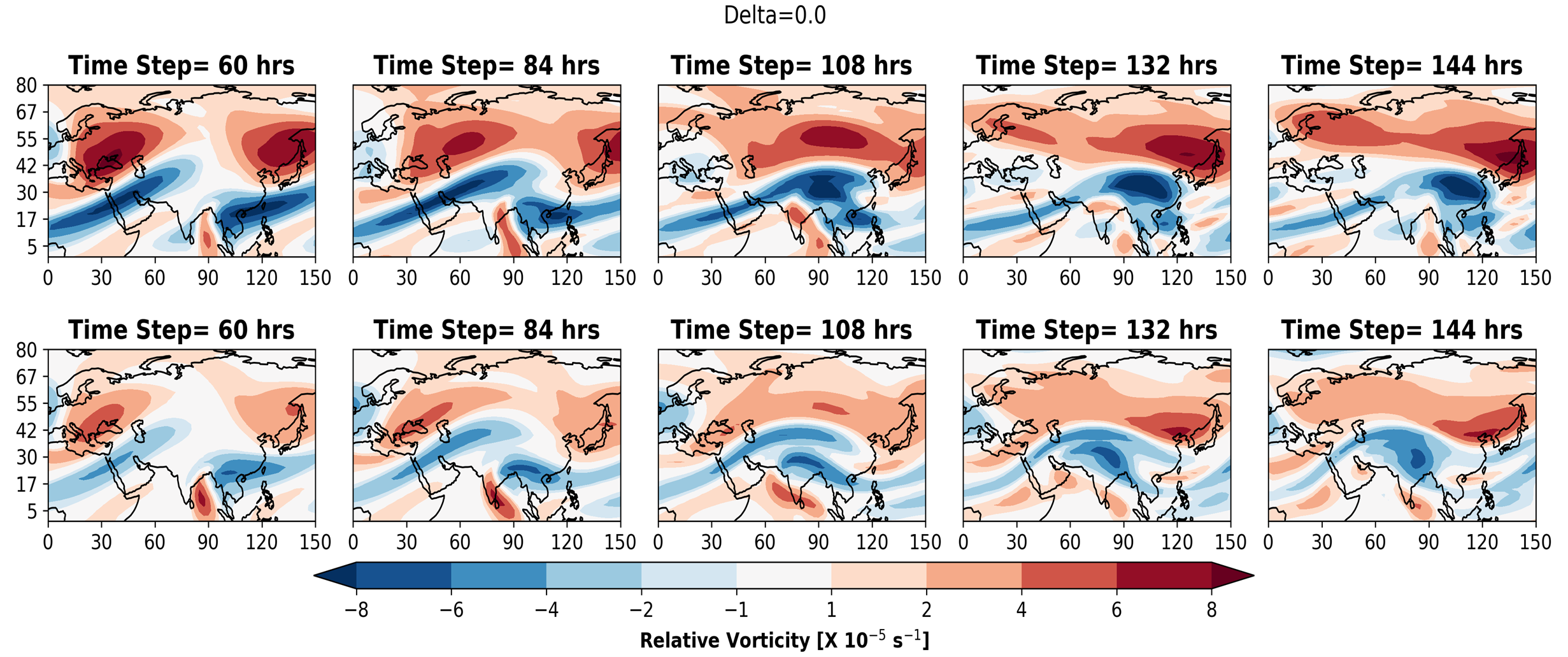}
\caption{Spatial pattern of superposition TR (4,34) =[T(4,34) +R (4,34)] as given in Table 1 in text compared with I (4,34) run as seen in Fig.8b for $\delta$=0.0.  All the runs have large-scale forcing z at t=0. The top panels show the TR (4,34) run at different model time steps, while the bottom panels show I (4,34) for the same time steps. Only a few steps are shown. Refer to Table 1 in the text for experiment nomenclature. }

\label{fig_wide}
\end{figure*}
\begin{figure*}[hbt!]
\vspace{-0.5cm}
\includegraphics[width=0.95\linewidth,height=0.4\textheight]{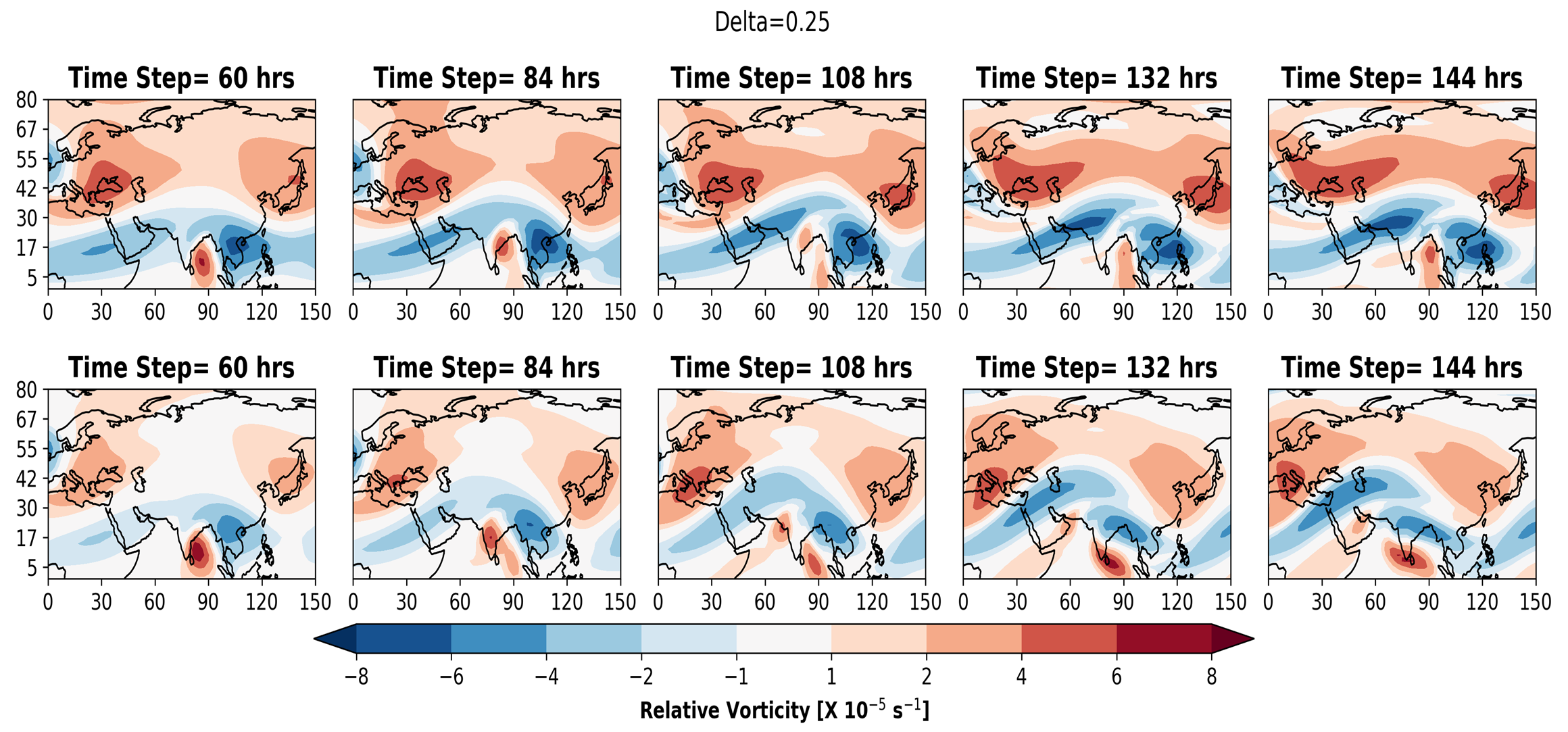}
\caption{ Same as Fig. SW-9 for $\delta$=0.25}

\label{fig_wide}
\end{figure*}

\begin{figure*}[hbt!]
\vspace{-0.5cm}
\includegraphics[width=0.95\linewidth,height=0.4\textheight]{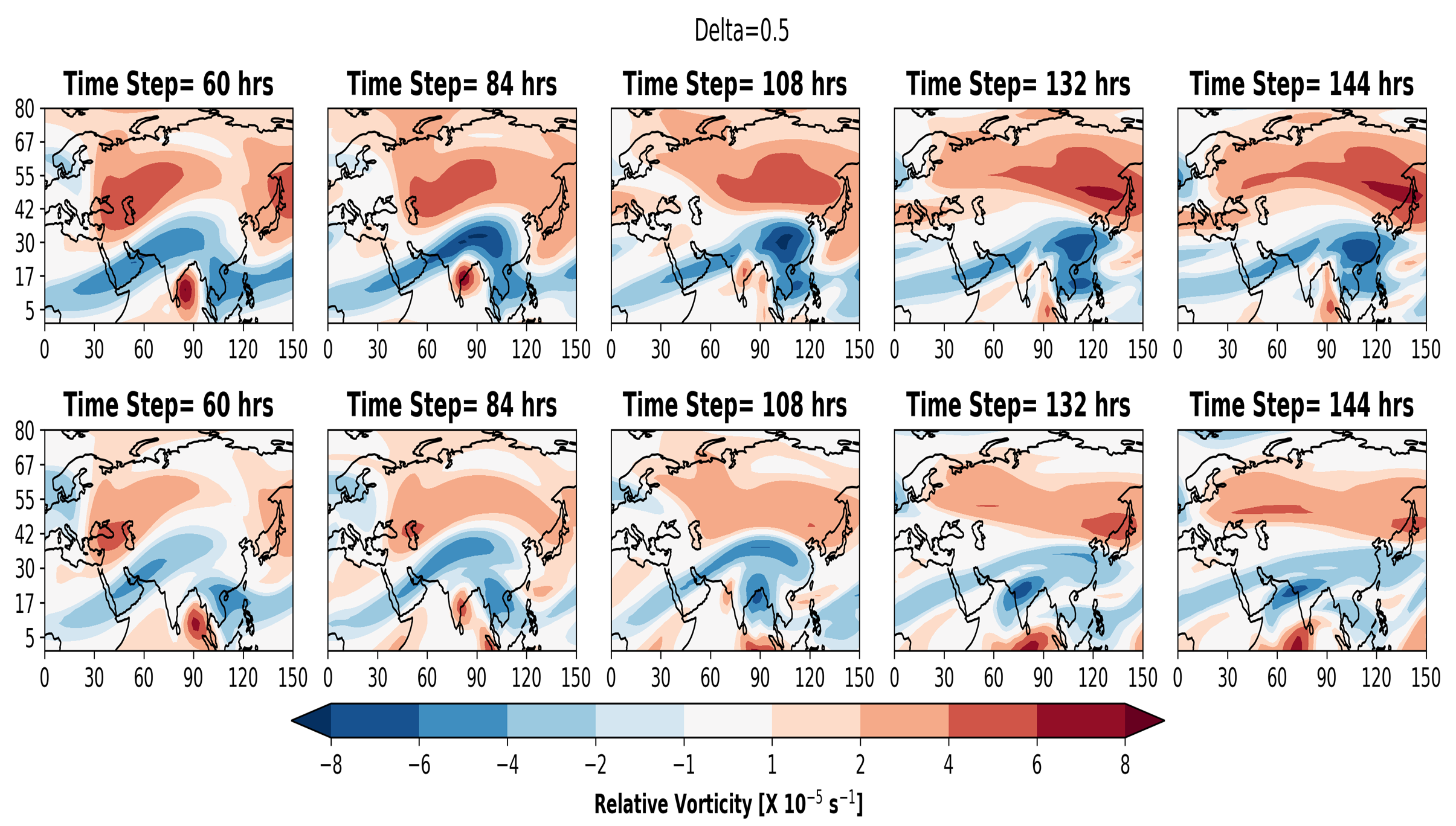}
\caption{Same as Fig. SW-9, for $\delta$=0.5}
\label{fig_wide}
\end{figure*}

\begin{figure*}[hbt!]
\vspace{-0.5cm}
\includegraphics[width=0.95\linewidth,height=0.4\textheight]{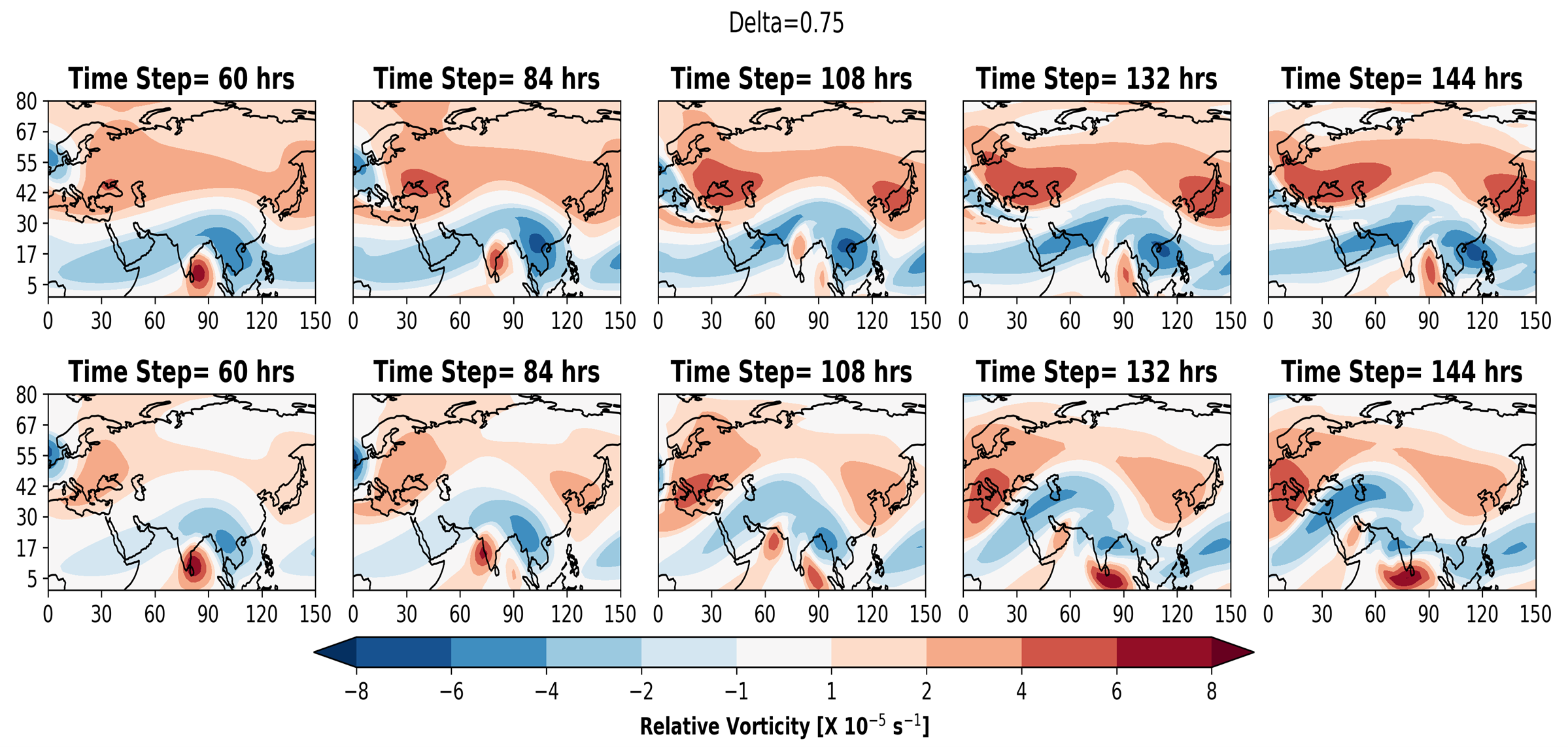}
\caption{ Same as Fig. SW-9, for $\delta$=0.75}
\label{fig_wide}
\end{figure*}

\begin{figure*}[hbt!]
\vspace{-0.5cm}
\includegraphics[width=0.95\linewidth]{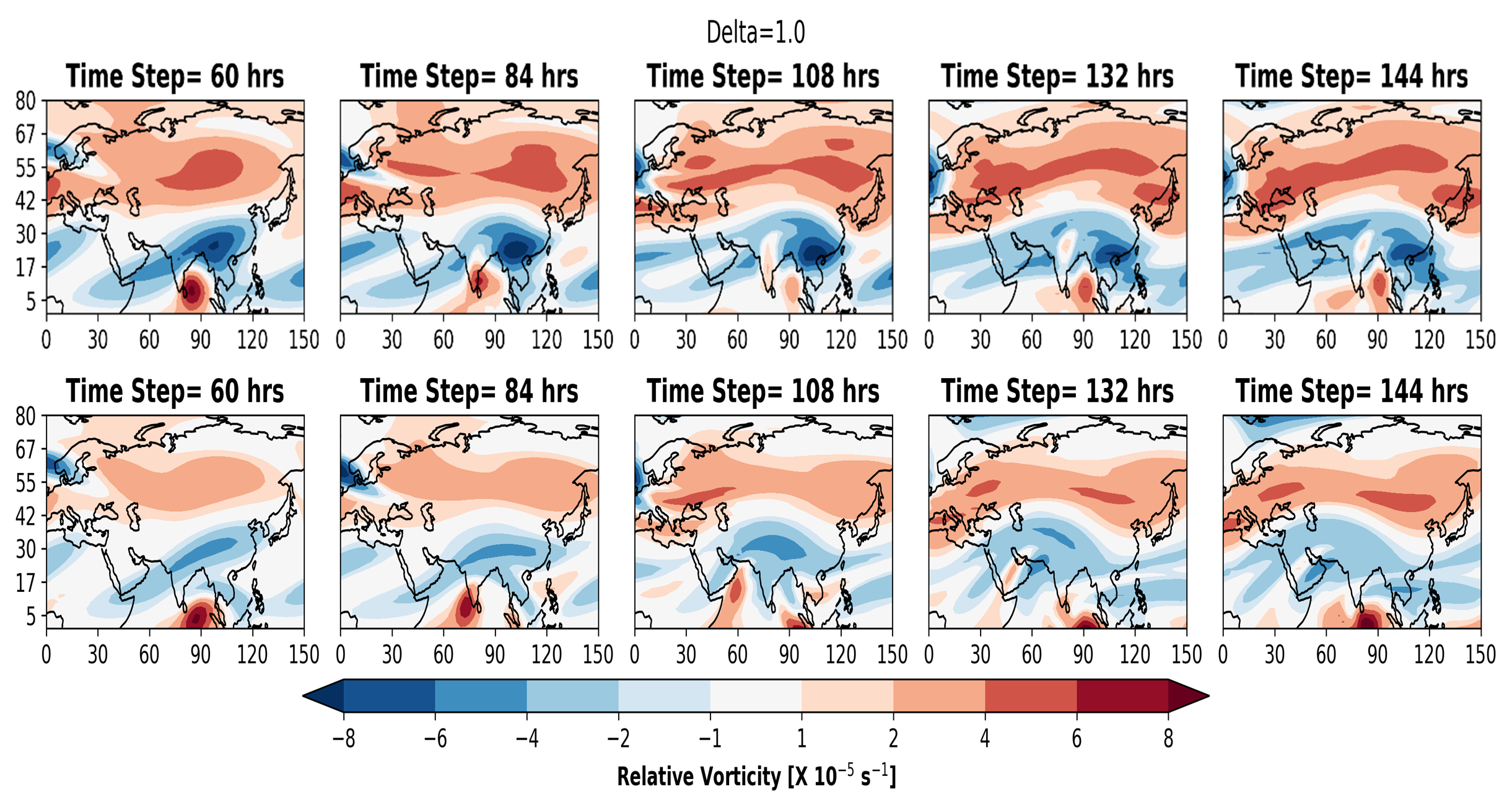}
\caption{ Same as Fig. SW-9, for $\delta$=1.}
\label{fig_wide}
\end{figure*}
\onecolumn
\section*{2.3	Discussion on Superposition   }
The figures Fig. SW-9 to Fig. SW-13 show the superposition under different phase shifts for the sum of (T+R) and intensification run, I. Let TR  be [ T(4,34) + I(4,34)]. In these runs, several forcings are present as shown in the main text (Table 1): Initial Vorticity (Z, t=0), forcing F1 and F2. Several combinations of these forcings are possible, in which models can be forced with different combinations. The spatial plots, Fig. SW-8 to Fig. SW-11, suggest superposition of waves as the model evolves. A few time steps [t=60,84,108,132,144] are shown in the above spatial plots. To what extent have these models evolved, forcing linear evolution over the Indian region?  Linear superposition, in principle, is different from Quasi-resonant amplification or blocking. To what extent is linear superposition valid?  To understand this, we plot area-averaged relative vorticity over India for model time steps in Fig. SW-13 forced with different combinations of forcings as given below:
(i)	No vorticity sources, but with the initial wave 
(ii)	Vorticity source F1 only, without the initial wave
(iii)	Vorticity source F2 only, without the initial wave
(iv)	Vorticity source F1 together with the initial wave (T(4,34) experiment)
(v)	Vorticity source F2 together with the initial wave (R (4,34))
(vi)	Vorticity source F1+F2 together with the initial wave (I(4,34)) 

The results, as shown in the plot below, show time evolution for these runs only. Also, we have given a Table to compute the correlation coefficient and RMSE for these different runs for the model time-steps.

\begin{figure*}[hbt!]
\vspace{-0.5cm}
\includegraphics[width=0.95\linewidth]{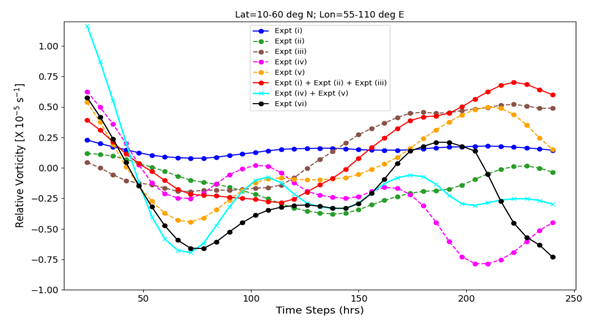}
\caption{ Comparison of temporal evolution of area-averaged vorticity for different experiments (i)-(vi) as mentioned above. Two extra combinations are added as given in the legends.}
\label{fig_wide}
\end{figure*}

\section*{\centering{Table-SW-1}}
\begin{table}[h]
\begin{center}
\captionsetup{labelformat=empty}
\caption{Correlation and RMSE of temporal evolution of all the experiments with I(4,34) (Expt.vi)}
\begin{tabular}{|l|l|l|}
\hline
Experiment &  \begin{tabular}{c}Correlation Coefficient \\ with Experiment (vi) $I(4,34)$ \end{tabular} &\begin{tabular}{c} RMSE with \\Experiment (vi) $I(4,34)$ \end{tabular}\\
\hline
(i) No vorticity sources, but with the initial wave & 0.697 & 0.475 \\
\hline
(ii) Vorticity source F1 only, without the initial wave & 0.184 & 0.343 \\
\hline
(iii) Vorticity source F2 only, without the initial wave & 0.285 & 0.515 \\
\hline
iv) Vorticity source F1 together with the initial wave $(\mathbf{T}(\mathbf{4 , 3 4})$ experiment) & 0.262 & 0.398 \\
\hline
v) Vorticity source F2 together with the initial wave ( $\mathbf{R}(\mathbf{4 , 3 4})$ ) & 0.63 & 0.376 \\
\hline
$\operatorname{Expt}(i)+\operatorname{Expt}(i i)+\operatorname{Expt}(i i i)$ & 0.388 & 0.52 \\
\hline
Expt (iv) + Expt (v) i.e. TR(4,34) $[\mathrm{T}(4,34)+\mathrm{R}(4,34)]$ & 0.74 & 0.255 \\
\hline
\end{tabular}
\end{center}
\end{table}
\section*{2.4 	Sensitivity of the amplitude of area-averaged vorticity over the Indian region (70°-100°E,20°-40°N) due to the initial stationary wave under different background conditions (jet speed and wavenumber) and comparison with other experiments.}

To what extent does the initial vorticity forcing show the amplitude response in the Indian region under different jet speed conditions? Or, in other words, is it possible for a stationary wave alone to show a very large anticyclonic vorticity response over India for different jet speeds and different wavenumbers? We provide the experiments below:\\

\textbf{(i) Experiments with different jet speeds (26,34,43m/s) with fixed wavenumber (z=4)}
SW-15 (a) and (b) show some plots from the experiments to give an idea. Panel(a) in Fig. SW-15 shows the comparison of amplitude for different experiments with different combinations involving the initial vorticity forcing $\zeta^{\prime} (z=4,t=0)$, F1, F2, and mean zonal wind ( $\overline{u}$). The blue bar shows the result when only ( $\overline{u}$)    (jmax) is present and other forcings are zero, i.e. $\zeta^{\prime}=0$, F1=0, F2=0.  This ($\overline{u}$) is the most basic background forcing in the model. As expected, a negligible amplitude response is seen (blue bar value is $\sim$0). The orange bar shows the result for  $\overline{u}$(jmax) = 34m/sec and the model is run with only with the initial vorticity spatial pattern $\zeta^{\prime}$ as given in Fig.SW-1 at t=0, z=4. It is clear that the amplitude of anticyclonic relative vorticity (orange bar) is much less when compared with other runs (Expt(ii)-(vi)), which have the same ( $\overline{u}$)=34m/s   as shown in other bars. The experiment numbers (i)-(vi) follow the previous Table-SW-1.

\begin{figure*}[hbt!]
\vspace{-0.5cm}
\includegraphics[width=0.95\linewidth]{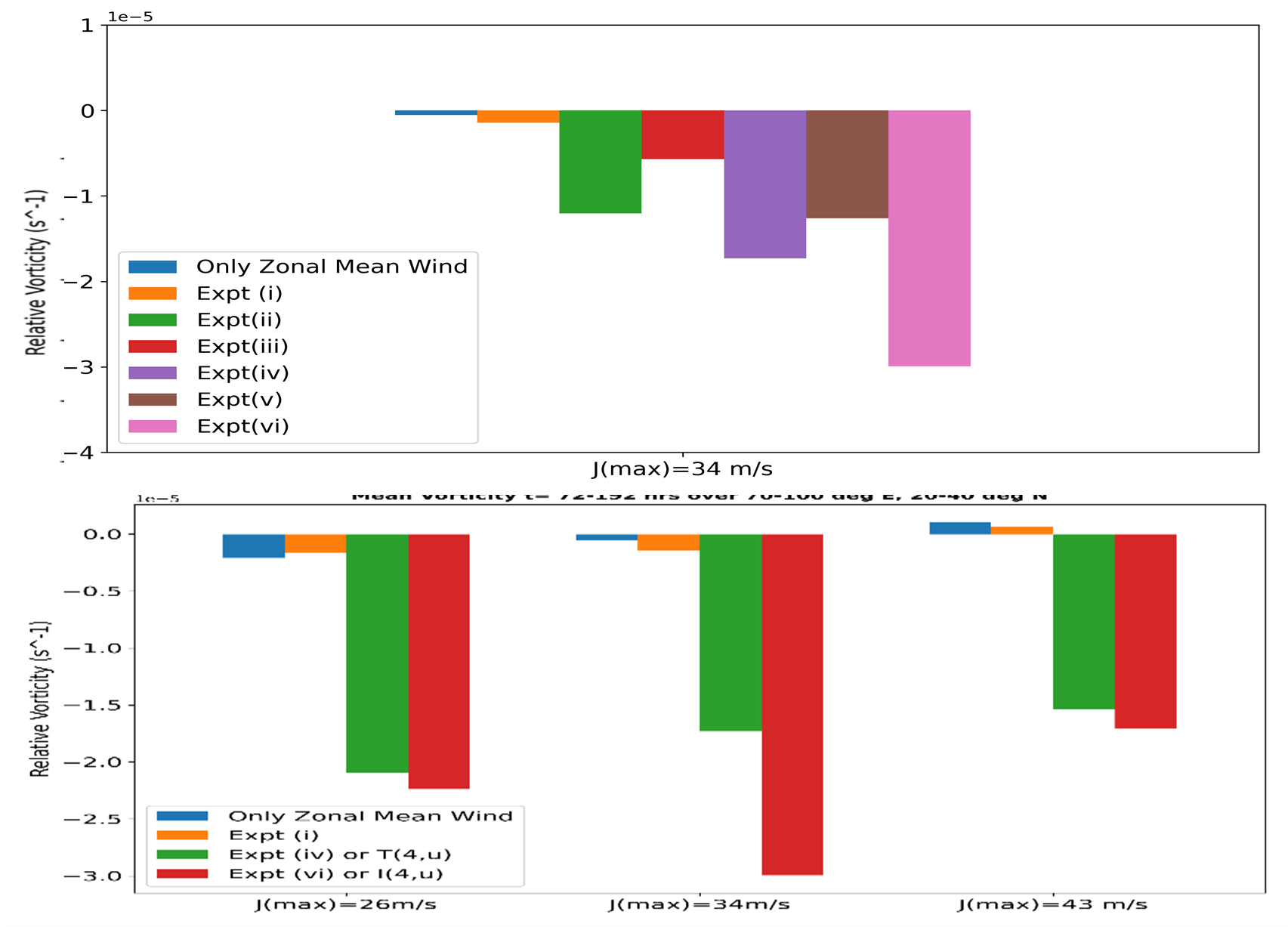}
\caption{ a) Comparison of relative vorticity amplitude over the Indian region for different forcing as discussed in sec. 3 (last section), for the same jet speed and z=4 (b) Comparison of relative vorticity amplitude of selected runs for different jet speeds (mentioned in x-axis) with z=4. Averaging Region:20°-40°N; 70°-100°E and averaging time-step: 72-192 model time step.}
\label{fig_wide}
\end{figure*}
Fig.SW-15b shows a comparison of results for different jet speeds (jmax) for the T (4,$\overline{u}$=26,34,43 m/s) runs (green bar), I (4,$\overline{u}$=26,34,43 m/s) runs (red bar) with the ($\overline{u}$=(blue bar) and ( $\overline{u}+\zeta{^\prime}(z=4,t=0)$ (orange bar) runs. Thus, we conclude from the above plots that for different jet speeds, the stationary wave contribution alone is negligible for teleconnection and intensification. \\

\textbf{(ii) Experiments with different phase shifts for different ( $\overline{u}$ (jmax) and same wavenumber z=4}\\

\begin{figure*}[hbt!]
\vspace{-0.5cm}
\includegraphics[width=0.95\linewidth]{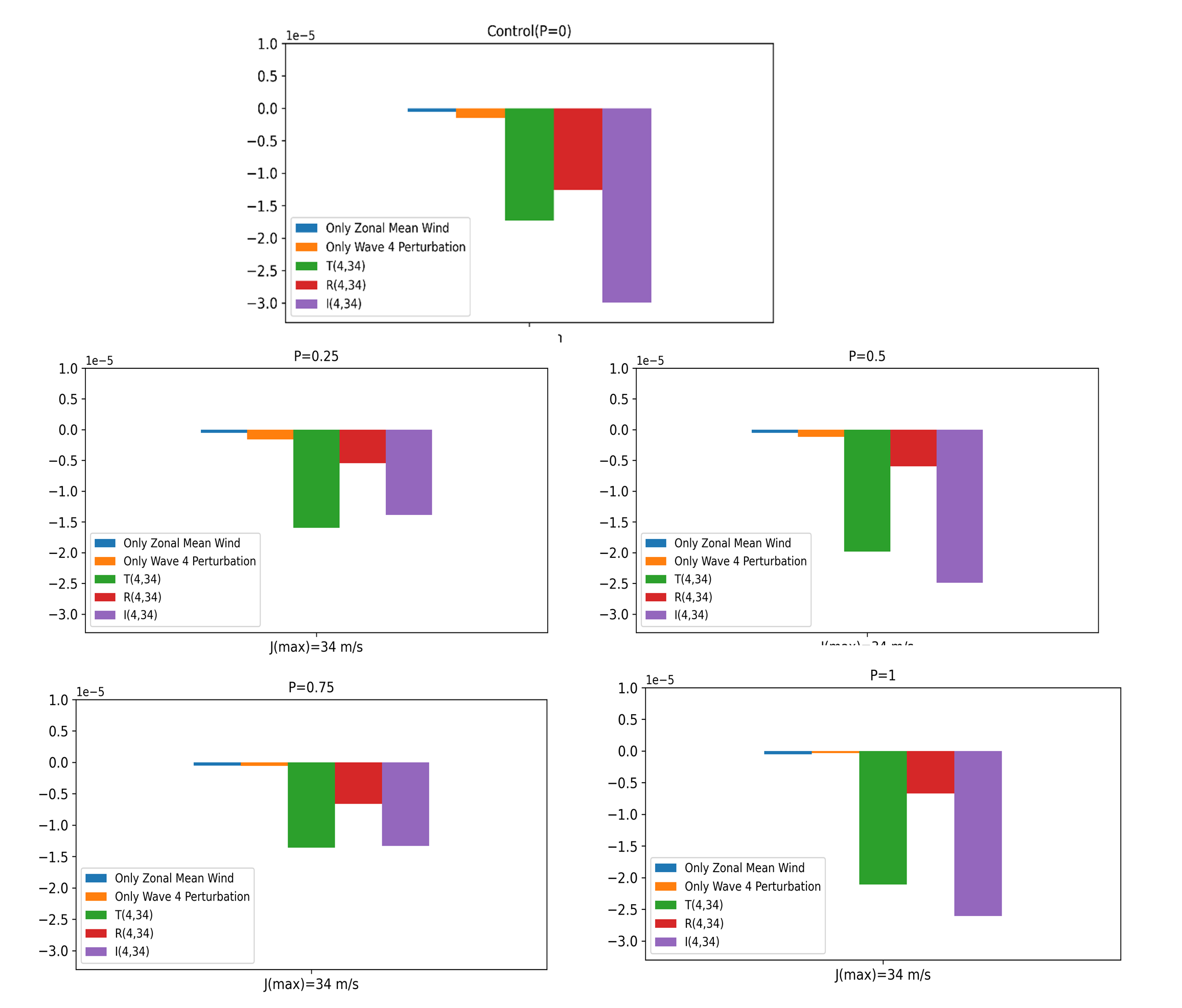}
\caption{ Figure showing the area-averaged and time-averaged relative vorticity amplitude (y-axis) of the large-scale forcing $\zeta{^\prime}$ (z=4,t=0) (orange bar) with respect to ( $\overline{u}$) forcing runs (blue), T(4,34) (green),R(4,34)(red) and I(4,34) (purple) runs for different phase ships $\delta$ =0,0.25.5.0.75, and 1. Averaging region: 70\degree-100\degree E;20\degree-40\degree N. Averaging time steps 72-144 model time steps.}
\label{fig_wide}
\end{figure*}
\twocolumn
The above plots show relative vorticity amplitude over the Indian region for different forcing runs. It may be seen that approximate linearity is valid, i.e., $T+R \approx I$ for (z=4,( $\overline{u}$=34m/s)) run over a larger area-average and time-average sense. However, this result is not general, and we find a proportionate linear response is not valid at all phase shifts (e.g. 0.25 and 0.5). For these two phases amplification occurs at different time steps outside the range time range 72-192.At a large-scale average and time-average, such linearity is valid at times. Also, results differ outside the Indian region and for other wave numbers. Intensification occurs as a result of superposition and depends on the location of the constructive interference when the phase overlap of multiple waves occurs at a domain (our model has multiple waves of different spatial scales, as there are multiple wave forcings). Those waves can cancel or reinforce in a domain. Only where reinforcement occurs in the superposition region is it likely to lead to intensification. For the I (4,34) run, the intensification as a result of superposition is over the Indian region, and it is approximately a linear sum of T+R runs for different jet speeds and phase shifts. We would like to highlight the similarity in the spatial evolution of the superposition pattern more than the magnitude of amplification, and also the possibility of non-linear interaction of waves is not entirely neglected. \\ 
	
\textbf{(iii) Experiments with different wavenumber (z=2,3,4,5) and same jet speed (34 m/s)} \\
 
We have done an experiment where we switched off the t=0 wavenumber (z=4) forcing (we refer to as no-stationary-wave or NSW experiments), i.e., we include only the other two forcings: (i) zonal mean flow background $\overline{u}$ and F1: i.e., the teleconnection experiment (which is a control experiment like in the text), and we refer to it as NSW\_T(34), and (ii) the intensification experiment,  forcing F1 and F2 superimposed on the zonal mean flow $\overline{u}$, as in the manuscript, we call it NSW\_I(34). \\
To demonstrate the importance of wave forcing at t=0, we compare intensification results discussed in the manuscript for the NSW\_T(34) and NSW\_I(34) experiments, and with the corresponding T(4,34) and I(4,34) experiments for different wavenumbers mentioned in each panel. The plot shows that in the presence of an initial large-scale forcing, the amplitude of I runs is larger than T runs, suggesting amplifications. Amplification for z=4 and 5 is prominent. Also, $NSW\_T \sim NSW\_I$  suggest that in the absence of a large-scale background at t=0, F1 and F2 don’t amplify proportionately over the Indian region for any wave number.

\begin{figure*}[hbt!]
\vspace{-0.5cm}
\includegraphics[width=0.95\linewidth]{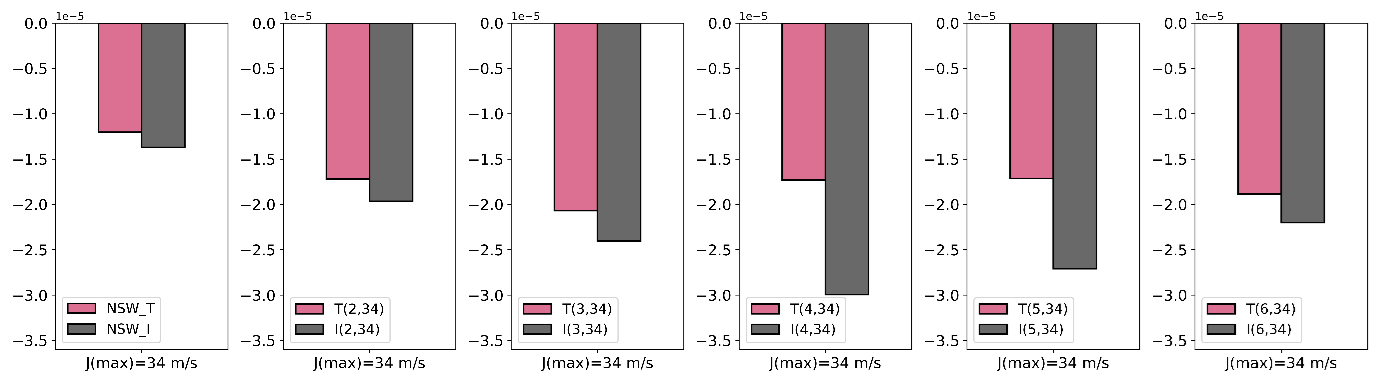}
\caption{Comparison of amplitude response for different wavenumber runs over the Indian region. Averaging region: 70\degree-100\degree E;20\degree-40\degree N. Averaging time steps 72-144 model time steps }
\label{fig_wide}
\end{figure*}

\section*{2.5 Further Discussion on Linearity   }

The barotropic vorticity equation itself is non-linear due to the presence of Jacobian and biharmonic diffusion terms. Therefore, all the runs viz. T, R, and I run evolve non-linearly in the grid-scale. The solution of this equation for T runs shows that the forcing F1 establishes the teleconnection pathway in the presence of a large-scale background z(at t=0). The forcing F2, when added (I runs), i.e., (background $\zeta^{\prime}$ (z,t=0) +F1+F2), gives intensification, as shown in Fig. 9a and Fig.9b in the main text or Fig.SW-3(right column). Fig. 9a shows the mean of the PDF shifted from $\sim$ -1 to -4 units. The T run is done following the prescription of Ambrizzi and Hoskins paper as mentioned in the text. 

Now, Fig.8b shows the explicit superposition of the signals largely from T and R experiments. Also, it may be recalled that T(z,( $\overline{u}$)) are runs with forcing F1 and $\zeta^{\prime}$ (z,t=0). R(z,( $\overline{u}$)) are runs with forcing F2 and $\zeta{^\prime}$ (z,t=0).  Intensification I (z,($\overline{u}$)) are runs with forcing F1, F2, and$\zeta{^\prime}$.  What we are asking is: To what extent is the sum of the net amplitude of response from T runs, and due to R runs, close to that of the I run?  i.e., to what extent $T+R \approx I$ ?. We are interested in this linear sum as it is important to understand if the role of Indian local forcing F2 is important in the intensification, i.e., whether it is F2 that plays a dominant role in the intensification of the teleconnection pathway, or the presence of the extratropical forcings has a more dominant impact through non-linearity?  Stated otherwise, if the remote response signal from F1 over the Indian region is proportionately intensified due to a superposition of the two oppositely propagating signals from F1 and F2. The F2 indicates a regional forcing, and a linear response is easy to interpret in a cause-and-effect sense for climate-change applications rather than non-linear impacts.

To answer the above question, we refer to Fig. SW-14 first. The cyan curve evolution, which is the amplitude of TR(4,34)=[T(4,34)+R(4,34)] run, approximately follows the black curve for  I(4,34)  run for most of the model time steps between 50-160. In Table SW 1, we quantified this more by calculating the correlation between I (4,34) and the TR (4,34) for the model time steps.  The table shows that the linear superposition sum TR (last row, marked in red) evolves more closely (high correlation and low RMSE as marked in red) with I(4,34) than other runs. Therefore, the amplitude of the superposed wave over the Indian region is close to I(4,34), and the sum evolves closely with I (4,34). Therefore, the comparability of amplitude (low RMSE) and phase (high correlation) suggests approximate linear evolution in time with amplitude superposition.
Is it valid for all combinations? i.e.:\\

 TR(z,($\overline{u}$)=[T(z,$\overline{u}$)+R(z,( $\overline{u}$)]=I(z, $\overline{u}$),for all z and  $\overline{u}$?

As shown in Fig. SW-15 and Fig.SW-16, the initial large-scale wave shows negligible contribution over India (orange Bar) as compared to T and I runs, for the cases shown with different jet speeds or wavenumber. Results suggest that the sum T+I will be approximately equal to I, when the initial wave (z) or mean wind contribution is negligible over the Indian region in terms of amplitude. However, Figs. SW-15 and SW-16 are not figures of all possible conditions. There can be many combinations of wavenumber, phase shifts, and jet speed. Also, the results differ for different areas and area-averaging.

 Irrespective of the fact that whether the evolution is linear or not, Fig.12 in the text shows the intensification heatmap for many such combinations by keeping the forcing amplitude and the area/time step averaging fixed. Some combinations of waves do not amplify. Some combinations, intensifying with large amplitudes, do not show linear superposition. Hence, the linearity or quasi-linearity is not a generalization; rather, it is a proof of the superposition of opposite propagating wave transients generated by F1 and F2, and some of those superpositions can be linear.  Whether linear or Non-linear, the results suggest an important role played by regional heating F2 in the presence of F1 and large-scale background in the heatwave intensification. This also suggests a difference in the evolution or origin of the Indian heatwave based on synoptic transients as compared to the blocking or quasi-resonance mechanism, in which the midlatitude stationary wave has a dominant role in the spatial and temporal evolution. \\
 
\section*{2.6 	Experiments with Initial Amplitude: }

The initial amplitude A0 [0.6*1E$^-5$] is prescribed in the model. It is taken from an observational study [Fig.8a in Lekshmi and Chattopadhyay, 2022, as referred in the manuscript]. How does a change in A0 change the amplification? A lot of combinations can be made. We show results for maximum jet speed =34m/s. The top panel reproduces the result in manuscript for jet speed =34m/s in Fig.12[Say CTRL experiment here]. The middle panels show the results for 0.6A0 and 0.8A0 [decreased amplitude with respect to CTRL], and the bottom panel shows the results for 1.2A0 and 1.4A0 [increased amplitude with respect to the CTRL]

\begin{figure}[hbt!]
\vspace{-1.5cm}
\hspace{-2.5cm}
\includegraphics[width=1.25\linewidth, height=0.6\textheight]{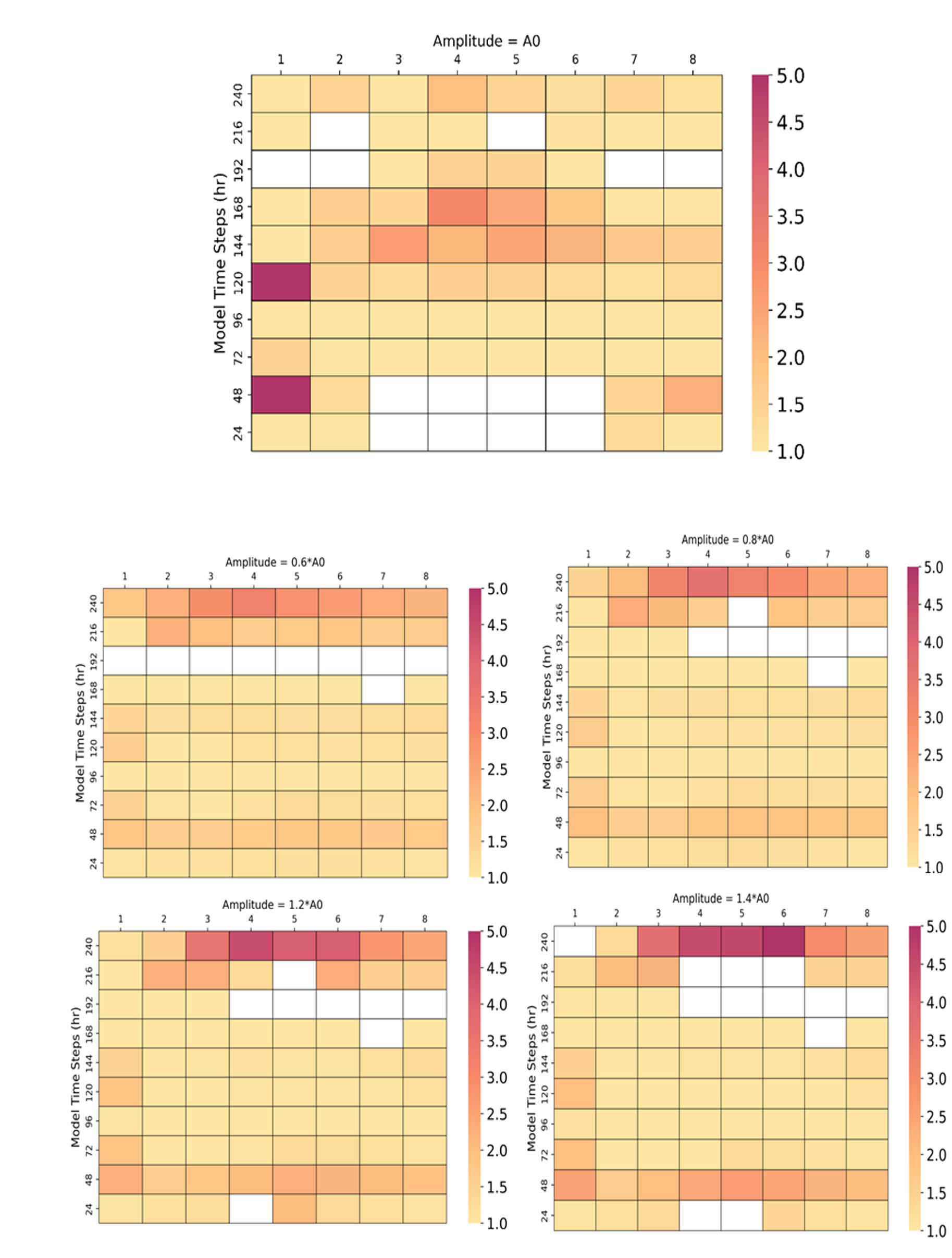}
\caption{Same as Fig. 12 in the main text, but for a different amplitude of the initial wave Z(t=0) with max jet wave speed 34. The top panel shows the result for the case with A0 taken in the main experiment as in Fig.12. }
\label{fig_wide}
\end{figure}

\end{document}